\documentclass[conf]{new-aiaa}

\title{Buffet Alleviation via Linear Stability Adjoint}

\author{Rohit Sunil Kanchi\footnote{PhD Candidate, Department of Mechanical and Aerospace Biomedical Engineering, AIAA Member.}}
\affil{University of Tennessee, Knoxville, United States}
\author{Sicheng He\footnote{Assistant Professor, Department of Mechanical and Aerospace Engineering, AIAA Member.}}
\affil{University of Tennessee, Knoxville, United States}
\author{Eirikur Jonsson\footnote{Postdoc, Department of Aerospace Engineering, AIAA Member.}}
\affil{University of Michigan, Ann Arbor, Michigan, United States}
\author{Joaquim R. R. A. Martins\footnote{Pauline M. Sherman Collegiate Professor, Department of Aerospace Engineering, AIAA Fellow.}}
\affil{University of Michigan, Ann Arbor, Michigan, United States}

\usepackage{fullpage}
\usepackage{xcolor}
\usepackage{commath}
\usepackage{amsmath}
\usepackage{amstext}
\usepackage{graphicx}
\usepackage{subcaption}
\usepackage{url}
\usepackage{varioref}%  smart page, figure, table, and equation referencing
\usepackage{threeparttable}% tables with footnotes
\usepackage{dcolumn}%   decimal-aligned tabular math columns
\newcolumntype{d}{D{.}{.}{-1}}
\usepackage{doi}
\usepackage{bm}
\usepackage{esdiff}
\usepackage{booktabs}
\usepackage{multirow}
\usepackage{float}
\usepackage{longtable,tabularx}
\usepackage{placeins}
\usepackage{afterpage}
\usepackage{dirtytalk}
\usepackage{mathtools}
\usepackage{tikz}
\usepackage{algorithm}
\usepackage{algpseudocode}
\usepackage{bbding}
\usepackage{pifont}
\usepackage{wasysym}
\usepackage{cleveref}
\microtypesetup{expansion=false}
\usetikzlibrary{shapes,arrows}
\usetikzlibrary{positioning}
\newcommand{\f}[2]{\frac{#1}{#2}}
\newcommand{\mb}[1]{\mathbf{#1}}

\DeclareMathAlphabet\mathbfcal{OMS}{cmsy}{b}{n}
\renewcommand{\d}{\mathop{}\!\mathrm{d}} % total derivative
\newcommand{\p}{\partial}
\usepackage{hyperref}
\colorlet{mblue}{blue!40!black}
\hypersetup{bookmarksnumbered=true,colorlinks=true,linkcolor=mblue,citecolor=mblue,urlcolor=mblue}

\usepackage{soul}
\definecolor{bananamania}{rgb}{0.98, 0.91, 0.71}
\sethlcolor{bananamania}
\soulregister\cite7
\soulregister\citep7
\soulregister\citet7
\soulregister\ref7
\soulregister\eqnref7
\soulregister\eqref7

\begin{document}

\maketitle

\begin{abstract}
Transonic buffet, self--sustained shock and shear--layer oscillations, imposes hard limits on the cruise envelope of modern transport aircraft, and avoiding it is a primary design driver.
State-of-the-art buffet-onset criteria used in design, such as the $\Delta\alpha = 0.1^\circ$ criterion and separation--sensor methods, are empirical surrogates rather than first--principle predictors, and can yield either overly conservative or unsafe designs.
Linear stability analysis (LST) predicts buffet onset directly from the spectrum of the linearized operator about the steady base flow, but using it as an aerodynamic shape optimization constraint has been bottlenecked by the cost of differentiating an eigenvalue with respect to many design variables.
In this paper, we develop a coupled adjoint method that efficiently computes the sensitivity of the dominant LST eigenvalue with respect to a large number of shape design variables, by reusing the steady CFD adjoint within a top and bottom level decomposition of the eigenproblem.
We verify the eigensolver and adjoint against the canonical cylinder vortex--shedding benchmark, then verify the LST predictions on the OAT15A supercritical airfoil at $M=0.73$, $Re=3.2\times 10^{6}$ against published eigenspectra and against the linear growth phase of a URANS run.
Using the resulting gradients, a single-point buffet-constrained drag minimization of the OAT15A achieves a $22.4\%$ drag reduction while satisfying the LST-based buffet constraint.
Finally, we present preliminary three-dimensional results on the wing only NASA common research model (CRM) at $M=0.85$, $Re=5\times 10^{6}$, recovering buffet onset at $\alpha \approx 4.0^\circ$ from a sweep of warm--started URANS runs and providing a stepping stone toward three-dimensional buffet-constrained wing optimization with the present adjoint.
\end{abstract}

\section{Nomenclature}
 {\renewcommand\arraystretch{1.0}
  \noindent\begin{longtable*}{@{}l @{\quad=\quad} l@{}}
      \multicolumn{2}{@{}l}{\emph{Optimization}} \\
      $\mb{x}$                          & vector of design variables \\
      $\mb{f}(\mb{x})$                  & generic objective function \\
      $\mb{g}(\mb{x})$                  & generic constraint \\
      $\alpha$                          & angle of attack \\
      $\chi$                            & Kenway--Martins separation metric \\
      $\beta$                           & buffet factor \\
      \multicolumn{2}{@{}l}{} \\
      \multicolumn{2}{@{}l}{\emph{Flow state and residuals}} \\
      $\mb{w}$                          & flow state vector (primitive variables) \\
      $\mb{w}_0$                        & steady (equilibrium) state \\
      $\mb{U}(\mb{w})$                  & conservative state vector \\
      $\mb{r}$                          & bottom-level (CFD) residual \\
      $\hat{\mb{r}}$                    & top-level (eigenvalue problem) residual \\
      $\mb{J}$                          & RANS Jacobian, $\partial\mb{r}/\partial\mb{w}_0$ \\
      $\mb{M}$                          & mass matrix, $\partial\mb{U}/\partial\mb{w}$ \\
      \multicolumn{2}{@{}l}{} \\
      \multicolumn{2}{@{}l}{\emph{Linear stability}} \\
      $\lambda$                         & complex eigenvalue \\
      $\lambda_r,\,\lambda_i$           & real and imaginary parts of $\lambda$ \\
      $\mb{q}$                          & complex eigenvector \\
      $\mb{q}_r,\,\mb{q}_i$             & real and imaginary parts of $\mb{q}$ \\
      $\mb{v}$                          & top-level state, $[\mb{q}_r^\intercal,\mb{q}_i^\intercal,\lambda_r,\lambda_i]^\intercal$ \\
      $\mb{e}_k$                        & $k$-th canonical basis vector \\
      $\sigma$                          & spectral shift (shift--invert) or growth rate (in physical units) \\
      $\omega$                          & angular frequency \\
      $S_L$                             & Strouhal number \\
      $\delta\rho$                      & density component of an LST eigenmode \\
      \multicolumn{2}{@{}l}{} \\
      \multicolumn{2}{@{}l}{\emph{Eigensolver}} \\
      $\mb{T}$                          & Cayley/Crank--Nicolson propagator \\
      $\mb{P}$                          & shifted operator $\mb{M} + (\Delta t/2)\mb{J}$ \\
      \multicolumn{2}{@{}l}{} \\
      \multicolumn{2}{@{}l}{\emph{Aerodynamics}} \\
      $C_L,\,C_D,\,C_M$                 & lift, drag, and moment coefficients \\
      $C_p$                             & pressure coefficient \\
      $M$                               & freestream Mach number \\
      $Re$                              & freestream Reynolds number \\
      $T_\infty$                        & freestream static temperature \\
      $\rho,\,p,\,T$                    & density, pressure, temperature \\
      $\tilde{\nu}$                     & Spalart--Allmaras working variable \\
      \multicolumn{2}{@{}l}{} \\
      \multicolumn{2}{@{}l}{\emph{Acronyms}} \\
      AD                                & algorithmic differentiation \\
      ANK/NK                            & approximate Newton--Krylov / Newton--Krylov \\
      CFD                               & computational fluid dynamics \\
      CN                                & Crank--Nicolson \\
      CRM                               & Common Research Model \\
      CSRAD                             & complex--step reverse algorithmic differentiation \\
      EVP, GEVP                         & (generalized) eigenvalue problem \\
      FFD                               & free--form deformation \\
      LCO                               & limit--cycle oscillation \\
      LST                               & linear stability theory \\
      RANS, URANS                       & (unsteady) Reynolds--averaged Navier--Stokes \\
      SLSQP                             & sequential least--squares quadratic programming \\
  \end{longtable*}}

  % \newpage
\section{Introduction}
\label{sec:introduction}

%%% Buffet phenomenon and why it matters for aircraft design
The occurrence of transonic buffet arises from the separation of airflow induced by shock waves, typically found at the base of these waves.
As the lift coefficient or Mach number increases, the intensity of shock waves amplifies, leading to the gradual development of buffet. 
The point at which buffet first manifests is referred to as buffet onset. 
Buffet is undesirable due to its adverse effects on aircraft control, passenger comfort, and structural integrity. 
Regulatory bodies such as the Federal Aviation Administration (FAA) and the European Union Aviation Safety Agency (EASA) mandate a 30\% lift margin from the cruise operating condition to the buffet-onset boundary for commercial aircraft. 
This safety margin allows aircraft to navigate and handle turbulence effectively. 
Consequently, quantifying buffet onset and incorporating corresponding constraints in aerodynamic shape optimization becomes imperative, as recently illustrated by the NASA/Boeing Transonic Truss--Braced Wing buffet--onset study of~\citet{Browne2025}.
% [x] TODO HS-: add ref.

%%% Survey of buffet simulation methods (sets up the comparison table)
It is essential to accurately and efficiently predict buffet onset and incorporate this information in aircraft design.
There are four methods to simulate buffet: (1). Time accurate method, (2). Time spectral method (or harmonic balance method), (3). Linear stability analysis, and (4). Heuristic methods.
We review the methods and the status of design optimization using these methods.
Examining buffet directly often requires employing high-fidelity, time-accurate simulations utilizing computational fluid dynamics (CFD), such as unsteady Reynolds-averaged Navier--Stokes (URANS) \cite{Thiery2005,Iovnovich2012a,Iovnovich2015}, hybrid RANS/LES approaches such as ZDES \cite{Plante2019}, and large-eddy simulation (LES) \cite{Dandois2018,Song2025}.
A representative recent example on a full aircraft is the Boeing Transonic Truss--Braced Wing study of~\citet{Browne2025}, in which URANS and zonal--DES (HRLES) simulations on the same configuration produced buffet--onset predictions in close agreement with each other and with the NASA Ames 11-by-11 ft transonic wind--tunnel data, illustrating that scale--resolving and URANS approaches can both be used to actually solve the buffet problem on practical configurations.
However, these methods prove excessively computationally demanding when applied to aircraft shape optimization tasks, as simulations may need to be executed numerous times, even for a single-point optimization problem.
In efforts to mitigate computational expenses, researchers have been exploring alternative approaches that assess buffet onset through steady CFD simulations.
Buffet has been simulated using time--accurate URANS, including the early forced--motion oscillating--airfoil studies of Raveh~\cite{Raveh2009aa}, the three--dimensional swept--wing simulations of Iovnovich and Raveh~\cite{Iovnovich2015}, the modal--analysis decomposition of two--dimensional shock--buffet by Poplingher, Raveh, and Dowell~\cite{Poplingher2019}, and more recent unsteady simulations~\cite{Church2019}.
It was demonstrated that the URANS-based simulation can predict buffet onset with high accuracy.
However, the computational cost of URANS is high and to conduct design optimization requires the development of unsteady adjoint, which can require significant development effort and incur high memory cost.
Besides, buffet is a Hopf bifurcation problem and the stability of the bifurcation can also be an important design characteristics. 
The URANS-based methods are not able to predict an unstable bifurcation (or subcritical bifurcation).

%%% Method 2: Time-spectral / harmonic balance --- still costlier than LST
% [] TODO HS-: high level ts+ lco then focus on buffet.
Time-spectral methods have also been used for periodic unsteady aerodynamic simulations relevant to buffet and post-buffet response \cite{Thomas2020,He2021c}.
The time spectral method is more efficient than the time accurate method in simulating periodic problem.
In addition, it can also be used to predict subcritical response \cite{Thomas2002nx,Li2018c,He2021c}.
The equation is solved using Newton method which requires an initial solution that is in the convergent domain of the Newton solver.
The initial solution can be obtained by solving a linear stability equation \cite{Thomas2002nx}.
Thus, its computational cost is usually higher than the linear stability equation-based solver.

%%% Method 4: Heuristic and data-driven buffet constraints (Kenway-Martins, Li) and their limits
In the industry, an alternative heuristic approach to buffet-onset analysis relies solely on global aerodynamic coefficients, such as the lift-curve break method \cite{Li2022c}.
This method can be implemented using the $\Delta\alpha = 0.1^\circ$ formulation, where the buffet-onset point is estimated by the intersection of the lift curve and an auxiliary line formed by offsetting the linear portion of the lift curve to the right by $0.1^\circ$.
Despite being computationally more economical than the URANS-based method, it still incurs a substantial cost, requiring a few RANS simulations for buffet onset evaluation.
To make the constraint compatible with gradient--based shape optimization, \citet{Kenway2017b} proposed a separation--based metric $\chi$ that approximates the $\Delta\alpha = 0.1^\circ$ buffet boundary using a single steady RANS solution and is differentiable through ADflow, and this Kenway--Martins formulation has since become the de facto state--of--the--art buffet--onset constraint in CFD--based aerodynamic shape optimization in both ADflow and SU2.
\citet{Li2022c} further reduces the cost and improves generalization across airfoil and wing shapes by training a physics--based data--driven model on the surface pressure and friction distributions, mapping the shock and separation features directly to a buffet factor.
These heuristic and data--driven constraints are attractive for their low cost and have enabled large--scale wing and aircraft shape optimizations, but they are not derived from first principles: the separation--metric cutoff, the auxiliary--line offset, and the training--data domain are all fitted to a particular class of geometries and flow conditions, and they do not in general generalize to novel configurations or off--design points.
Furthermore, none of these heuristics resolve the buffet onset as a Hopf bifurcation of the underlying mean flow, so they cannot distinguish a stable design from a marginally unstable one and they provide no information about the buffet frequency or mode shape.
% [x] TODO HS-: thorough check -: - connect regular two words e.g. off--design, -- connnect two names, Navier--stokes. You need to teach claude this...

%%% Method 3: Linear stability analysis --- our chosen first-principles route
A first--principles alternative is to view buffet as a global instability of the steady mean flow~\cite{Crouch2009,Sartor2015} and to track its onset through the rightmost eigenvalue of the RANS Jacobian linearized about the mean flow.
Recently, this linear stability equation has been solved at large scale using shift--and--invert Krylov methods to simulate buffet onset on full--aircraft configurations including the NASA CRM wing~\cite{Timme2020a}, and the algorithm has been shown to capture buffet onset reliably on realistic geometries.
As mentioned above, the computational cost of linear stability analysis is expected to be lower than time accurate and time spectral methods.
Although the linear stability method is not able to capture the subcritical buffet, by including higher order coefficients (second and third order Jacobian matrix), the subcritical response can be captured \cite{Kuznetsov2004} and it can be efficiently used for optimization \cite{He2026z}.
Due to its low computational cost, we select linear stability method to simulate buffet onset.
A detailed comparison of different methods is given in \Cref{tab:methods}.
% [x] TODO HS-: thorough check on cref for table etc. ,make sure consistent with the paper. table usually is upper case in AIAAJ?

% [] TODO HS-; the logic flow here is a bit strange... Broadly the review should be about: 1. modeling, 2. optimization then under optimization discuss sensitivity analysis. sens analysis should be discussed with more details--explain current fd, direct work and also the adjoint work--explain there limitation in detail. then our work. 

%%% Sensitivity bottleneck: prior eigenvalue-adjoint work (Marquet, Mettot, Heuveline) and its limitations
One major bottleneck for the adoption of large scale design optimization using the linear stability analysis is the efficient sensitivity computation.
Marquet et al.~\cite{Marquet2008} introduced a sensitivity framework that computes the derivative of a linear--stability eigenvalue with respect to base--flow modifications, originally demonstrated on the cylinder vortex--shedding problem.
Mettot et al.~\cite{Mettot2014} subsequently recast the same idea in a fully discrete framework that operates directly on the linearized residual Jacobian, and the resulting discrete machinery has been used to compute eigenvalue sensitivities for the transonic buffet problem~\cite{Sartor2015}.
Both works build a sequential adjoint that computes the derivative in two stages:
First, compute the derivative of the eigenvalue with respect to the Jacobian matrix entries.
Then, compute the derivative of the nonlinear equation at the equilibrium point.
The proposed algorithm scales well with the number of design variables.
However, the mathematical derivation uses two simplifications:
First, the design variable (the forcing term) only affects the underlying nonlinear system.
A more general design variable that affect both the underlying nonlinear system and the eigenvalue problem can not be handled by this approach.
Second, the function of interest is limited to eigenvalues.
In some applications, such as the laminar-turbulent transition flow modeling, the eigenvector (mode shape) derivatives is required~\cite{Shi2020a}.
An early demonstration of using eigenvalue--based stability as a shape--optimization constraint was given by Heuveline and Strauß~\cite{Heuveline2009}, who performed gradient--based shape optimization of a two--dimensional channel--with--body benchmark with the rightmost eigenvalue of the linearized Navier--Stokes operator as the stability measure.
Their eigenvalue gradients were computed by \emph{finite differences} rather than by an adjoint, requiring one additional generalized eigensolve per design variable~\cite[Sec.~4.2 and Algorithm~2]{Heuveline2009}, which limited the demonstration to a handful of B\'ezier shape parameters and is precisely the bottleneck that an adjoint approach removes.
Thus, in contrast to prior work that uses these discrete sensitivity frameworks for open--loop control maps based on base--flow or source--term perturbations, we propose to extend the methods to embed the buffet--eigenvalue sensitivity directly into a gradient--based aerodynamic shape--optimization problem in which the design variables modify the airfoil/wing geometry and angle of attack and the stability constraint is enforced simultaneously with the aerodynamic performance constraints.

%%% Our approach: hybrid coupled + sequential adjoint (linear stability coupled)
Alternatively, the coupled adjoint equation~\cite{Kenway2014a} can be used to compute the derivative of general design variables and functions of interest.
This method combines all the individual coupled equations into a system of equations, and a global adjoint equation is constructed to compute the derivatives.
However, blindly applying the coupled adjoint method will result in a high computational cost.
Recently, \citet{Boulle2022b} proposed using the continuous coupled adjoint to solve the derivative extending the results of their previous paper \cite{Boulle2022a}.
In this paper, we combine the idea of the coupled adjoint method and the sequential adjoint method in the following way:
The coupled adjoint method provides the global adjoint equation that can be used to compute the derivative for general design variables and functions of interest.
Then, leveraging the feed-forward structure of the residuals, we use an approach similar to the sequential adjoint to solve two individual adjoint equations to obtain a solution to the coupled adjoint equation.
To extend the method to eigenvector derivative computation, we apply the general adjoint equation developed by He et al.~\cite{He2022ua}.

%%% Novelty: position the contribution between buffet stability analysis and buffet shape optimization
% [] TODO HS-: this paragraph should be before Our approach?
On one hand, direct and adjoint global stability analyses have been applied extensively to transonic buffet, identifying the dominant unstable mode, its frequency, and its spatial sensitivity structure on two--dimensional airfoils~\cite{Iorio2014,Sartor2015,Paladini2019b} and on swept wings and full--aircraft configurations~\cite{Paladini2019c,Crouch2019a,Sansica2023}; and for surface--geometry perturbations~\cite{MartinezCava2020}.
These studies establish that the buffet eigenvalue carries a meaningful sensitivity structure, but the sensitivity fields are typically interpreted to explain the physics or to suggest passive/active control mechanisms rather than embedded inside a full aerodynamic shape optimization loop.

On the other hand, aerodynamic shape optimization for buffet suppression has been pursued through several routes that avoid the global--stability eigenvalue: separation-- and lift--curve--break heuristics applied as differentiable steady--RANS constraints~\cite{Kenway2017b,Li2022c}, unsteady--adjoint formulations that minimize lift--fluctuation amplitudes~\cite{Chen2022}, and surrogate-- or data--driven searches over design space~\cite{Xu2019,Gong2024,Ma2026}.
These approaches demonstrate that buffet behavior can be modified by shape, but they either avoid the global--stability eigenvalue entirely, rely on long unsteady simulations during optimization, or use surrogate stability proxies rather than a fully gradient--based eigenvalue--constrained formulation.
The present work closes this gap by using the real part of the rightmost LST eigenvalue as a differentiable design constraint and by supplying the optimizer with its sensitivity to shape and angle--of--attack design variables through the proposed linear stability coupled adjoint in ADflow~\cite{Mader2020a}; to our knowledge, this is among the first aerodynamic shape optimization frameworks to use adjoint eigen--sensitivity of a buffet--relevant global mode as a gradient--based buffet constraint for transonic airfoil design.

% [x] TODO HS-: merge to where it fiurst discussed, and put it near the end of THAT paragraph.

%%% Numbered list of contributions
In lieu of these research gaps, we summarize our contributions as follows:
1) We extend the result of~\citet{Mettot2014} to consider arbitrary design variables.
This is the first scalable adjoint based linear stability coupled eigenvalue derivative computation and optimization performed on a transonic buffet case;
2) We propose the methods for optimization using the linear stability constraint which is fundamental and from first principles, instead of a fully time--accurate method, time--spectral method or heuristic buffet constraints;
3) In addition to the eigenvalue derivative, the eigenvector derivative is also taken into account.
We apply a more general eigenvalue problem adjoint proposed by He et al.~\cite{He2022ua} that can compute both the eigenvalue and eigenvector derivatives in a scalable way;
4) We propose an algorithmic differentiation (AD) and complex-step~\cite{Martins2003a} mixed method to compute the derivative of the Jacobian-vector product.
This combination gives machine-precision accuracy and is superior to the less accurate finite differences-based method found in the literature~\cite{Heuveline2009};
5) We demonstrate the linear stability coupled adjoint end--to--end on a single--point buffet--constrained drag minimization of the OAT15A airfoil that achieves $22.4\%$ drag reduction while satisfying the LST--based buffet constraint.

%%% Roadmap of the paper, merged with the OAT15A demonstration result and the CRM-wing preview
The paper is organized as follows.
In~\Cref{sec:analysis} we present the residual form for the equilibrium point solution and the linear stability problem; in~\Cref{sec:ls_adjoint} we derive the proposed linear stability coupled adjoint for the sensitivity of the rightmost eigenvalue with respect to arbitrary design variables; and in~\Cref{sec:solver} we describe the eigenvalue solver used for the LST: the shift--invert spectral transformation and the Cayley/Crank--Nicolson time--stepper Arnoldi method, including the mode--selection criterion, the inner linear solver, and the algorithms.
Results are then presented in~\Cref{sec:results}: a cylinder vortex--shedding benchmark with a Hopf--bifurcation sweep against literature, the OAT15A transonic--buffet case with a steady RANS grid--convergence study, an unsteady simulation, an LST eigenspectrum sweep against published data and a verification of the dominant LST eigenvalue against the linear--growth phase of the URANS run, a CFD adjoint derivative verification on the cylinder, and a single--point gradient--based buffet--constrained drag minimization of the OAT15A airfoil at $M = 0.73$, $Re = 3.2\times 10^{6}$ that demonstrates the full pipeline end--to--end.

\section{Linear stability analysis}
\label{sec:analysis}

In this section, we describe the analysis of both the underlying nonlinear dynamical system equations (the \emph{bottom-level} problem) and the linear stability equation of the tangential dynamical system (the \emph{top-level} problem) evaluated at a solution of the bottom-level problem.
Then we combine these two blocks to obtain the governing equations for stability analysis.
Finally, we present the forms of functions that can be evaluated by this method.

% [] TODO HS-: https://par.nsf.gov/servlets/purl/10629984 we use upper and lower level to describe it check. Also check how we name the adjoint method.

\subsection{Bottom-level}
We consider problems of the form
\begin{equation}
\label{eq:dynamic_governing}
\f{\p {\mb{w}}}{\p t} + \mb{r}(\mb{w}; \mb{x}) = 0,
\end{equation}
where $\mb{r}\in\mathbb{R}^{n}$ is the residual, $\mb{w}\in\mathbb{R}^{n}$ is the state variable, $\mb{x}\in\mathbb{R}^{n_{\mb{x}}}$ is the design variable, $n$ is the problem size, and $n_{\mb{x}}$ is the design variable dimension.
The semi-colon in the residual form $\mb{r}(\mb{w}; \mb{x})$ distinguishes state variables from the design variables and other parameters.
For an equilibrium point, $\mb{w}_0$, we have
\begin{equation}
\label{eq:steady_governing}
\mb{r}(\mb{w}_0; \mb{x}) = 0,
\end{equation}
where $\mb{w}_0$ is no longer a function of time.

\subsection{Top-level}
The equilibrium point solution Eq.~\eqref{eq:steady_governing} is usually sufficient for engineering application.
However, there are problems that we must consider the time derivative, e.g., the buffet onset problem encountered in aircraft wing design.
Specifically, in some problems, in addition to the equilibrium point performance, we need to consider the stability of the equilibrium point.
This is captured by the eigenvalue of the following eigenvalue problem.

First, we assume that the solution takes an infinitesimal perturbation around the equilibrium point  $\mb{w}_0$, so we can write
\begin{equation}
\label{eq:w_expansion}
\mb{w} = \mb{w}_0 + \epsilon \mb{q}e^{\lambda t} + \mathcal{O}(\epsilon^2),
\end{equation}
where $\mb{q}\in \mathbb{C}^{n}$ is a normalized eigenvector, $\lambda\in\mathbb{C}$ is the eigenvalue, and $\epsilon$ is the perturbation magnitude.
The eigenvalue and the eigenvector are from the Jacobian of the dynamical system Eq.~\eqref{eq:dynamic_governing} at the equilibrium point, $\mb{w}_0$.
There are multiple eigenvalue and eigenvector pairs available.
Because the real part of the eigenvalues determines the stability of the linearized dynamical system, We choose the one whose eigenvalue has the largest real part.

The eigenvector is normalized for both its magnitude and phases
\begin{equation}
\label{eq:normalization}
\begin{aligned}
\mb{q}^* \mb{q} &= 1, \\
\mb{e}_k^\intercal \mb{q} &= 0,
\end{aligned}
\end{equation}
where ``$*$'' is the complex adjoint operator and $k$ is the index of the entry with the largest magnitude, i.e.,
\begin{equation}
k = \mathrm{argmax}_{l} \norm{\mb{q}[l]}_2,
\end{equation}
where $\mb{q}[l]$ denotes the $l^{\mathrm{th}}$ entry of $\mb{q}$.
The vector $\mb{e}^k$ has all zero entries besides the $k^{\mathrm{th}}$ entry that is equal to $1$.
The normalization condition is not unique.

Inserting Eq.~\eqref{eq:w_expansion} into Eq.~\eqref{eq:dynamic_governing} and linearizing about the equilibrium point, neglecting high order terms for $\epsilon$, and using the fact that $\mb{r}(\mb{w}_0; \mb{x})=0$, we obtain,
\begin{equation}
\label{eq:eigen_govern}
\f{\p \mb{r}}{\p \mb{w}_0} (\mb{w}_0; \mb{x})\mb{q} = \lambda \mb{q}.
\end{equation}
Thus, the information passed from the bottom to the top level is the Jacobian matrix,
\begin{equation}
\mb{J} = \f{\p \mb{r}}{\p \mb{w}_0} (\mb{w}_0; \mb{x}),
\end{equation}
where $\mb{J}\in\mathbb{R}^{n \times n}$.

Since the eigenvalue and eigenvector are complex, the complex equations Eqs.~\eqref{eq:eigen_govern} and \eqref{eq:normalization} can be decomposed into real and imaginary components.
The resulting real system of equations is written as
\begin{equation}
\label{eq:eigen_govern_expanded}
\hat{\mb{r}}(\mb{v}; \mb{w}_0, \mb{x})
=
\hat{\mb{r}}(\mb{v}; \mb{J}(\mb{w}_0; \mb{x}))
=
0,
\end{equation}
where the state variable $\mb{v} = [\mb{q}_r^\intercal, \mb{q}_i^\intercal, \lambda_r, \lambda_i]^\intercal$ collects the real and imaginary parts of the eigenvector and eigenvalue.
The detailed expansion is given in~\Cref{app:real_formulation}.

For the solution to be locally stable, we need to enforce that
\begin{equation}
\lambda_r < 0.
\end{equation}
For more details about the linear stability criterion, we refer the reader to Section 3.3 of the textbook by Slotine and Li~\cite{Slotine1991}.
Recall that the eigenvalue with the largest real part is the one we consider here.
Computing this eigenvalue efficiently is still a challenging problem.
This is because most iterative solvers (e.g., the Lanczos method) can only efficiently compute the eigenvalue with the largest magnitude that is not necessarily the one with the largest real part.
The solution to this special eigenvalue problem is not the focus of the paper; there are various other references that detail methods for computing this eigenvalue~\cite{Elman2012,Timme2012a,Timme2020a}.
Also, the equation does not track the Hopf bifurcation onset point but rather only gives information about whether the solution is stable.
For the Hopf onset point solution method, see the paper by Roose and Hlava{\v{c}}ek~\cite{Roose1985}.

\subsection{Governing equation for the linear stability analysis}
Combining the bottom and the top-level residual forms, we have the governing equation for the linear stability problem
\begin{equation}
\begin{aligned}
\mb{r}(\mb{w}_0; \mb{x}) &= 0, \\
\hat{\mb{r}}(\mb{v}; \mb{w}_0, \mb{x}) &= 0. \\
\end{aligned}
\end{equation}
There is no feedback from the top level to the bottom level.
Later, we will leverage this observation to avoid a coupled adjoint solution.

\subsection{Function of interest evaluation}
The formulation is general because we can select any function of interest $f$.
We define the general form of $f$ that we can evaluate and compute derivatives for as
\begin{equation}
f = f\left(\mb{v}, \mb{w}_0; \mb{x}\right).
\end{equation}

Specifically, when computing the linear stability, the real part of the eigenvalue is of interest; that is,
\begin{equation}
\label{eq:obj_eigval}
f = \lambda_r.
\end{equation}
Further, in some problems, we need to evaluate and compute the derivative of a function with respect to the eigenvectors.
For example, we may need to compute the weighted eigenvector derivative where the function of interest is defined as
\begin{equation}
\label{eq:obj_eigvec}
f = \tilde{\mb{q}}_r^\intercal \mb{q}_r + \tilde{\mb{q}}_i^\intercal\mb{q}_i,
\end{equation}
where $\tilde{\mb{q}}_r$ and $\tilde{\mb{q}}_i \in\mathbb{R}^n$ are constant weight vectors.
In the aerodynamic shape optimization with the laminar-turbulent flow transition model, the eigenvector needs to be differentiated in this manner.

In the derivation presented in this paper, the general definition of the function of interest is used.
The associated partial derivatives need to be computed based on the underlying function of interest.

% In this paper, when we compute the dominant eigenvalue real component derivative, we set $\p f / \p \mb{v}$ according to Eq.~\eqref{eq:pItop_pv}.

\subsection{Algorithm}

\begin{figure}[!h]
\centering
\includegraphics[width=12.9cm]{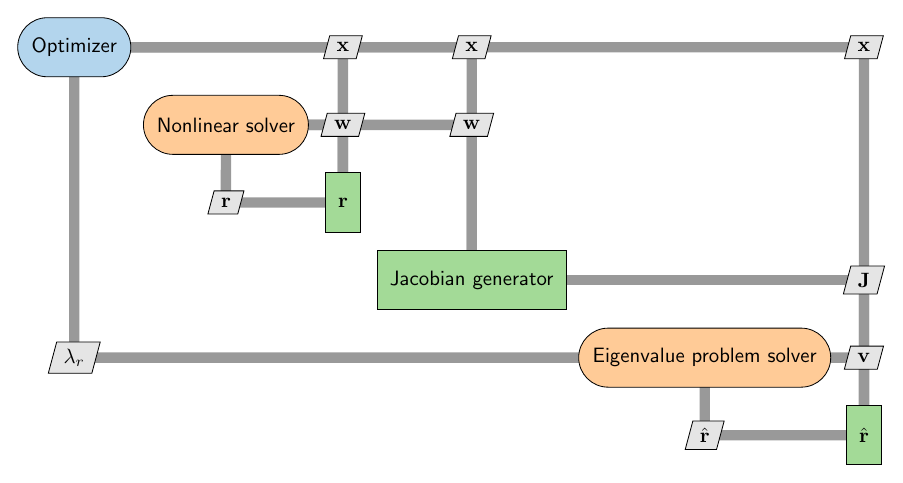}
\caption{XDSM for linear stability optimization.}
\label{fig:XDSM}
\end{figure}

The procedures discussed in this section are presented in Fig.~\ref{fig:XDSM} using an extended design structure matrix (XDSM) format~\cite{Lambe2012a}.
The algorithm is defined in Algorithm~\ref{alg.analysis}.
For the bottom-level and top-level equation solutions, the user has the freedom to choose their own solvers.

\begin{algorithm}[H]
\begin{spacing}{1.5}
\caption{Linear stability analysis}
\label{alg.analysis}
\begin{algorithmic}[1]
\Function{$\mathbf{f}_{\mathrm{stab}}$}{$\mb{x}$}
\State $\mb{w}_0\leftarrow \mb{r}(\mb{w}_0; \mb{x}) = 0$ \Comment{Bottom-level state variable solution.}
\State $\mb{J} = \f{\p \mb{r}}{\p \mb{w}_0}(\mb{w}_0; \mb{x})$ \Comment{Construct Jacobian.}
\State $\mb{v}\leftarrow \hat{\mb{r}}(\mb{v}; \mb{J}(\mb{w_0}; \mb{x})) = 0$ \Comment{Top-level state variable solution.}
\State $\lambda_r \leftarrow \mb{v}$ \Comment{Extract real part of the eigenvalue.}\\
\Return $\lambda_r$
\EndFunction
\end{algorithmic}
\end{spacing}
\end{algorithm}

\section{Linear stability adjoint sensitivity}
\label{sec:ls_adjoint}
This section discusses how to differentiate the linear stability measure, i.e., the maximum real part of all the eigenvalues, to obtain the derivative.
% [] TODO HS-: more details... mirror the other sections
% The section is organized in reverse order of the previous section, which corresponds to the order of the derivative computations.

\subsection{Coupled adjoint method}

To compute the derivative, we propose using the adjoint method due to its favorable scaling with many design variables.
Since we have two sets of equations, we need to solve the coupled adjoint equation.
The total derivative equation and the coupled adjoint equation are written as follows:
\begin{equation}
\label{eq:coupled_adjoint}
\begin{aligned}
&\frac{\d f}{\d \mb{x}} =
\frac{\partial f}{\partial \mb{x}}
-
\begin{bmatrix}
\pmb{\psi}_{\mb{r}}^\intercal &\pmb{\psi}_{\hat{\mb{r}}}^\intercal
\end{bmatrix}
\begin{bmatrix}
\frac{\partial \mb{r}}{\partial \mb{x}}\\
\frac{\partial \hat{\mb{r}}}{\partial \mb{x}}
\end{bmatrix}, \\
&\begin{bmatrix}
\frac{\partial \mb{r}}{\partial \mb{w}_0}^\intercal & \frac{\partial \hat{\mb{r}}}{\partial \mb{w}_0}^\intercal \\
\frac{\partial \mb{r}}{\partial \mb{v}}^\intercal & \frac{\partial \hat{\mb{r}}}{\partial \mb{v}}^\intercal \\
\end{bmatrix}
\begin{bmatrix}
\pmb{\psi}_{\mb{r}}\\
\pmb{\psi}_{\hat{\mb{r}}}
\end{bmatrix}
=
\begin{bmatrix}
\frac{\partial f}{\partial \mb{w}_0}^\intercal \\
\frac{\partial f}{\partial \mb{v}}^\intercal \\
\end{bmatrix},
\end{aligned}
\end{equation}
where the second equation is the coupled adjoint equation, and $\pmb{\psi}_{\mb{r}}, \pmb{\psi}_{\hat{\mb{r}}}$ are the adjoint vectors.
As mentioned in the previous section, the stability problem has a special feature that the bottom-level residual form does not depend on the top-level state variables, that is,
\begin{equation}
\frac{\p \mb{r}^\intercal}{\p \mb{v}} = 0.
\end{equation}
Thus, we can apply back-substitution by solving the second-row equation from Eq.~\eqref{eq:coupled_adjoint} first and then solve the equation of the first row.
Then, we have the top level adjoint equation
\begin{equation}
\label{eq:cadjoint_top}
\pmb{\psi}_{\hat{\mb{r}}} = \left(\frac{\partial \hat{\mb{r}}}{\partial \mb{v}}\right)^{-\intercal}\frac{\partial f}{\partial \mb{v}}^\intercal,
\end{equation}
and the bottom-level adjoint equation
\begin{equation}
\label{eq:cadjoint_bot}
\pmb{\psi}_{\mb{r}} = \left(\frac{\partial \mb{r}}{\partial \mb{w}_0}\right)^{-\intercal}\left(\frac{\partial f}{\partial \mb{w}_0}^\intercal - \frac{\partial \hat{\mb{r}}}{\partial \mb{w}_0}^\intercal\pmb{\psi}_{\hat{\mb{r}}} \right).
\end{equation}
By reducing the solution process into two separated adjoint equations, we avoid having to solve a larger coupled adjoint equation directly.
The benefit of this approach is that the individual adjoint equation algorithms and solvers are more well-developed than those of the coupled adjoint equation.
In addition, the two segregated adjoint equations are about a third and two thirds the size of the coupled one, respectively, making them cheaper to solve.

\subsection{Top level}
\label{sec:adj_top}

When the function of interest is a component of the eigenvalue (e.g., $f = \lambda_r$ or $f = \lambda_i$), the top-level adjoint Eq.~\eqref{eq:cadjoint_top} admits an analytic solution expressed in terms of the left eigenvector $\mb{u}$ of $\mb{J}$~\cite{He2022ua}.
For $f = \lambda_r$, the adjoint solution is
\begin{equation}
\label{eq:analytic_top_sol_1}
\pmb{\psi}_{\hat{\mb{r}}} =
\begin{bmatrix}
{\mb{u}_r}^\intercal &
{\mb{u}_i}^\intercal &
0 &
0
\end{bmatrix}^\intercal,
\end{equation}
where $\mb{u}$ is the left eigenvector satisfying $\mb{J}^\intercal \mb{u} = \lambda^* \mb{u}$ with the normalization $\mb{u}^* \mb{q} = -1$.
We recommend using this analytic solution rather than solving Eq.~\eqref{eq:cadjoint_top} directly, since the same eigensolver can be reused for the left eigenvector.
The detailed coefficient matrix, the analytic solutions for other functions of interest, and the normalization procedures are given in~\Cref{app:top_adjoint}.

\subsection{Bottom level}
\label{sec:der_bot}

The bottom-level adjoint equation Eq.~\eqref{eq:cadjoint_bot} requires computing the term $\left({\partial \hat{\mb{r}}}/{\partial \mb{w}_0}\right)^\intercal\pmb{\psi}_{\hat{\mb{r}}}$.
This reduces to computing the derivative of the pattern
\begin{equation}
\label{eq:derivVecJacVec}
\f{\p \mb{r}_2^\intercal\mb{J}^\intercal\mb{r}_1}{\p\mb{w}},
\end{equation}
which is a second-order derivative involving the Jacobian (see~\Cref{app:bot_adjoint} for the detailed expansion).
We propose three methods to compute it using reverse algorithmic differentiation (RAD)~\cite{Linnainmaa1976,Martins2022}:
\begin{equation}
\label{eq:jac_der_FD}
\mathrm{(FDRAD):} \f{\p \mb{r}_2^\intercal\mb{J}^\intercal\mb{r}_1}{\p\mb{w}}
\approx\left(\f{\f{\p\mb{r}(\mb{w}_0 + h \mb{r}_2)}{\p \mb{w}}^\intercal \mb{r}_1 - \f{\p\mb{r}(\mb{w}_0)}{\p \mb{w}}^\intercal \mb{r}_1}{h}\right)^\intercal,
\end{equation}
\begin{equation}
\label{eq:jac_der_CD}
\mathrm{(CDRAD):} \f{\p \mb{r}_2^\intercal\mb{J}^\intercal\mb{r}_1}{\p\mb{w}}
\approx\left(\f{\f{\p\mb{r}(\mb{w}_0 + h \mb{r}_2)}{\p \mb{w}}^\intercal \mb{r}_1 - \f{\p\mb{r}(\mb{w}_0 - h\mb{r}_2)}{\p \mb{w}}^\intercal \mb{r}_1}{2h}\right)^\intercal,
\end{equation}
\begin{equation}
\label{eq:jac_der_CS}
\mathrm{(CSRAD):} \f{\p \mb{r}_2^\intercal\mb{J}^\intercal\mb{r}_1}{\p\mb{w}}
=\f{\mathrm{Im}\left(\f{\p\mb{r}(\mb{w}_0 + ih \mb{r}_2)}{\p \mb{w}}^\intercal \mb{r}_1\right)}{h},
\end{equation}
where FDRAD, CDRAD, and CSRAD denote the forward difference, central difference, and complex-step of a RAD formula, respectively.
CSRAD achieves machine precision by selecting a small $h$ (e.g., $h = 10^{-200}$), making it superior to the finite difference-based alternatives.
However, it requires the differentiated code to be complexified, which may require more development effort.
We compare these three methods in Section~\ref{sec:cfd_adjoint_verify}.

\subsection{Total derivative assembly}
\label{sec:total_derivative}

After solving the top-level and bottom-level adjoints, the total derivative is assembled as
\begin{equation}
\label{eq:total_derivative_lamr}
\f{\d \lambda_r}{\d \mb{x}} = -\pmb{\psi}_{\mb{r}}^\intercal \f{\p \mb{r}}{\p \mb{x}} - \pmb{\psi}_{\hat{\mb{r}}}^\intercal \f{\p \hat{\mb{r}}}{\p \mb{x}},
\end{equation}
where the first term is the bottom-level residual term and the second is the top-level residual term.
The ADflow-specific implementation, including the analytic mass-matrix contribution that arises in the discrete operator, is presented in~\Cref{sec:lst_mass_matrix}.

\subsection{Algorithm}

We summarize and organize the previously discussed operations into Algorithm~\ref{alg.derivative}.
In Algorithm~\ref{alg.derivative}, we assume $f = \lambda_r$ and omit zero terms.

\begin{algorithm}[H]
\begin{spacing}{1.5}
\caption{Local stability derivative using the block back-substitution method.}
\label{alg.derivative}
\begin{algorithmic}[1]
\Function{$\mathbf{g}_{\mathrm{stab}}$}{$\mb{x}$}
\State $\f{\p \lambda_r}{\p \mb{v}} = \begin{bmatrix} 0 & 0 & 1 & 0 \end{bmatrix}^\intercal$ \Comment{Set the RHS for the top-level adjoint equation.}
\State $\pmb{\psi}_{\hat{\mb{r}}}\leftarrow \f{\p \hat{\mb{r}}}{\p \mb{v}}^\intercal \pmb{\psi}_{\hat{\mb{r}}} = \f{\p \lambda_r}{\p \mb{v}}$ \Comment{Solve the top-level adjoint by direct solution Eq.~\eqref{eq:cadjoint_top} or analytic solution Eq.~\eqref{eq:analytic_top_sol_1}.}
\State $\mb{b} \leftarrow \frac{\partial \hat{\mb{r}}}{\partial \mb{w}_0}^\intercal\pmb{\psi}_{\hat{\mb{r}}}$ \Comment{Evaluate the partial derivative using one of Eqs.~\eqref{eq:jac_der_FD}, \eqref{eq:jac_der_CD}, and \eqref{eq:jac_der_CS}.}
\State $\pmb{\psi}_{\mb{r}}\leftarrow \f{\p {\mb{r}}}{\p \mb{w}_0}^\intercal \pmb{\psi}_{\mb{r}} = \mb{b}$ \Comment{Solve the bottom-level adjoint Eq.~\eqref{eq:cadjoint_bot}.}
\State $\f{\d \lambda_r}{\d \mb{x}} \rightarrow -\pmb{\psi}_{\mb{r}}^\intercal \f{\p \mb{r}}{\p \mb{x}} - \pmb{\psi}_{\hat{\mb{r}}}^\intercal \f{\p \hat{\mb{r}}}{\p \mb{x}}$ \Comment{Assemble the total derivative, Eq.~\eqref{eq:total_derivative_lamr}.}
\\
\Return $\f{\d \lambda_r}{\d \mb{x}}$
\EndFunction
\end{algorithmic}
\end{spacing}
\end{algorithm}

Algorithm~\ref{alg.derivative} and the equations from which it is derived are written for the standard eigenvalue problem.
For the generalized eigenvalue problem that arises when the discretization carries a mass matrix, we refer the reader to Algorithm~\ref{alg.derivative_mm} in~\Cref{sec:lst_mass_matrix}, which uses the modified operator $\mb{J}_\mathrm{mod}=-\mb{M}^{-1}\mb{J}$ and poses it as a standard eigenvalue problem.
From Algorithm~\ref{alg.derivative}, we find that no matter how many design variables we have, we only need to solve the coupled adjoint equation once.
For more details about a comparison of different optimization strategies, such as using gradient-free and gradient-based together with the finite differences, we refer the reader to the paper by~\citet{Lyu2014f} in the context of PDE-constrained optimization.
%               With a focus on large-scale problems. For example, we may not explicitly evaluate the Jacobian as Jacobian-free is better...

One advantage of the block back-substitution method is that it requires much lower implementation effort than the direct solution method.
This is because the coupled system is never formed explicitly, and existing implementation of the top-level eigenvalue adjoint equation solver and the bottom-level adjoint equation solver can be applied directly.

Besides the more straightforward implementation, the block back-substitution method is also more efficient than the direct solution method.
For large-scale PDE-based problems, the adjoint equations are high-dimensional and usually sparse.
In this case, we may never form the coefficient matrices explicitly and use a Jacobian-free approach to solve the adjoint equations.
An iterative solver, such as Krylov subspace methods, can be used together with a preconditioner to solve such equations.
One example is the aerostructural optimization coupled-adjoint solver proposed by Kenway et al.~\cite{Kenway2014a}.
More general coupled-adjoint formulations are also possible with the OpenMDAO framework~\cite{Gray2019a}, which has been used in many applications, including PDE-constrained problems~\cite{Yildirim2021c}.
A well-constructed preconditioner is essential for the efficiency of the overall solution.
The preconditioner for a coupled problem can be constructed by putting the segregated preconditioners on the diagonal.
This diagonal preconditioner may work to some extent but is likely not comparable to a full preconditioner.
More computational studies need to be conducted for the adjoint solvers in the future to validate the discussion.

\section{Eigenvalue computation methodology}
\label{sec:solver}

The eigenvalue with the largest real part determines the linear stability of the equilibrium point and is the quantity of interest for the LST constraint.
Most iterative eigensolvers, such as the implicitly restarted Arnoldi method, efficiently extract the eigenvalues of largest magnitude, which are not in general the rightmost in the complex plane.
This section presents the spectral transformations and Krylov solver used to converge to the rightmost generalized eigenpair $(\lambda,\mb{q})$ of $\mb{J}\mb{q}=-\lambda\mb{M}\mb{q}$ at the cost of one inner linear solve per eigensolver matvec.
% [] TODO HS-: J overloaded... probably a better way is to start with the generalized eigenvalue problem at the beginning...

\subsection{Shift-invert spectral transformation}
\label{sec:shift_invert}

The shift-invert spectral transformation maps the eigenvalue nearest to a chosen shift $\sigma\in\mathbb{C}$ to the eigenvalue of largest magnitude of the transformed operator,
\begin{equation}
\label{eq:shift_invert}
(\mb{J}+\sigma\mb{M})^{-1}\mb{M}\,\mb{q} = \f{1}{\lambda - \sigma}\,\mb{q}.
\end{equation}
A Krylov--Schur eigensolver~\cite{Stewart2002,Hernandez2005} targeting the eigenvalue of largest modulus then converges to the eigenvalue closest to $\sigma$.
The cost is dominated by one global LU factorization of $\mb{J}+\sigma\mb{M}$ in complex arithmetic, after which each shell-mult is a single back-solve.
The drawback is that $\sigma$ must be selected close to the eigenvalue of interest a priori; an unfortunate choice converges to a different, possibly stable mode.

The complete procedure is given in~\Cref{alg:shift_invert}.

\begin{algorithm}[H]
\begin{spacing}{1.3}
\caption{Shift-invert Krylov--Schur for the eigenpair of $\mb{J}\mb{q}=-\lambda\mb{M}\mb{q}$ closest to a chosen shift $\sigma\in\mathbb{C}$.}
\label{alg:shift_invert}
\begin{algorithmic}[1]
\Require Equilibrium state $\mb{w}_0$; spectral shift $\sigma$; Krylov dimension $m$; outer tolerance $\eta_\mathrm{out}$; number of requested eigenpairs $k$.
\Ensure $k$ eigenpairs $\{(\lambda_j,\mb{q}_j)\}_{j=1}^k$ closest to $\sigma$.
\State $\mb{J} \leftarrow \p \mb{r}/\p \mb{w}|_{\mb{w}_0}$ \Comment{Assemble linearized residual.}
\State $\mb{S} \leftarrow \mb{J} + \sigma\mb{M}$ \Comment{Shifted operator,~\cref{eq:shift_invert}.}
\State $(\mb{L},\mb{U}) \leftarrow \mathrm{LU}(\mb{S})$ \Comment{One-time global factorization in complex arithmetic.} % [x] TODO HS-: LU or ILU here?
\State $\mathrm{EPS} \leftarrow \mathrm{KS}(\mb{S}^{-1}\mb{M};\, m,\,\eta_\mathrm{out},\,\mathrm{LM})$ \Comment{Configure Krylov--Schur with restart $m$, outer tolerance, and largest-modulus target.}
\Repeat
\State $\mb{v}_{j+1} \leftarrow \mathrm{Arnoldi}(\mb{V}_j)$ \Comment{Next basis vector (orthogonalization done by EPS).}
\State $\mb{z} \leftarrow (\mb{L},\mb{U})^{-1}\mb{M}\mb{v}$ \Comment{Back-solve with the cached factorization.}
\Until{$k$ eigenpairs of $\mb{S}^{-1}\mb{M}$ have converged at tolerance $\eta_\mathrm{out}$}
\For{each converged $(\theta_j,\mb{q}_j)$ of $\mb{S}^{-1}\mb{M}$}
\State $\lambda_j \leftarrow \sigma + 1/\theta_j$ \Comment{Shift-invert back-map.}
\EndFor
\State \Return $\{(\lambda_j,\mb{q}_j)\}_{j=1}^k$
\end{algorithmic}
\end{spacing}
\end{algorithm}
% [x] TODO HS-: for alg try math symbol on the left and comments etc on the right.

\subsection{Cayley/Crank--Nicolson time-stepper Arnoldi}
\label{sec:cayley}

To eliminate the need to guess a shift, we instead apply a time-stepper Arnoldi method~\cite{Goldhirsch1987,Edwards1994,Tuckerman2000,Bagheri2009} based on the Cayley transform.
Discretizing the linearized dynamics $\mb{M}\,\d(\delta\mb{w})/\d t = -\mb{J}\,\delta\mb{w}$ with the Crank--Nicolson rule~\cite{CrankNicolson1947} over a step $\Delta t$ yields the discrete propagator
\begin{equation}
\label{eq:cayley_T}
\mb{T}(\Delta t) = \left(\mb{I} + \tfrac{\Delta t}{2}\mb{J}_\mathrm{mod}\right)\left(\mb{I} - \tfrac{\Delta t}{2}\mb{J}_\mathrm{mod}\right)^{-1},
\quad \mb{J}_\mathrm{mod} \coloneqq -\mb{M}^{-1}\mb{J},
\end{equation}
% [x] TODO HS-: A or Jmod. unify notation.
which is the Cayley transform of $\Delta t\,\mb{J}_\mathrm{mod}$.
Eigenpairs of $\mb{J}_\mathrm{mod}$ and $\mb{T}$ are related by the M\"obius map
\begin{equation}
\label{eq:moebius}
\mu = \f{1 + \Delta t\,\lambda/2}{1 - \Delta t\,\lambda/2},
\quad
\lambda = \f{2}{\Delta t}\,\f{\mu - 1}{\mu + 1}.
\end{equation}
This map sends the imaginary axis $\mathrm{Re}\,\lambda = 0$ to the unit circle $|\mu|=1$, the right half-plane to the exterior of the unit disk $|\mu|>1$, and the left half-plane to the interior $|\mu|<1$.
Working with $\mb{T}$ instead of $\mb{J}_\mathrm{mod}$ therefore turns the open-ended question of finding the rightmost eigenvalue into the closed one of finding the eigenvalue of $\mb{T}$ closest to or outside the unit circle, with no shift required.

\subsubsection{Eigenvalue selection: largest-modulus versus largest-real-part target}
\label{sec:lm_vs_lr}

The squared magnitude under the M\"obius map is
\begin{equation}
\label{eq:mu_mag}
|\mu|^2 = \f{1 + \mathrm{Re}\,\lambda \cdot \Delta t + (\Delta t/2)^2 |\lambda|^2}{1 - \mathrm{Re}\,\lambda \cdot \Delta t + (\Delta t/2)^2 |\lambda|^2},
\end{equation}
which is monotonically increasing in $\mathrm{Re}\,\lambda$ at fixed $|\lambda|$.
When at least one eigenvalue is unstable, that eigenvalue maps to $|\mu|>1$ while every stable mode lies strictly inside the unit disk, so the largest-modulus selection criterion on $\mb{T}$ converges directly to the rightmost continuous-time eigenvalue~\cite{Meerbergen1997,Garratt1991}.
When the spectrum is fully stable, every $|\mu|<1$, and the largest-modulus criterion can target the wrong mode: from~\cref{eq:mu_mag}, a high-frequency stable eigenvalue with large $|\lambda|$ but very negative $\mathrm{Re}\,\lambda$ may still produce a $|\mu|$ that exceeds the $|\mu|$ of the true rightmost mode, since the monotonicity in $\mathrm{Re}\,\lambda$ holds only at fixed $|\lambda|$.
In this regime we instead target the eigenvalue of $\mb{T}$ with largest $\mathrm{Re}\,\mu$, which under~\cref{eq:moebius} remains aligned with the rightmost $\lambda$ across the dominant modes of the spectrum.
The largest-real-part criterion therefore covers exactly the regime in which the largest-modulus criterion fails: when no mode lies in the right half-plane.

\subsubsection{Linear solution}
\label{sec:inner_solve}

The shell operator $\mb{T}$ is applied without ever assembling $\mb{T}$ or $\mb{B} \coloneqq \mb{I} - (\Delta t/2)\mb{J}_\mathrm{mod}$ explicitly.
Substituting $\mb{J}_\mathrm{mod}=-\mb{M}^{-1}\mb{J}$ in $\mb{B}$ and multiplying by $\mb{M}$ gives
\begin{equation}
\label{eq:Pdef}
\mb{M}\mb{B} = \mb{P} \coloneqq \mb{M} + \tfrac{\Delta t}{2}\mb{J},
\quad
\mb{B}^{-1} = \mb{P}^{-1}\mb{M},
\end{equation}
so each shell-mult requires solving $\mb{P}\mb{z} = \mb{M}\mb{v}$ in real arithmetic.
The inner solve uses a direct MUMPS~\cite{petsc-user-ref} LU factorization of $\mb{P}$ computed once at solver setup, after which every shell--mult is a back--solve and inner GMRES converges in one or two iterations on every call.
To achieve better scaling on larger problems, the direct LU factorization must be replaced with iterative methods, and is left for future work on larger--scale systems.

After the eigensolver converges, the eigenvector is polished by one step of complex-shift inverse iteration applied at the converged $\lambda$, which reuses the same factorization machinery and reduces the relative GEVP residual $\|\mb{J}\mb{q}+\lambda\mb{M}\mb{q}\|/(|\lambda|\,\|\mb{M}\mb{q}\|)$ from order one to the floating-point floor of MUMPS LU on the converged primal in a single iteration.

\subsubsection{Algorithm}
\label{sec:cayley_algo}

The complete procedure for the rightmost eigenpair using the Cayley/Crank--Nicolson time-stepper Arnoldi with direct LU Linear solution is given in~\Cref{alg:cayley_cn}.

\begin{algorithm}[!htb]
\begin{spacing}{1.3}
\caption{Cayley/Crank--Nicolson time-stepper Arnoldi for the rightmost generalized eigenpair.}
\label{alg:cayley_cn}
\begin{algorithmic}[1]
\Require Equilibrium state $\mb{w}_0$; Cayley step $\Delta t$; Krylov dimension $m$; outer tolerance $\eta_\mathrm{out}$; inner GMRES tolerance $\eta_\mathrm{in}$; number of requested eigenpairs $k$.
\Ensure $k$ rightmost eigenpairs $\{(\lambda_j,\mb{q}_j)\}_{j=1}^k$ of $\mb{J}\mb{q}=-\lambda\mb{M}\mb{q}$.
\State $\mb{J} \leftarrow \p \mb{r}/\p \mb{w}|_{\mb{w}_0}$ \Comment{Assemble linearized residual.}
\State $\mb{P} \leftarrow \mb{M} + (\Delta t/2)\,\mb{J}$ \Comment{Inner-solve operator,~\cref{eq:Pdef}.}
\State $(\mb{L},\mb{U}) \leftarrow \mathrm{LU}(\mb{P})$ \Comment{One-time global factorization.}
\State $\mathrm{EPS} \leftarrow \mathrm{KS}(\mb{T};\, m,\,\eta_\mathrm{out},\,\mathrm{LM})$ \Comment{Configure Krylov--Schur with restart $m$, outer tolerance, and largest-modulus target (use LR when no unstable mode is expected; see~\cref{sec:lm_vs_lr}).}
\Repeat
\State $\mb{v}_{j+1} \leftarrow \mathrm{Arnoldi}(\mb{V}_j)$ \Comment{Next basis vector (orthogonalization done by EPS).}
\State $\mb{z} \leftarrow \mathrm{GMRES}(\mb{B}, \mb{v};\, \mathrm{PC} = \mb{P}^{-1}\mb{M},\, \eta_\mathrm{in})$ \Comment{Inner solve in the unpreconditioned residual norm.}
\State $\mb{u} \leftarrow -\mb{M}^{-1}(\mb{J}\mb{z})$
\State $\mb{T}\mb{v} \leftarrow \mb{z} + (\Delta t/2)\,\mb{u}$
\Until{$k$ eigenpairs of $\mb{T}$ have converged at tolerance $\eta_\mathrm{out}$}
\For{each converged $(\mu_j,\mb{q}_j)$}
\State $\lambda_j \leftarrow (2/\Delta t)(\mu_j-1)/(\mu_j+1)$ \Comment{M\"obius back-map,~\cref{eq:moebius}.}
\EndFor
\State $(\lambda_j, \mb{q}_j) \leftarrow \mathrm{InvIter}(\mb{J},\mb{M};\, \sigma=\lambda_j,\, \mb{q}_j)$ \Comment{One step of complex-shift inverse-iteration polish.}
\State \Return $\{(\lambda_j,\mb{q}_j)\}_{j=1}^k$
\end{algorithmic}
\end{spacing}
\end{algorithm}
% [] TODO HS-: descrobe algs in words...

\section{Results}
\label{sec:results}
In this section, we present the results for the bottom stage ADflow\footnote{\url{https://github.com/mdolab/adflow}} primal runs, top stage eigen--computation primal runs and finally discuss the optimization results.
In the first subsection, we benchmark our results against literature for the canonical cylinder vortex shedding case.
Next, we present the steady state results and unsteady results for the OAT15A transonic buffet case, along with the eigen-analysis of the steady state unstable base flow.

The eigen computation results for the OAT15A transonic buffet case are verified and validated by comparing our eigenspectra against the eigenspectra from~\citet{Sartor2015}, and comparing the unsteady lift coefficient temporal response in the linear identified region against the growth rate and frequency from the unstable eigenvalue.

Next, we present the verification of the gradients obtained from the adjoint method to compute $\d \lambda_r / \d \mb{x}$ by comparing the results against gradients obtained by finite differences on the cylinder vortex--shedding case.
The full pipeline is implemented within the MACH--Aero framework\footnote{\url{https://mdolab-mach-aero.readthedocs-hosted.com/en/latest/}}, which couples ADflow with a set of geometry, mesh, and optimization tools developed by the same group.
The FFD volume that parameterizes the surface mesh and supplies the design variables to both the verification and the optimization is constructed and manipulated with pyGeo~\cite{Hajdik2023c}\footnote{\url{https://github.com/mdolab/pygeo}}, and the resulting surface displacements are propagated to the volume mesh using the IDWarp\footnote{\url{https://github.com/mdolab/idwarp}} mesh deformation library, which also provides the analytic derivatives of volume node positions with respect to design variables that are needed by the chain rule assembly of $\d\lambda_r/\d\mb{x}$.
Finally, we perform gradient--based optimization using the pyOptSparse~\cite{Wu2020a} optimizer interface with the SLSQP~\cite{Kraft1988} sequential least--squares quadratic programming algorithm and discuss the shape optimization results in the final subsection.
% [x] TODO HS-: if show idwarp link show the rest... consistency! 

\subsection{Cylinder vortex shedding benchmark}
\label{sec:cylinder}

ADflow has not been previously used for linear stability analysis.
We therefore use the canonical cylinder vortex shedding case to verify our LST implementation in ADflow against published benchmarks.
Because ADflow stores the flow state in primitive-like variables while integrating the conservative form, the linearized semi-discrete operator carries a non-trivial mass matrix and the resulting LST problem is a generalized eigenvalue problem; the corresponding derivation and equations are detailed in~\Cref{app:mass-matrix}.
For unstable flow exhibiting the Hopf bifurcation and entering into saturated limit-cycle oscillations, we expect a complex conjugate pair of eigenvalues with positive real part~\cite{Sahin2004,Marquet2008}.

\subsubsection{Steady state and unsteady simulation results}
\label{sec:cyl_steady_unsteady}

It is a fact that \href{https://github.com/mdolab/adflow}{ADflow} non--dimensionalizes the governing equations.
The eigenvalues sought are therefore non--dimensional.
To re--dimensionalize the eigenvalues to help compare with literature, we must scale them by the factor
\begin{equation}
  \lambda_s = \f{\lambda}{t_{\mathrm{ref}}},
\end{equation}
where $\lambda_s$ is the scaled eigenvalue and 
\begin{equation}
  t_{\mathrm{ref}} = \mathrm{L}_\mathrm{ref}\f{1}{\sqrt{\mathrm{RT}}},
\end{equation}
where $\mathrm{R}$ is the gas constant 287 $\mathrm{J/kg K}$, $\mathrm{T}$ is the reference temperature 300 $\mathrm{K}$ and $\mathrm{L}_\mathrm{ref}$ is set to a value of 1 $\mathrm{m}$.
$\lambda_s$ is in $\mathrm{rad/s}$.
To convert it into physical frequency, we divide it by $2\pi$.

\begin{figure}[H]
  \centering

  \begin{subfigure}[t]{0.48\textwidth}
    \centering
    \includegraphics[width=\textwidth]{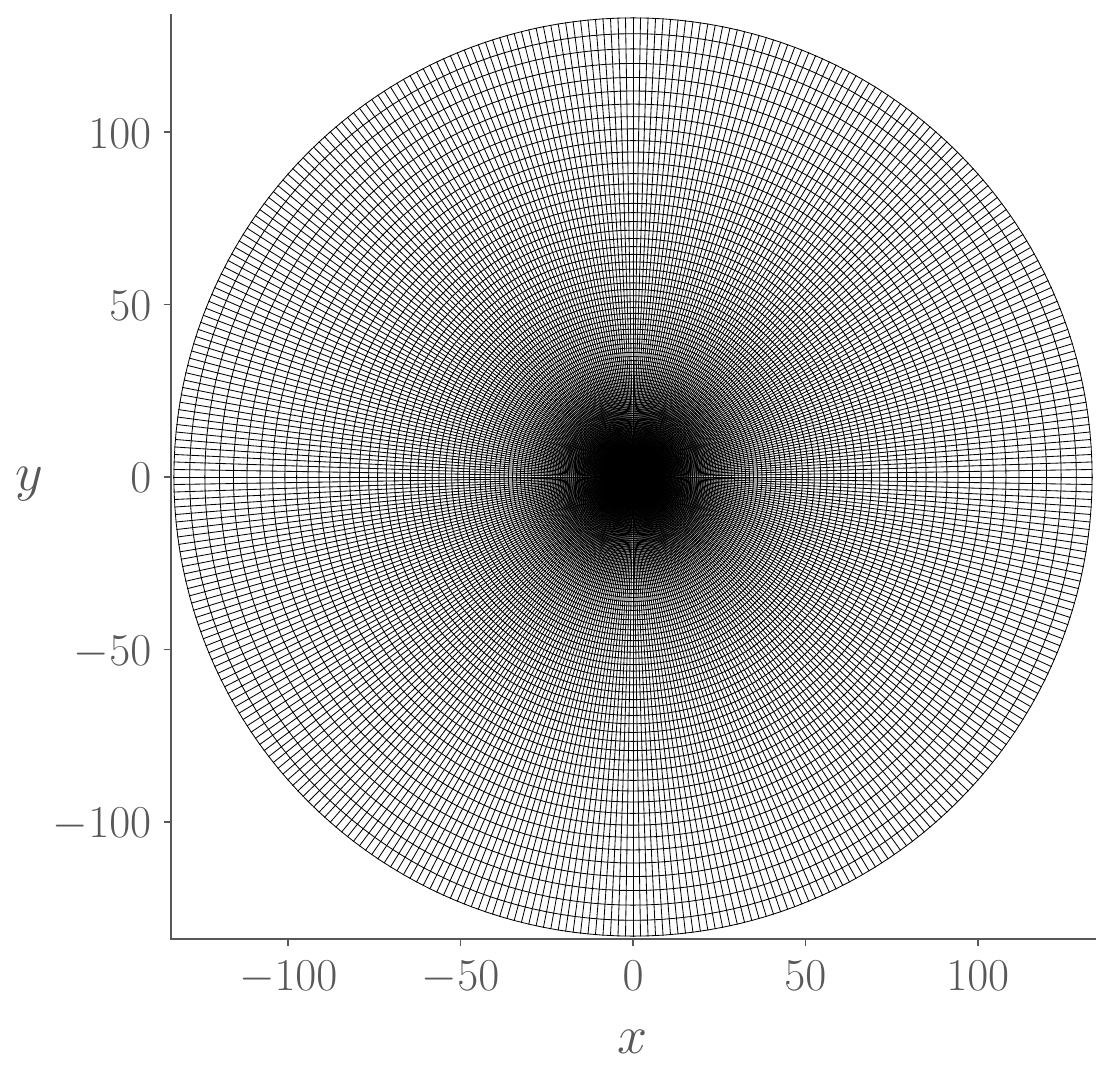}
  \end{subfigure}
  \hfill
  \begin{subfigure}[t]{0.48\textwidth}
    \centering
    \includegraphics[width=\textwidth]{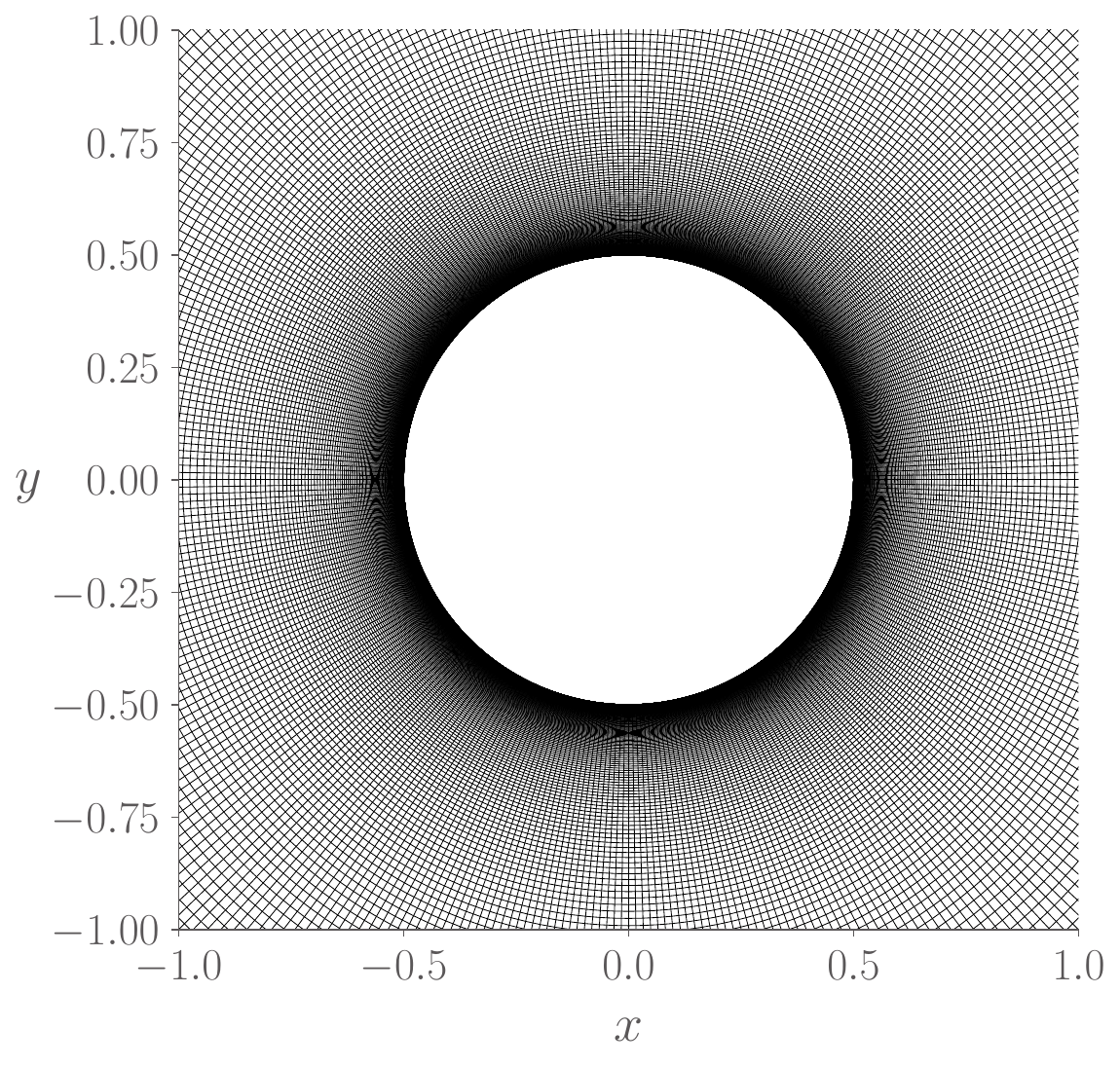}
  \end{subfigure}

  \caption{The full structured grid for the cylinder flow (left) and the grid near the cylinder wall region zoomed in (right).
  A boundary layer grid is made with the first cell normal to the cylinder wall $\Delta r = 10^{-5}\,\mathrm{m}$.
  The cylinder diameter is set to 1 m.}
  \label{fig:cyl_grid}
\end{figure}

A structured grid was generated for a cylinder of diameter 1 $\mathrm{m}$.
The grid is shown below in the first sub--figure (left) in~\Cref{fig:cyl_grid}.
The domain was discretized with 385 elements in both the circumferential and radial directions.
A boundary layer grid near to the cylinder wall was made, as shown in the second sub--figure (right) in~\Cref{fig:cyl_grid}.
The off-wall spacing of the first cell normal to the cylinder is $\Delta r = 10^{-5} \mathrm{m}$.
The grid is similar to that used in literature~\cite{Rajani2009,Franke1990}.

\begin{figure}[H]
  \centering
  \includegraphics[width=0.85\textwidth]{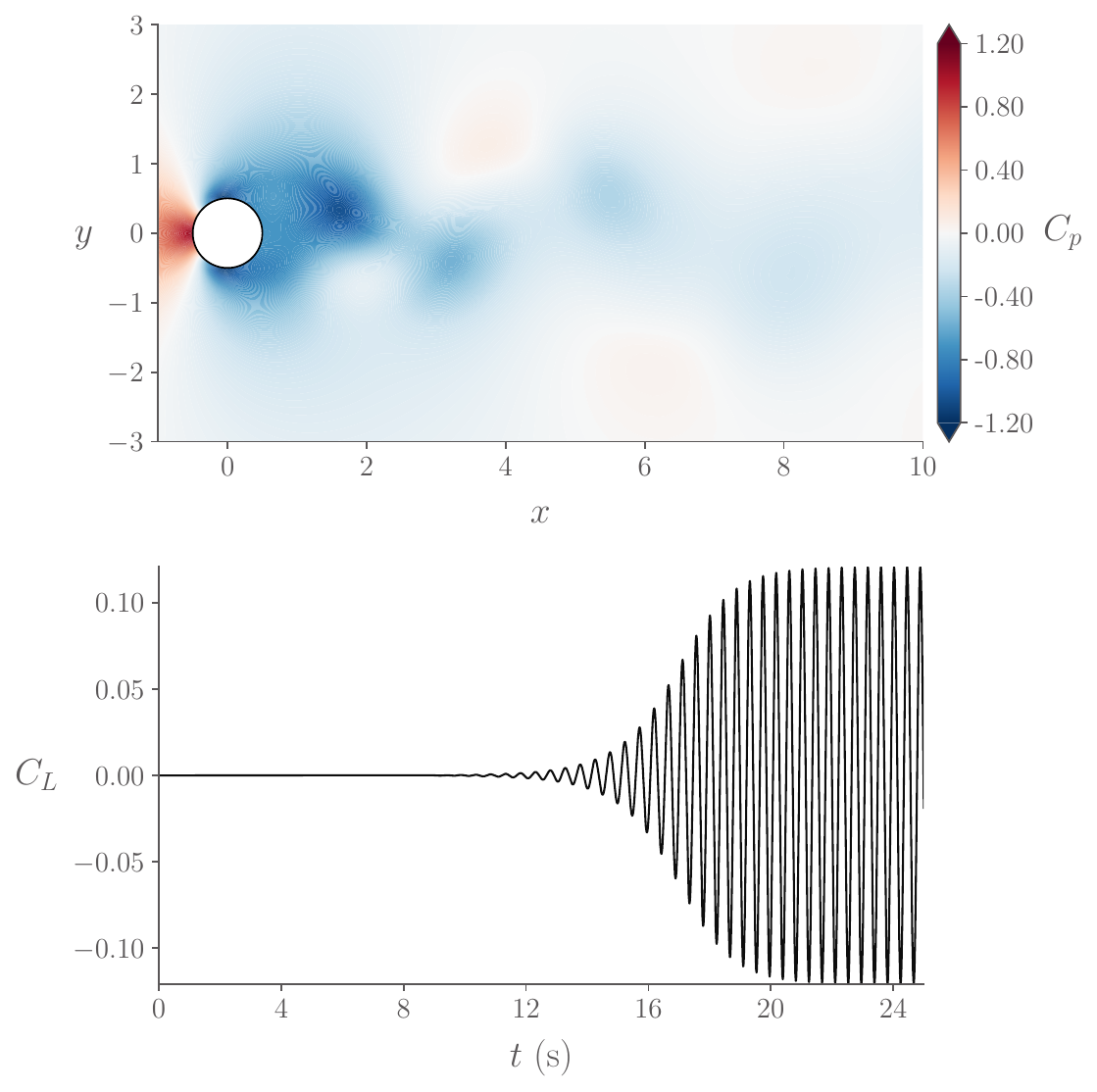}
  \caption{Unsteady vortex shedding at Reynolds number 60.
  Top: instantaneous pressure coefficient $C_p$ field, showing the Von--Karman vortex street formed inside the saturated limit cycle.
  Bottom: unsteady lift coefficient $C_L$ vs time, in which the small initial perturbation from a one--second cylinder rotation grows and saturates into limit cycle oscillations.}
  \label{fig:cyl_unsteady}
\end{figure}

Unsteady simulations were run with the laminar Navier--Stokes model.
The freestream conditions are Mach number of 0.1, Reynolds number in the range 46--120 and temperature of 300 $\mathrm{K}$.
A snapshot of the vortex shedding is shown in~\Cref{fig:cyl_unsteady}, for a Reynolds number of 60.
% [x] TODO HS-: von Karman
A classic von K\'arm\'an vortex street is seen behind the cylinder and the flow is in a saturated limit cycle, as seen in the lift--coefficient $C_L$ vs time plot in~\Cref{fig:cyl_unsteady}.
A small perturbation was made around the base flow, by rotating the cylinder five degrees in one second of physical time.
This pushes the system into the saturated limit cycle oscillations.

The Strouhal number is defined as 
\begin{equation}
  \mathrm{St} = \f{f}{\mathrm{L}\mathrm{U}},
\end{equation}
where $\mathrm{L}=\mathrm{L}_\mathrm{ref}=1 \mathrm{m}$, $\mathrm{U}=34.718\text{ } \mathrm{m/s}$ and $f$ is the vortex shedding frequency in Hz.
In our simulations, we measure this frequency as the frequency of one complete oscillation of the lift--coefficient when the flow has entered into a fully saturated limit cycle.

A plot of the Strouhal number vs Reynolds number is shown in~\Cref{fig:st_vs_re} below.
We compare our results with Williamson's results~\cite{Williamson1996}.
Our results match appreciably with the Hopf--bifurcation diagram from literature.

\begin{figure}[H]
  \centering
  \includegraphics[width=0.44\textwidth]{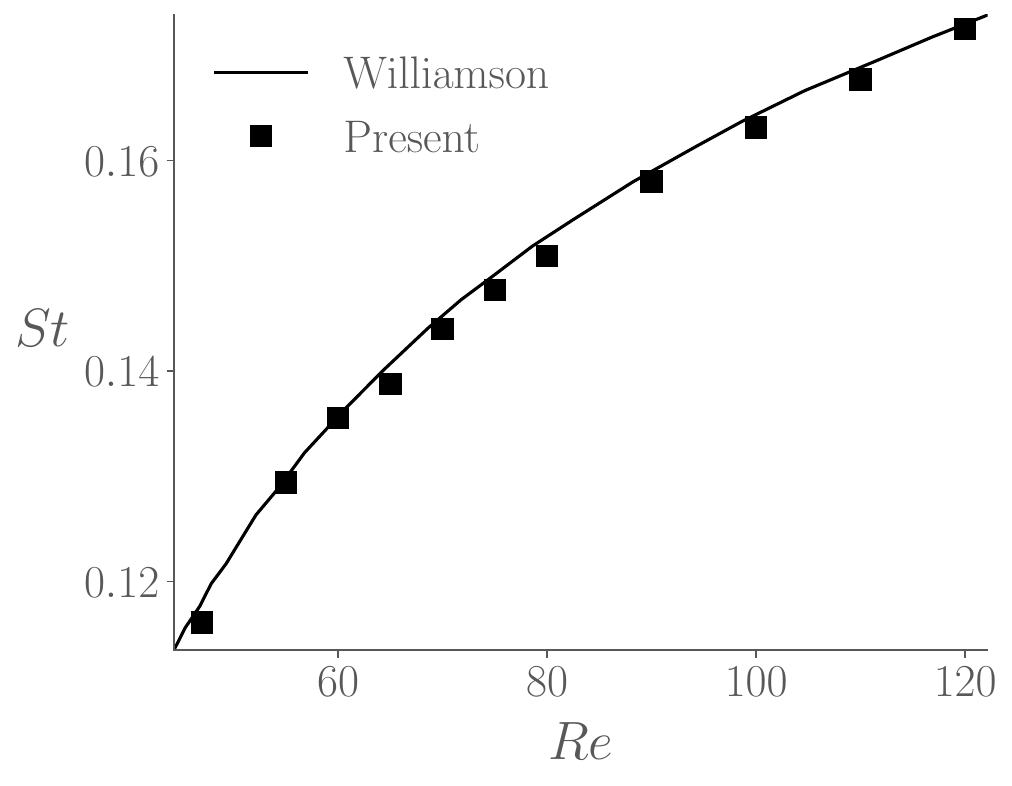}
  \caption{A plot of the Strouhal numbers evaluated at different Reynolds numbers for the cylinder vortex shedding case.
  A comparison is made against Williamson's results~\cite{Williamson1996} and the data agrees appreciably.}
  \label{fig:st_vs_re}
\end{figure}

\subsubsection{Linear stability analysis}
\label{sec:cyl_lst}

\begin{figure}[H]
  \centering

  \begin{subfigure}[t]{0.5\textwidth}
    \centering
    \includegraphics[width=\textwidth]{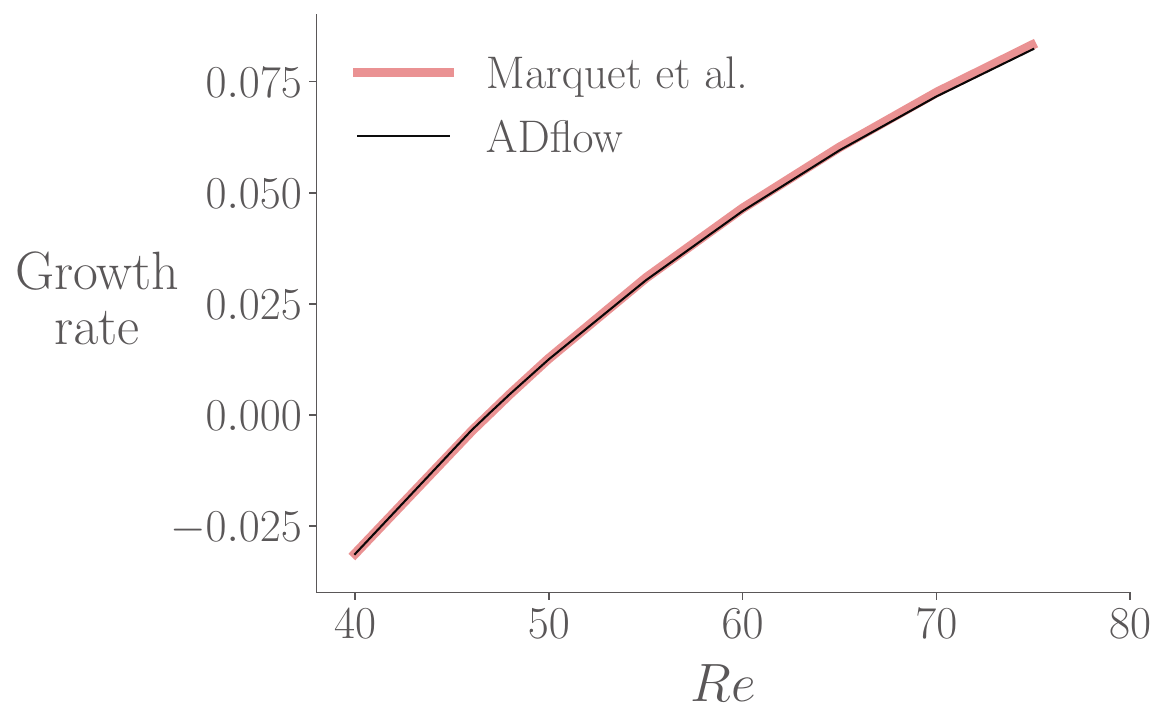}
  \end{subfigure}
  \hfill
  \begin{subfigure}[t]{0.45\textwidth}
    \centering
    \includegraphics[width=\textwidth]{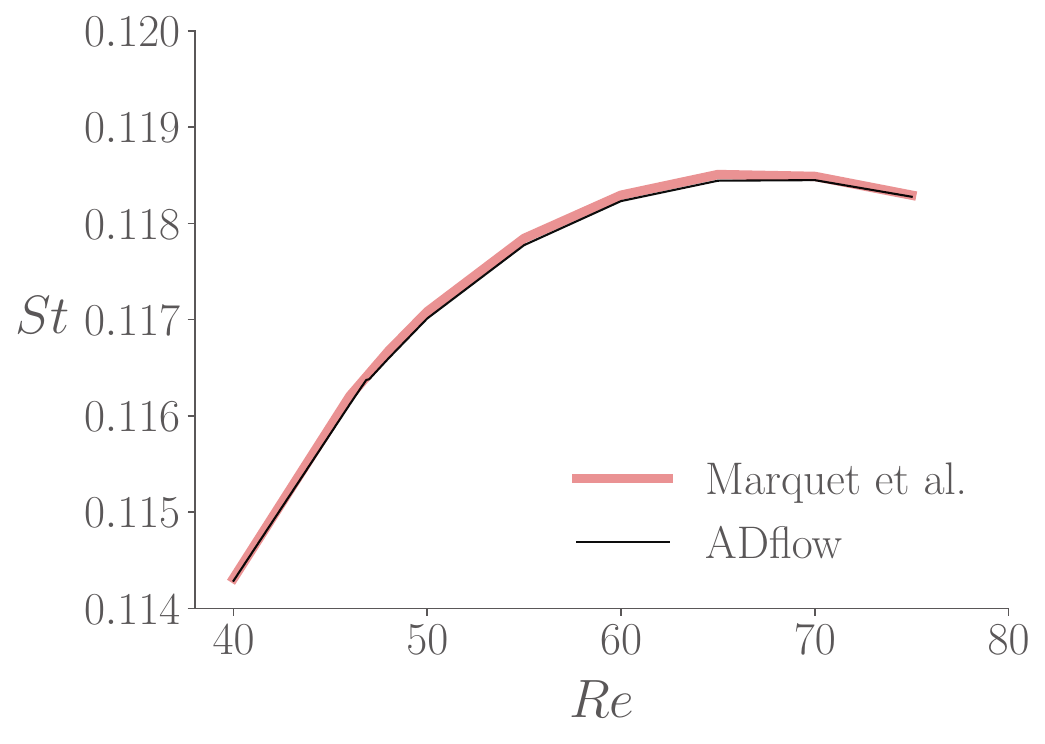}
  \end{subfigure}

  \caption{The growth rates (left) and the eigenfrequency derived Strouhal numbers (right) for the base flow computed at different Reynolds numbers.
  Results from ADflow are benchmarked against those of Marquet et al.~\cite{Marquet2008}, which seem to match appreciably.
  The critical Reynolds number is found to be 46.85.
  }
  \label{fig:lst_cyl_eigvals}
\end{figure}

\begin{figure}[H]
  \centering
  \includegraphics[width=0.9\textwidth]{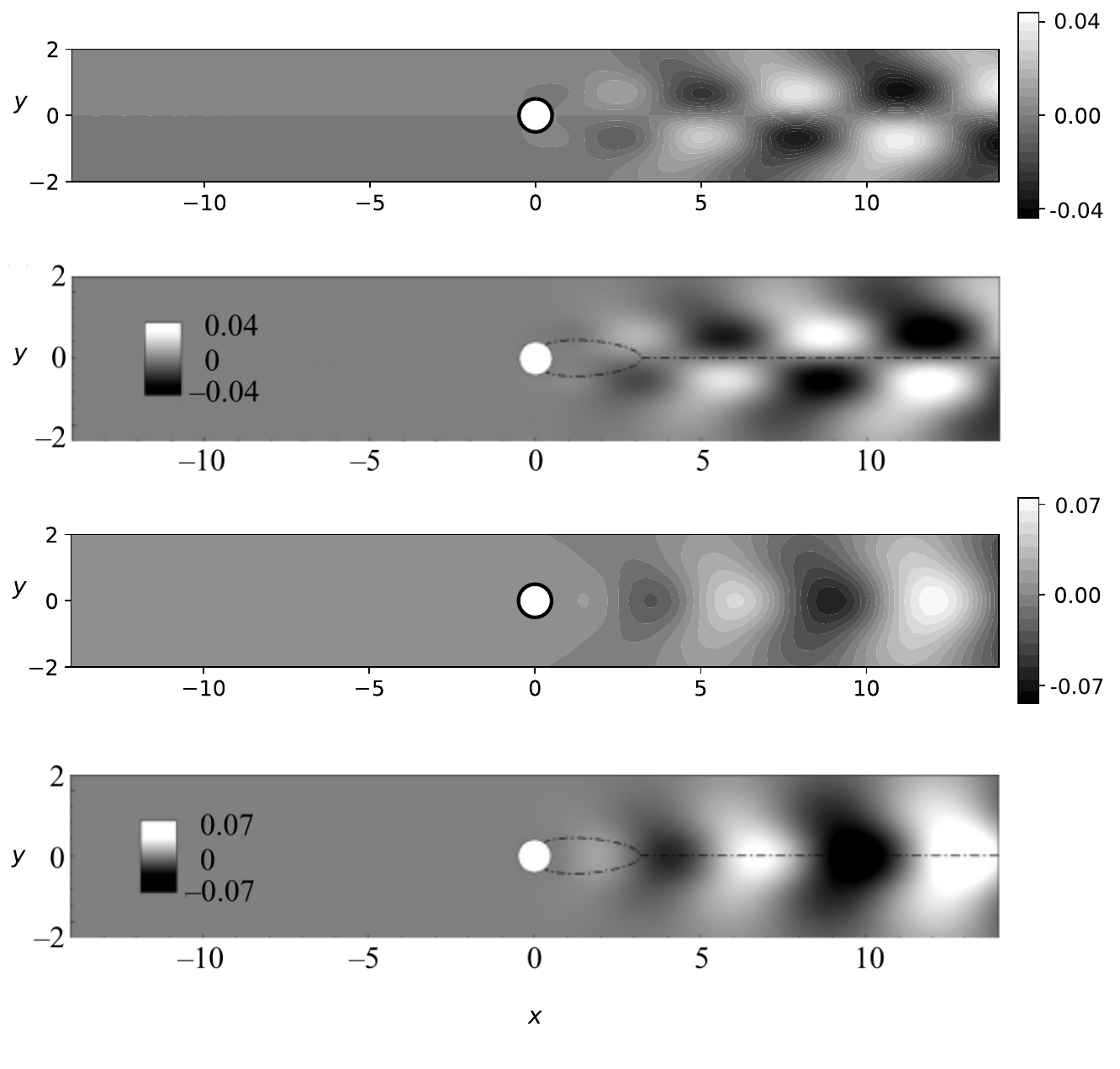}
  \caption{Unstable eigenmodes from ADflow (from top) are shown in the first and third figures for $\hat{u}$ and $\hat{v}$ respectively, at the critical Reynolds number 46.85 evaluated in the current study.
  The results from Marquet et al.~\cite{Marquet2008} are shown in the second and fourth figures for $\hat{u}$ and $\hat{v}$ respectively, at the critical Reynolds number 46.8 evaluated in their study.
  }
  \label{fig:modes_cyl}
\end{figure}

Finally, we present a plot of the eigenfrequencies from our LST with ADflow in~\Cref{fig:lst_cyl_eigvals}.
We compare our results with Marquet et al~\cite{Marquet2008}.
Our results match to almost machine precision with their results.
It should also be noted that the first unstable eigenpair was observed at the critical Reynolds number 46.85, which is very close to Marquet et al~\cite{Marquet2008} who reported the critical Reynolds number to be 46.8$\pm$0.05.
We also present the modes $\hat{u}$ and $\hat{v}$, the $x$ and $y$ velocity modes in~\Cref{fig:modes_cyl}.
Complex eigenmodes can be scaled and rotated in the complex plane for the same eigenvalues~\cite{He2023} and therefore are not expected to always match with the results from other solvers.
The difference in the modes is typically a result of normalization factor used, or whether the mode was at all normalized to begin with.
Nevertheless, the mode shapes match appreciably.

All simulations and results were converged to tolerance ($<10^{-12}$), including the eigensolver used--\href{https://slepc.upv.es/material/slides/tslepc-anl-1p.pdf}{SLEPc}, using the shift--invert strategy.
The complex part of eigenfrequencies (presented as Strouhal number) do not match with the unsteady Strouhal numbers as shown in~\Cref{fig:st_vs_re,fig:lst_cyl_eigvals}.
This is a classic case of mean flow and base flow being different, leading to different unstable unsteady frequencies and base flow eigenfrequencies~\cite{Barkley2006,Sipp2007}.

In terms of computational cost, the direct--LU CN--Arnoldi path of~\Cref{sec:cayley} converges to the $Re = 100$ eigenpair $\lambda = 0.004198 + 0.044210\,\mathrm{j}$ on the $n = 49{,}152$--dof cylinder mesh in $\sim 9$~s on $4$ MPI ranks at a final GEVP residual of $1.3\times 10^{-7}$ after one inverse--iteration polish step.
The polish step reduces the residual from order one to the floating--point floor of MUMPS LU on the converged primal in a single iteration on every Reynolds number tested, with $\Delta t = 10$ used throughout.

\newpage
\subsection{OAT15A transonic buffet}
\label{sec:oat15a}
We next present results for the OAT15A supercritical airfoil~\cite{Jacquin2009}, a standard two-dimensional benchmark for transonic-buffet computation.
The presentation proceeds in four stages: a steady-state RANS solution with a grid convergence study, an unsteady simulation on the converged mesh that captures buffet onset and limit-cycle saturation, a linear stability analysis of the steady base flow whose rightmost eigenpair is the constraint quantity used in the optimization of~\Cref{sec:crm_optimization}, and an internal verification of the LST eigenvalue against the linear-growth phase of the URANS run.
% [] TODO HS-: internal verification?

\subsubsection{Steady state and unsteady simulation results}
\label{sec:oat15a_steady}

\begin{figure}[H]
  \centering
  \begin{subfigure}[t]{0.48\textwidth}
    \centering
    \includegraphics[width=\textwidth]{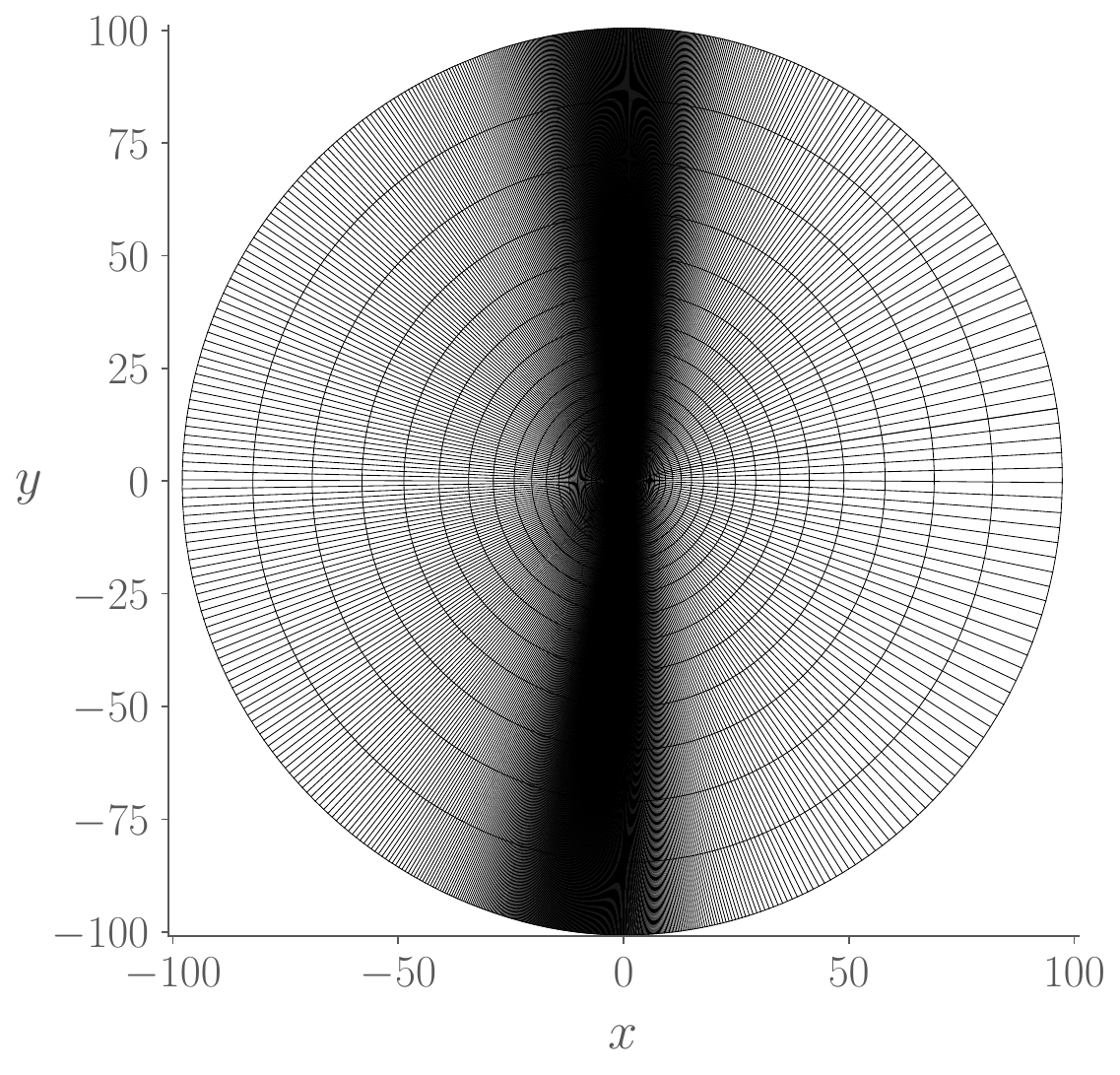}
  \end{subfigure}
  \hfill
  \begin{subfigure}[t]{0.48\textwidth}
    \centering
    \includegraphics[width=\textwidth]{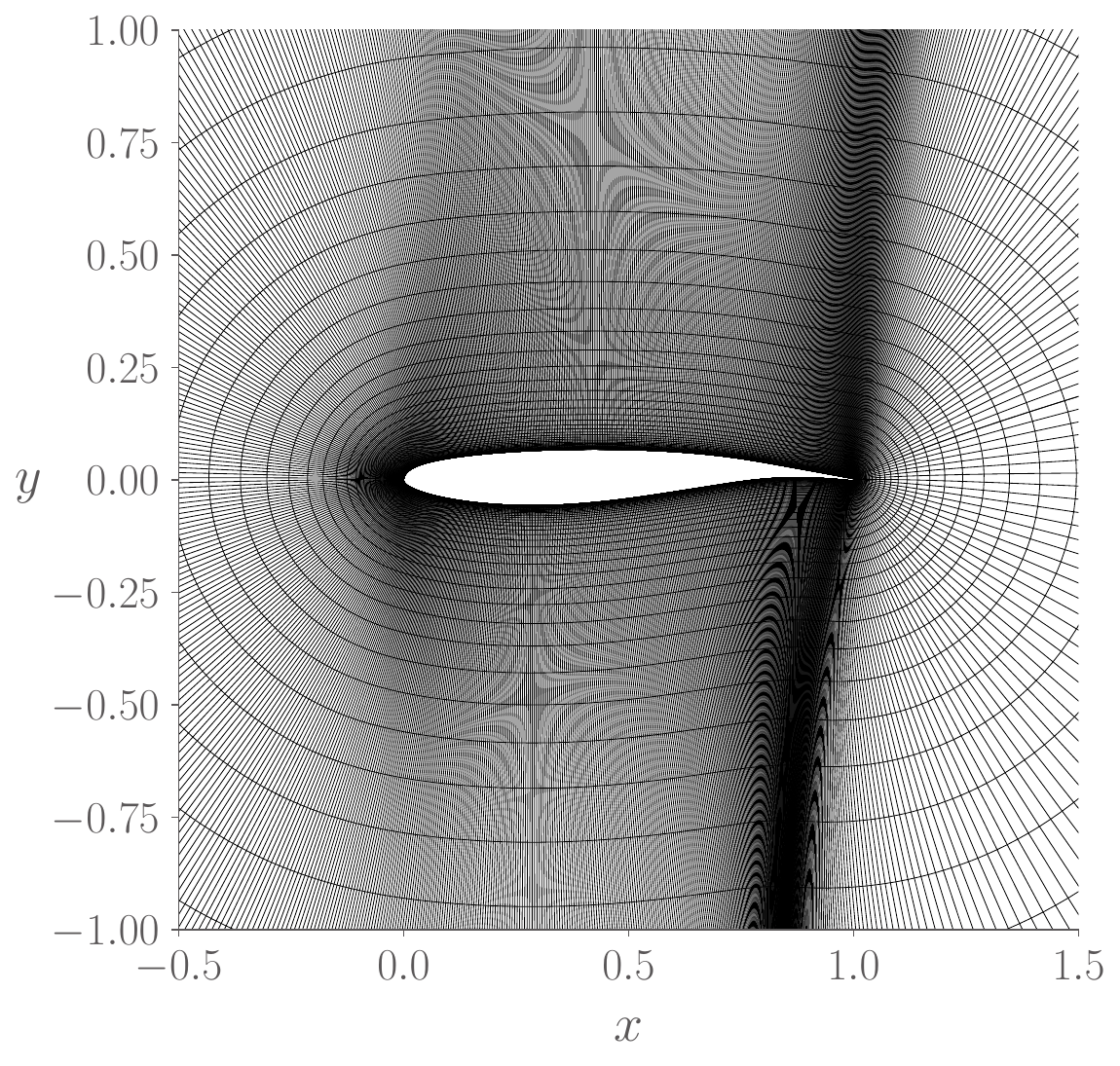}
  \end{subfigure}
  \caption{The full structured grid for the OAT15A transonic airfoil at the far--field domain (left) and zoomed in on the airfoil region (right).
  A boundary layer grid is made with the first cell normal to the airfoil wall $\Delta r = 10^{-6}\,\mathrm{m}$.
  The airfoil chord is set to 1 m.}
  \label{fig:oat15a_grid}
\end{figure}

The computational grid is presented in~\Cref{fig:oat15a_grid}.
The airfoil chord is $1 \mathrm{m}$.
This grid is generated using pyHyp, a hyperbolic structured grid generator~\cite{Secco2021a}.
The domain is discretized as a body fitted O-grid with symmetry boundary conditions into the plane of paper and no--slip adiabatic wall for the airfoil wall.
The outermost surface is set to a far--field boundary condition.
We discretize the fine mesh level with 912 points along the circumferential direction, which includes 15 points for the trailing edge discretization.
We source our geometry for OAT15A airfoil from the ONERA database\footnote{\url{https://www.aiaa-dpw.org/geometry.html}}.
Therefore, we have a blunt TE and which necessitates discretization.
The radial direction has 97 points with an inflation towards the wall as part of a boundary layer mesh and this is made with the first cell normal to the cylinder wall $\Delta r = 10^{-6}\,\mathrm{m}$.

\begin{figure}[H]
  \centering
  \includegraphics[width=0.95\textwidth]{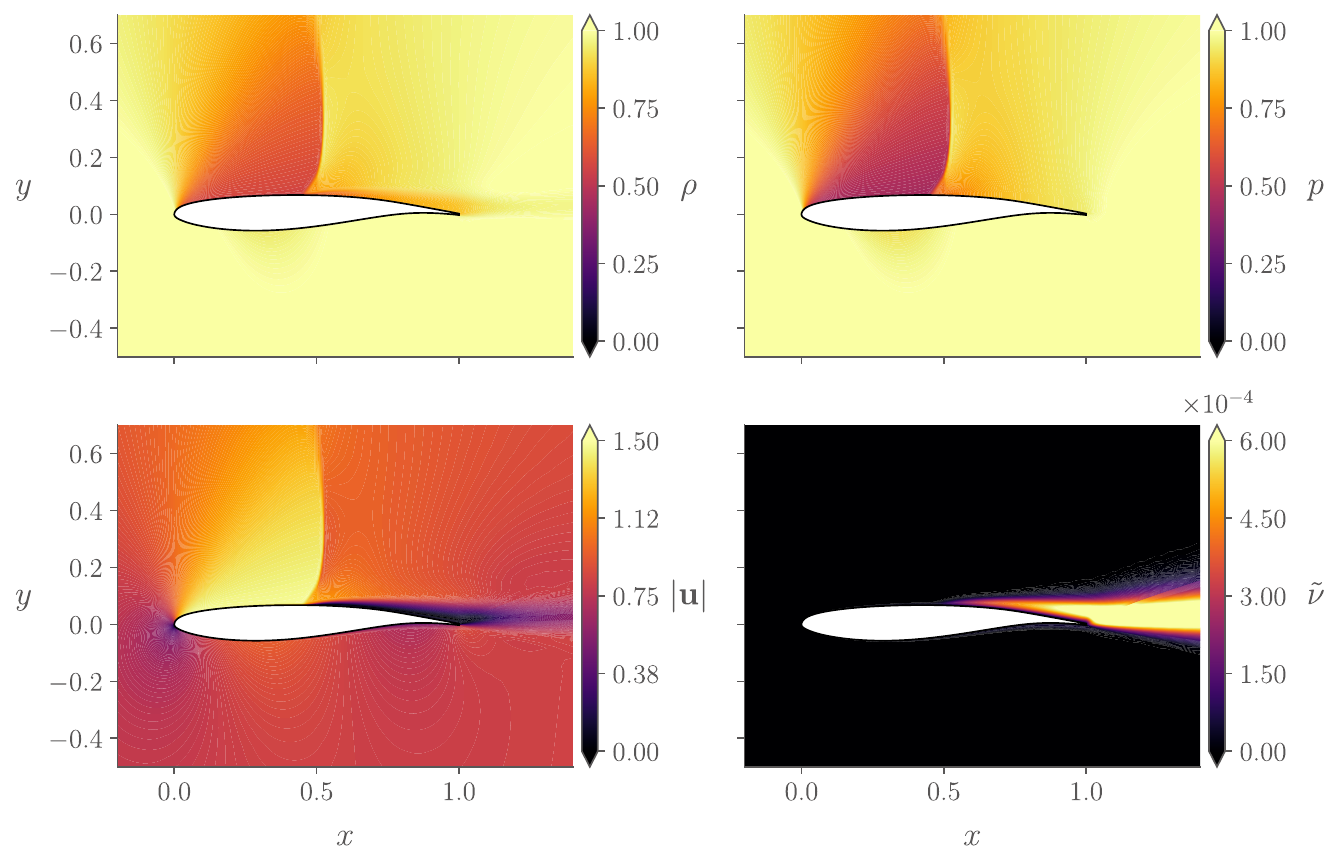}
  \caption{Steady-state contours of density $\rho$, pressure $p$, velocity magnitude $\lvert\mathbf{u}\rvert$, and Spalart--Allmaras working variable $\tilde{\nu}$ on the L0 mesh for the OAT15A airfoil at Mach number 0.73, Reynolds number $3.2\times 10^{6}$ and angle of attack $4^\circ$.
  Quantities are shown in non-dimensional ADflow units.}
  \label{fig:oat15a_steady_contour}
\end{figure}

We simulated it for $Re = 3.2\times 10^{6}$, $M = 0.73$, $\alpha = 4^\circ$, $T_\infty = 300$ K for standard air properties.
Air is the fluid medium and is treated as an ideal gas. 
The steady state results at this flight condition are shown in~\Cref{fig:oat15a_steady_contour}.
The flow over the suction side accelerates and experiences a shockwave as it moves past the sound barrier.
The shock location is at $x \approx 0.5$.
The flow just aft the shock foot is heavily separated, as is evident from the velocity field contour.
The steady state results correspond closely with those of~\citet{Sartor2015}.
It must be noted that~\citet{Sartor2015} used a sharpened TE tip instead of the blunt TE tip.
This is known to cause differences in results and also cause primal convergence difficulties~\cite{EldridgeAllegra2024}.
However, these difficulties were overcome when using the ANK/NK solver~\cite{Yildirim2019b,Mader2020a}, a state of the art linear solver that ADflow uses.

To establish confidence in the results, we conduct a formal grid convergence study following the $h$--refinement procedure documented by the MDO Lab.\footnote{\url{https://mdolab-mach-aero.readthedocs-hosted.com/en/latest/machAeroTutorials/airfoilanalysis_gridRefinementStudy.html}.}
The grid family is generated by uniformly coarsening the L0 mesh with the cgnsutilities ``coarsen'' utility, which halves the cell count along each curvilinear direction and yields the L1 and L2 meshes summarized in~\Cref{tab:oat15a_grids}.
The grid refinement ratio is $r=2$ in each direction, giving the four--fold cell count reduction between successive levels required for a classical Richardson extrapolation.

\begin{table}[h]
  \centering
  \caption{Grid levels used in the OAT15A grid convergence study; each level is a uniform $r = 2$ coarsening of the next finer mesh.}
  \label{tab:oat15a_grids}
  \begin{tabular}{lccc}
    \toprule
    Level
      & \multicolumn{1}{c}{Number of cells $N$}
      & \multicolumn{1}{c}{$1/N$}
      & \multicolumn{1}{c}{Refinement ratio $r$} \\
    \midrule
    L0 & \multicolumn{1}{r}{$87{,}552$} & \multicolumn{1}{r}{$1.14\times 10^{-5}$} & \multicolumn{1}{r}{---} \\
    L1 & \multicolumn{1}{r}{$21{,}888$} & \multicolumn{1}{r}{$4.57\times 10^{-5}$} & \multicolumn{1}{r}{$2$} \\
    L2 & \multicolumn{1}{r}{$5{,}472$}  & \multicolumn{1}{r}{$1.83\times 10^{-4}$} & \multicolumn{1}{r}{$2$} \\
    \bottomrule
  \end{tabular}
\end{table}

\begin{figure}[H]
  \centering
  \includegraphics[width=0.55\textwidth]{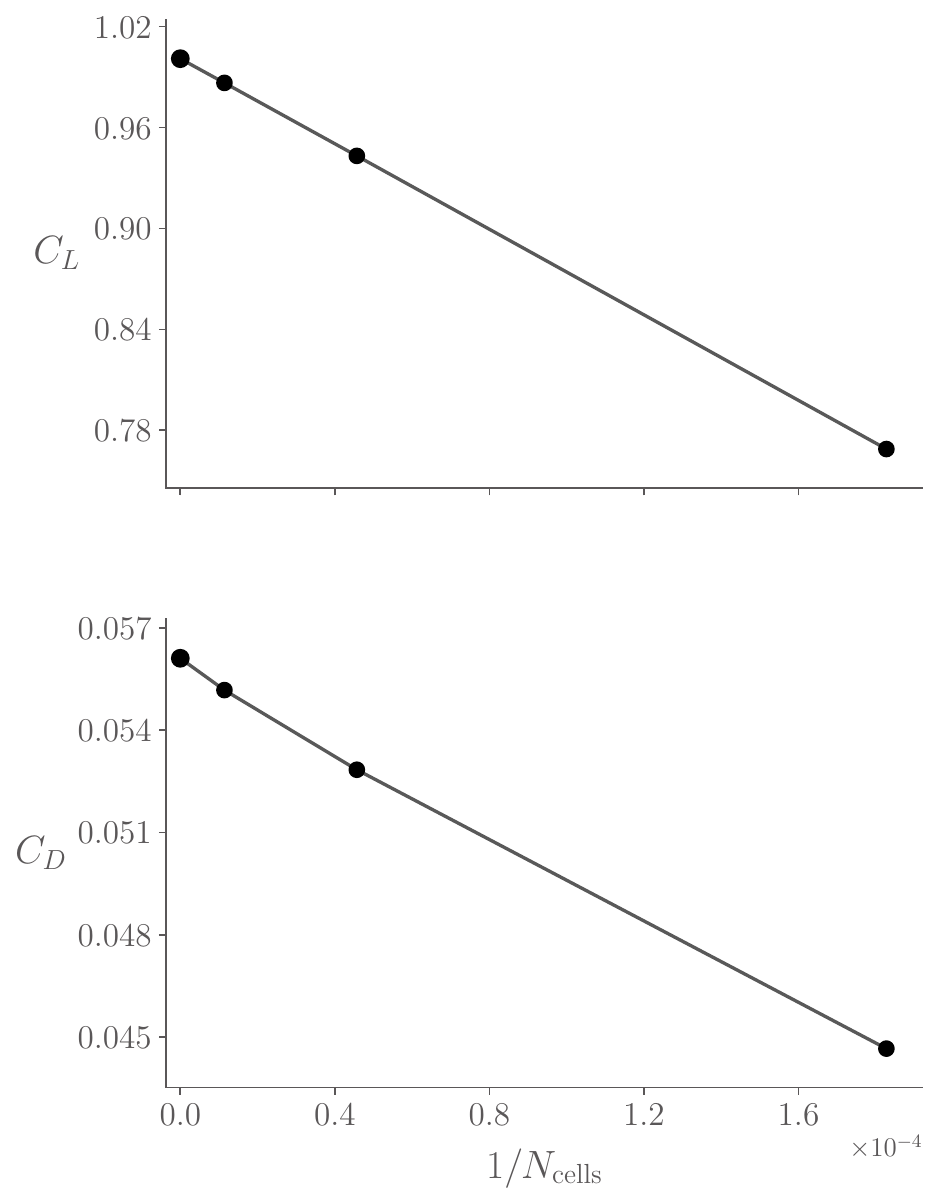}
  \caption{Grid convergence of the lift coefficient $C_L$ (top) and drag coefficient $C_D$ (bottom) for the OAT15A airfoil at $M=0.73$, $Re=3.2\times 10^{6}$, $\alpha=4^{\circ}$.
  The three markers at finite $1/N_{\mathrm{cells}}$ correspond to the L0, L1, and L2 meshes of~\Cref{tab:oat15a_grids}, and the marker at $1/N_{\mathrm{cells}}=0$ is the Richardson extrapolation.
  The achieved order of accuracy is $\hat{p}=2.01$ for $C_L$ and $\hat{p}=1.81$ for $C_D$.}
  \label{fig:oat15a_grid_refinement}
\end{figure}

The achieved order of accuracy is computed from the finest three solutions as
\begin{equation}
\label{eq:oat15a_phat}
\hat{p} = \ln\left(\f{f_{L2}-f_{L1}}{f_{L1}-f_{L0}}\right)\Big/\ln r,
\end{equation}
and the Richardson extrapolation to zero grid spacing is
\begin{equation}
\label{eq:oat15a_richardson}
f_{h=0} = f_{L0} + \f{f_{L0}-f_{L1}}{r^{\hat p}-1}.
\end{equation}
For $C_L$ we obtain $\hat{p}=2.01$ and $f_{h=0}=1.0009$, and for $C_D$ we obtain $\hat{p}=1.81$ and $f_{h=0}=0.0561$.
The achieved orders are within $10\%$ of the formal second--order accuracy of the finite volume scheme used in ADflow, and the L0/L1/L2 points lie on a straight line through the extrapolated value, confirming that the finest three grids are in the asymptotic range.
The convergence behavior is shown in~\Cref{fig:oat15a_grid_refinement}.

\begin{figure}[H]
  \centering
  \includegraphics[width=0.85\textwidth]{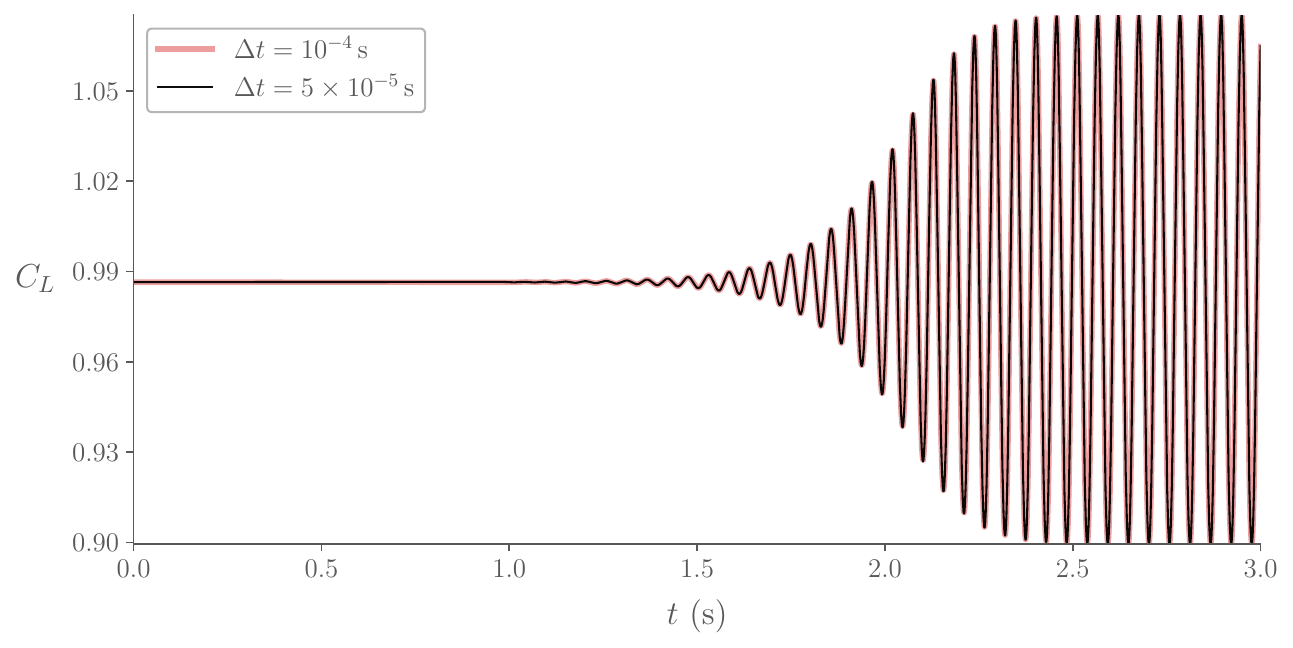}
  \caption{Unsteady $C_L$ history for the OAT15A at $M=0.73$, $Re=3.2\times 10^{6}$, $\alpha=4^\circ$, L0 grid; $\Delta t=10^{-4}$ s (thick translucent) and $\Delta t=5\times 10^{-5}$ s (thin) are indistinguishable, confirming time--step independence.}
  \label{fig:oat15a_cl_vs_time}
\end{figure}

\begin{figure}[H]
  \centering
  \includegraphics[width=1.0\textwidth]{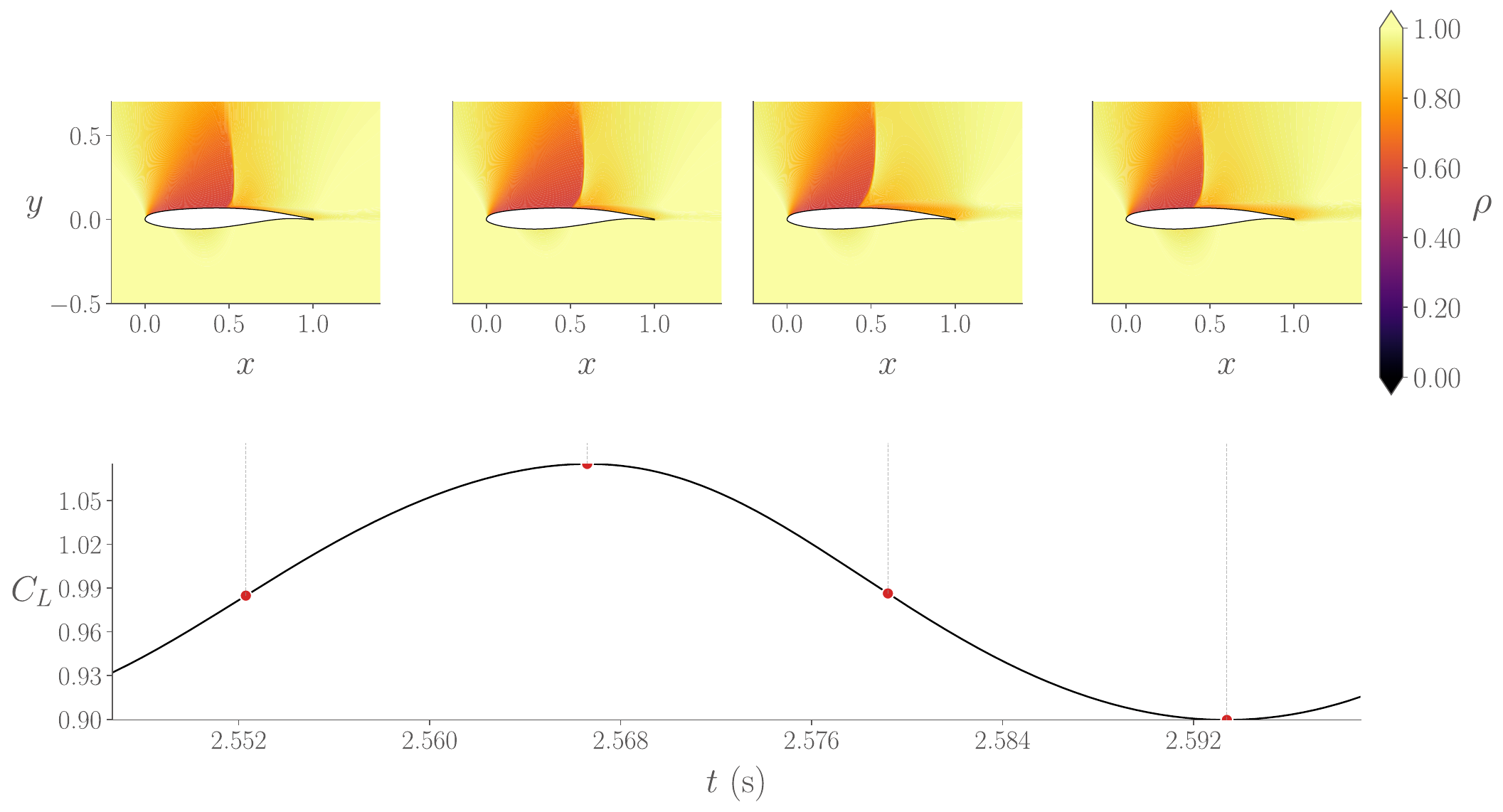}
  \caption{The unsteady simulation density contours over one cycle in the saturated limit cycle oscillations for the OAT15A transonic airfoil from a cold--started run.
  Freestream conditions are Mach number 0.73, Reynolds number $3.2 \times 10^6$ angle of attack $4.0^\circ$.
  Four plots have been made, each at distinct physical times on the limit cycle, indicated by vertical solid black lines from the sinusoidal wave.
  The Strouhal number for this unsteady transonic buffet is 0.074.}
  \label{fig:cl_plot_contour_plot}
\end{figure}

Unsteady simulations were performed on the L0 grid using the steady--state solution as the initial condition.
Two time steps were considered, $\Delta t = 10^{-4}$ s and $\Delta t = 5\times 10^{-5}$ s, at the same freestream conditions as the steady analysis.
The lift coefficient history for both runs is shown in~\Cref{fig:oat15a_cl_vs_time}.
After an initial damped transient, small numerical perturbations to the steady--state solution grow exponentially starting at $t \approx 1.3$ s, and the system saturates into a limit--cycle oscillation by $t \approx 2.5$ s.
The two time--step traces lie on top of each other across the full window, indicating that $\Delta t = 10^{-4}$ s is sufficient to resolve the buffet oscillation and that the response is time--step independent at this resolution.
The saturated limit--cycle frequency yields a Strouhal number $\mathrm{St} = f c / U_\infty = 0.074$, with chord $c=1$ m and freestream velocity $U_\infty = 253.4$ m/s, in close agreement with the value reported by~\citet{Sartor2015}.
Density contours over one period of the saturated limit cycle are shown in~\Cref{fig:cl_plot_contour_plot} at four phases of the oscillation, capturing the periodic shock motion that drives the lift modulation.

% [x] TODO HS-: check AIAAJ caption guideline... too long. check all fighures

\subsubsection{Linear stability analysis}
\label{sec:oat15a_lst}

LST was performed on the transonic buffet case.
The eigensolver used was the time--stepper Arnoldi method with inverse iteration polish, as described in~\Cref{sec:solver}.
The base flow steady--state contours are presented in~\Cref{fig:oat15a_steady_contour}, and the dominant unstable eigenmode is shown later as the baseline panel of~\Cref{fig:oat15a_opt_evec_compare} in the optimization results section.
The unstable non--dimensional eigenvalues for this base flow are $0.021206 \pm 0.391287\,\mathrm{j}$.
The Strouhal number from this eigenfrequency is computed to be 0.072, in close agreement with the unsteady value of 0.074 reported above and with the value reported by~\citet{Sartor2015}.
A grid convergence study of the dominant eigenvalue was also conducted at the same flight condition ($\alpha = 4^\circ$, $M = 0.73$, $Re = 3.2\times 10^{6}$) on four mesh levels: the L0 mesh of~\Cref{tab:oat15a_grids}, its two uniform coarsenings L1 and L2, and one additional refinement L--0.5 obtained by uniformly halving the cell spacing of L0 along each curvilinear direction.
Each eigenpair was returned by the shift--invert eigensolver of~\Cref{sec:shift_invert} and refined by inverse--iteration polishing where applicable, with the eigensolver relative error recorded for each level.
The results are reported in~\Cref{tab:oat15a_eig_grid}.

\begin{table}[h]
  \centering
  \caption{Grid convergence of the dominant unstable LST eigenvalue for the OAT15A airfoil at $\alpha = 4^\circ$, $M = 0.73$, $Re = 3.2\times 10^{6}$.}
  \label{tab:oat15a_eig_grid}
  \begin{tabular}{cccc}
    \toprule
    \multicolumn{1}{c}{Number of cells $N$}
      & \multicolumn{1}{c}{$\mathrm{Re}(\lambda)$}
      & \multicolumn{1}{c}{$\mathrm{Im}(\lambda)$}
      & \multicolumn{1}{c}{Relative error} \\
    \midrule
    \multicolumn{1}{c}{$175{,}576$} & \multicolumn{1}{r}{$0.00700$} & \multicolumn{1}{r}{$0.40109$} & \multicolumn{1}{r}{$2.5\times 10^{-8}$}  \\
    \multicolumn{1}{c}{$87{,}552$}  & \multicolumn{1}{r}{$0.02122$} & \multicolumn{1}{r}{$0.39120$} & \multicolumn{1}{r}{$1.26\times 10^{-8}$} \\
    \multicolumn{1}{c}{$21{,}888$}  & \multicolumn{1}{r}{$0.06286$} & \multicolumn{1}{r}{$0.37670$} & \multicolumn{1}{r}{$3.39\times 10^{-8}$} \\
    \multicolumn{1}{c}{$5{,}472$}   & \multicolumn{1}{r}{$0.10999$} & \multicolumn{1}{r}{$0.28257$} & \multicolumn{1}{r}{$3.28\times 10^{-7}$} \\
    \bottomrule
  \end{tabular}
\end{table}

The growth rate $\mathrm{Re}(\lambda)$ decreases monotonically from $0.110$ on L2 to $0.0070$ on L--0.5, while the angular frequency $\mathrm{Im}(\lambda)$ tightens to a band of $0.39$--$0.40$ from L1 onwards.
The L2 eigenvalue is an outlier in both components because L2 lies outside the asymptotic range for this quantity, consistent with the L2 lift and drag values reported in~\Cref{fig:oat15a_grid_refinement}.
The L0 mesh used throughout the LST results in this section is therefore in the asymptotic range for the dominant eigenpair, and the residual L0--vs--L--0.5 spread of $\sim 0.014$ in $\mathrm{Re}(\lambda)$ and $\sim 0.01$ in $\mathrm{Im}(\lambda)$ is taken as the discretization uncertainty of the reported eigenvalue.

\begin{figure}[H]
  \centering
  \includegraphics[width=\textwidth]{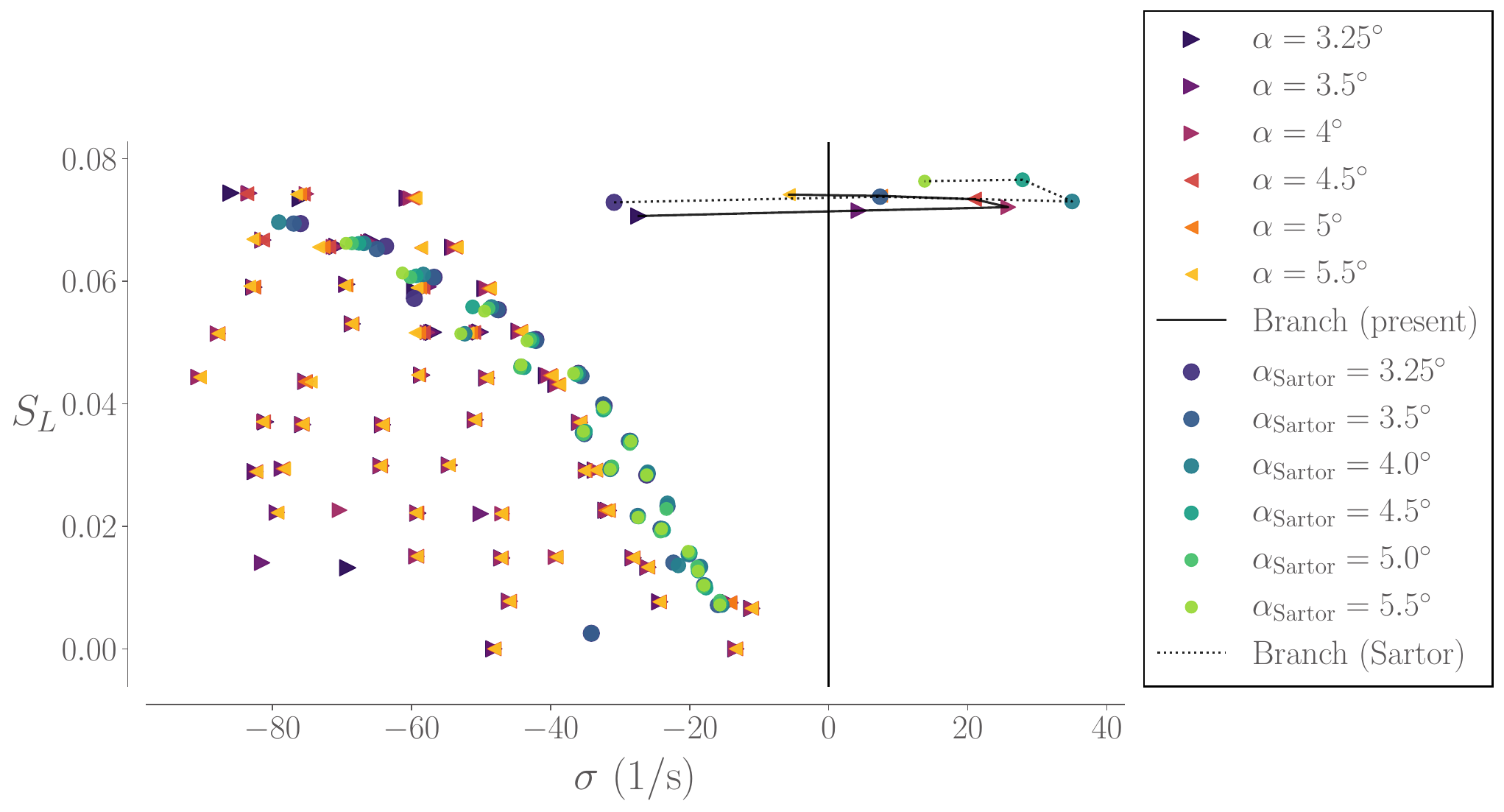}
  \caption{Leading eigenspectra of the OAT15A base flow at $M=0.73$, $Re=3.2\times 10^{6}$ for six angles of attack from $\alpha=3.25^\circ$ to $\alpha=5.5^\circ$, compared against the spectra reported by~\citet{Sartor2015}.
  Triangle markers are the present results ($\alpha$) and circles are the literature values ($\alpha_\mathrm{Sartor}$); each AoA pair shares a hue.
  The growth rate $\sigma$ is in $1/\mathrm{s}$ and the Strouhal number is $S_L$, following the convention of~\citet{Sartor2015}; the conversion from our nondimensional eigenvalues is detailed in~\Cref{app:eig_conversion}.
  The vertical line at $\sigma = 0$ separates stable modes (left) from unstable modes (right).}
  \label{fig:oat15a_lst_spectrum}
\end{figure}
% [x] TODO HS-: ours--infernal, literature--viridis?

To benchmark the spectrum more comprehensively, we sweep the angle of attack from $\alpha = 3.25^\circ$ to $\alpha = 5.5^\circ$ and compare the leading eigenvalues against the spectrum reported by~\citet{Sartor2015}.
The eigenvalues are obtained using the shift--invert method of~\Cref{sec:shift_invert} with multiple complex shifts placed across the spectrum, and every eigenpair returned by the solver is converged to a SLEPc relative residual below $10^{-9}$.
The two studies use different non--dimensionalization conventions, so our nondimensional eigenvalues are converted to the growth rate $\sigma$ in $1/\mathrm{s}$ and the Strouhal number $S_L$ used by~\citet{Sartor2015} following the procedure derived in~\Cref{app:eig_conversion}.
The resulting spectra are shown in~\Cref{fig:oat15a_lst_spectrum}.
Our spectrum tracks the literature data closely across all six angles of attack, including the rightmost mode that crosses the imaginary axis between $\alpha = 3.5^\circ$ and $\alpha = 4^\circ$, confirming that the LST implementation in ADflow reproduces the published transonic--buffet bifurcation behavior of the OAT15A airfoil.

For the OAT15A buffet case, the same direct--LU CN--Arnoldi path converges to the rightmost eigenvalue on the $n = 131{,}328$--dof baseline grid in $\sim 40$~s on $8$ MPI ranks ($\lambda = 0.062813 + 0.376739\,\mathrm{j}$, residual $5.2\times 10^{-8}$), and on the $n = 525{,}312$--dof fine grid in $\sim 179$~s ($\lambda = 0.0212060 + 0.3912872\,\mathrm{j}$, residual $4.96\times 10^{-8}$).
On the fine mesh we set $\Delta t = 5$ near the optimum $\Delta t^\star = 2/|\lambda|$ that maximizes the Cayley separation~\cref{eq:mu_mag}; the breakdown of the fine--mesh wall time is $3.21$~s of operator assembly, $66.6$~s of MUMPS LU factorization of $\mb{P}$, $27.0$~s of EPS Krylov--Schur, and $81.5$~s of inverse--iteration polish, with the LU factorization dominating as expected.

\Cref{tab:eigsolve_headtohead} compares the CN--Arnoldi path against the direct--LU complex shift--invert reference path of~\Cref{sec:shift_invert} on the same fine mesh.
The CN--Arnoldi path is $\sim 4.4\times$ faster end--to--end than the complex shift--invert reference; excluding the polish step, the LU--factor + Arnoldi stages alone total $66.6 + 27.0 = 93.6$~s versus $786$~s for the complex shift--invert, a $\sim 8.4\times$ speedup for the same algorithmic stage.
The two eigenvalues agree to $\sim 1.3\times 10^{-6}$ in absolute value, providing cross--validation of the CN--Arnoldi result against the complex shift--invert reference.

\begin{table}[h]
\centering
\caption{Direct-LU CN-Arnoldi versus direct-LU complex shift-invert on the OAT15A fine mesh ($n = 525{,}312$, $n_p = 8$).}
\label{tab:eigsolve_headtohead}
\begin{tabular}{lcc}
\toprule
Stage
  & \multicolumn{1}{c}{CN-Arnoldi}
  & \multicolumn{1}{c}{Shift-invert} \\
\midrule
Operator assembly      & \multicolumn{1}{r}{$3.21$~s}    & \multicolumn{1}{r}{---} \\
LU factor              & \multicolumn{1}{r}{$66.6$~s}    & \multicolumn{1}{r}{$736$~s} \\
Krylov--Schur Arnoldi  & \multicolumn{1}{r}{$27.0$~s}    & \multicolumn{1}{r}{$50$~s} \\
Polish                 & \multicolumn{1}{r}{$81.5$~s}    & \multicolumn{1}{r}{---} \\
\midrule
Total wall             & \multicolumn{1}{r}{$\sim 179$~s} & \multicolumn{1}{r}{$\sim 786$~s} \\
$\lambda_r$            & \multicolumn{1}{r}{$0.0212060$}  & \multicolumn{1}{r}{$0.0212051$} \\
$\lambda_i$            & \multicolumn{1}{r}{$0.3912872$}  & \multicolumn{1}{r}{$0.3912881$} \\
Final GEVP residual    & \multicolumn{1}{r}{$4.96\times 10^{-8}$} & \multicolumn{1}{r}{$5.49\times 10^{-9}$} \\
\bottomrule
\end{tabular}
\end{table}

For problem sizes where the global MUMPS factor of $\mb{P}$ no longer fits in memory ($n \gtrsim 5\times 10^{6}$ on the available hardware), the direct LU factorization must be replaced with iterative methods, and is left for future work on larger--scale systems.

\subsubsection{Verification against URANS}
\label{sec:oat15a_lst_verification}

\begin{figure}[H]
  \centering
  \includegraphics[width=0.75\textwidth]{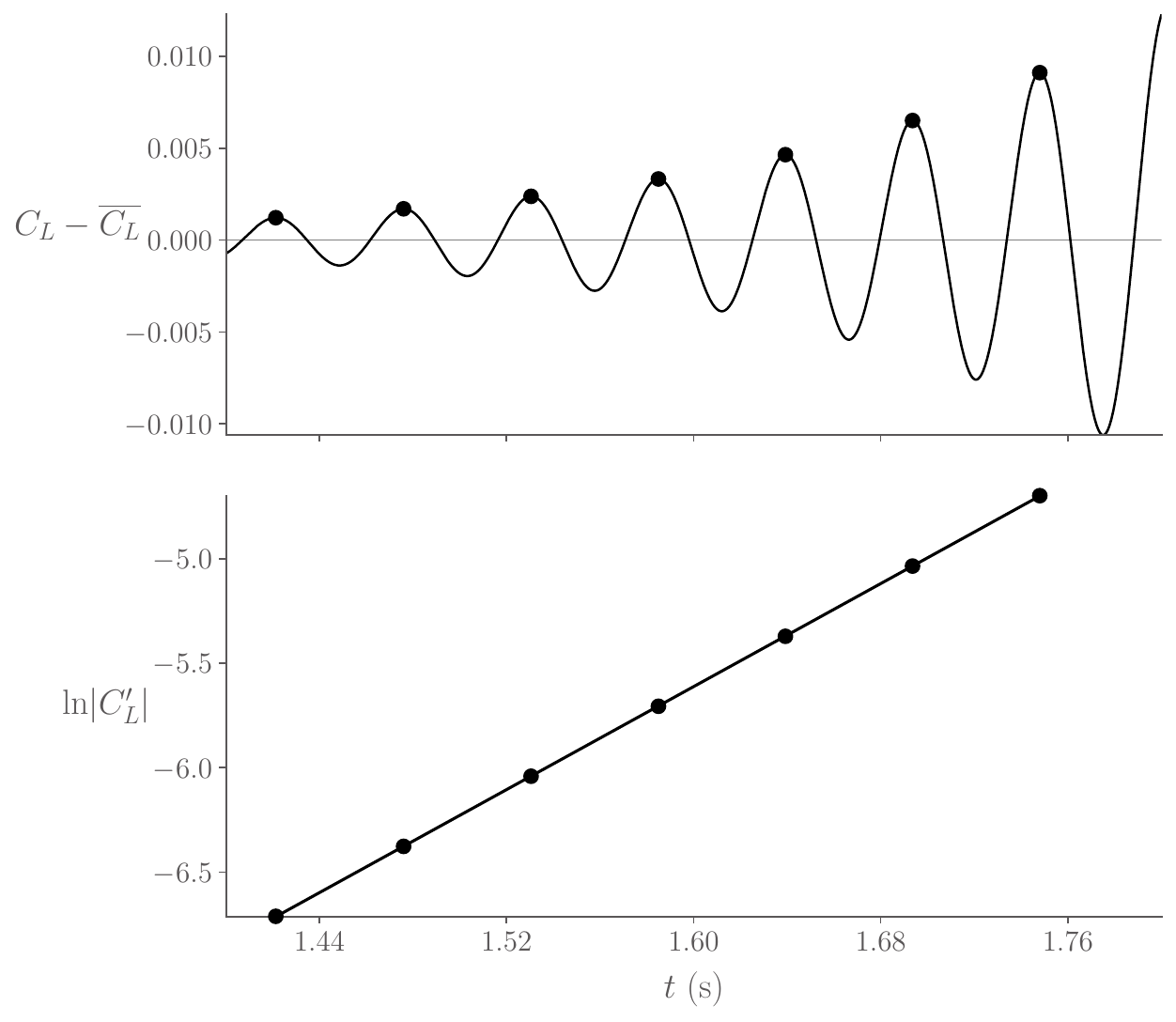}
  \caption{Verification of the LST eigenvalue at $\alpha = 4^\circ$ against the linear growth phase of the URANS run with $\Delta t = 5\times 10^{-5}$ s shown in~\Cref{fig:oat15a_cl_vs_time}.
  Top: demeaned lift coefficient $C_L^{\prime} = C_L - \overline{C_L}$ in the auto--selected linear--growth window $t \in [1.40,\,1.80]$ s, with detected peaks marked.
  Bottom: $\ln\lvert C_L^{\prime}\rvert$ at the seven peaks with the linear regression line whose slope equals the unsteady growth rate $\sigma_\mathrm{unsteady} = 6.16\ \mathrm{s}^{-1}$.
  The mean peak--to--peak period gives $f_\mathrm{unsteady} = 18.37$ Hz.}
  \label{fig:oat15a_lst_validation}
\end{figure}

As a further verification of the LST result, we compare the dominant eigenvalue at $\alpha = 4^\circ$ directly against the linear growth phase of the URANS time history of~\Cref{fig:oat15a_cl_vs_time}.
For a marginally unstable equilibrium, the lift fluctuation $C_L^{\prime}(t) \coloneqq C_L(t)-\overline{C_L}$ in the linear--regime portion of the response evolves as $C_L^{\prime}(t) = A_0\,\mathrm{e}^{\sigma t}\cos(\omega t + \phi)$, where $\sigma$ and $\omega$ are the real and imaginary parts of the dominant eigenvalue in dimensional units.
The growth rate $\sigma$ is therefore recovered as the slope of the upper envelope on a logarithmic scale, and the angular frequency $\omega = 2\pi f$ is recovered from the mean peak--to--peak period; the procedure is detailed in~\Cref{app:lst_unsteady_verification} and applied to the $\Delta t = 5\times 10^{-5}$ s URANS run of~\Cref{fig:oat15a_cl_vs_time}.
The auto--selected linear--growth window of $t \in [1.40,\,1.80]$ s contains seven peaks and yields $\sigma_\mathrm{unsteady} = 6.16\ \mathrm{s}^{-1}$ and $f_\mathrm{unsteady} = 18.37$ Hz from the peak fit shown in~\Cref{fig:oat15a_lst_validation}.
Converting the LST eigenvalue $\lambda = 0.021206 + 0.391287\,\mathrm{j}$ on the L0 mesh to physical units using~\Cref{eq:our_nondim} of~\Cref{app:eig_conversion} gives $\sigma_\mathrm{LST} = 6.23\ \mathrm{s}^{-1}$ and $f_\mathrm{LST} = 18.27$ Hz, agreeing with the unsteady values to within $1.1\%$ in growth rate and $0.5\%$ in frequency.
Combined with the literature spectrum overlay of~\Cref{fig:oat15a_lst_spectrum}, this peak--fit comparison serves as a verification--and--validation test of the LST implementation: the spectrum overlay is the validation against published data, and the unsteady peak fit is the internal verification that our eigenvalue solver, mass--matrix handling, and time--integrator agree on the same physical instability of the same base state.

It is a fact that the OAT15A transonic buffet phenomenon is weakly non--linear~\cite{Sartor2015,Crouch2009,Crouch2024}.
This means that the mean and base flows are almost identical, leading to the same Strouhal numbers computed from the unsteady limit cycle frequency and the base flow's LST derived eigenfrequency.
Our current focus is to investigate this with ADflow and benchmark the LST results for transonic buffet.
The dominant unstable eigenmode for this base flow is shown later in the optimization results section as the baseline panel of~\Cref{fig:oat15a_opt_evec_compare}.

\subsection{CFD adjoint derivative verification}
\label{sec:cfd_adjoint_verify}
% [] TODO HS-: move to the cylinder case?

\begin{figure}[H]
  \centering
  \includegraphics[width=0.95\textwidth]{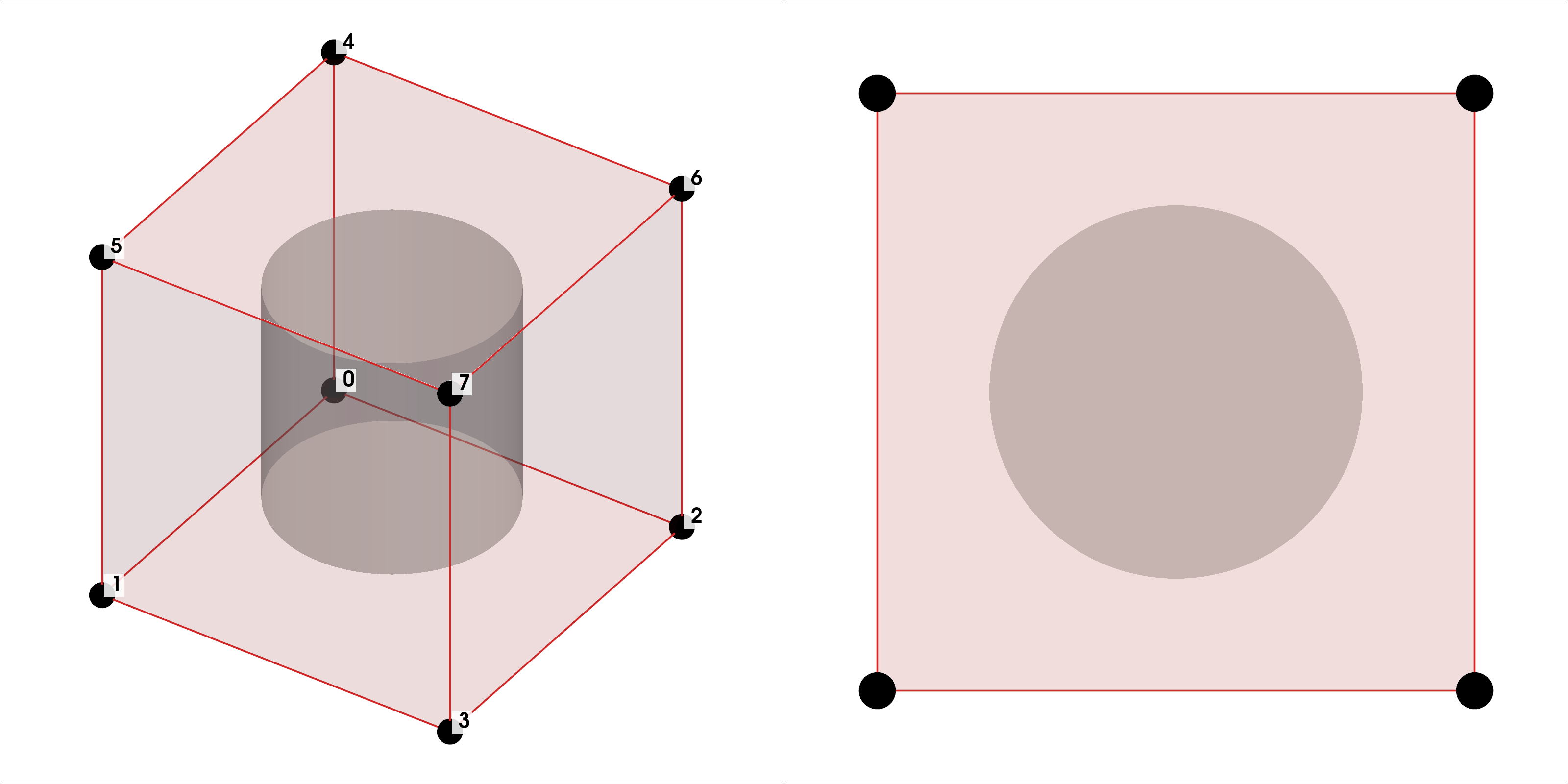}
  \caption{Free--form deformation (FFD) volume used to parameterize the cylinder mesh for the CFD adjoint verification, shown in isometric (left) and front (right) views.
  The translucent red box and red wireframe edges show the FFD volume, the eight black spheres are the FFD control points (with their indices labeled in the isometric view), and the gray solid cylinder is the body.}
  \label{fig:ffd_cylinder}
\end{figure}

To verify the linear stability coupled adjoint of~\Cref{sec:ls_adjoint} in a full CFD setting, we apply~\Cref{alg.derivative_mm} to the cylinder vortex--shedding benchmark of~\Cref{sec:cylinder} at the unstable condition $Re = 70$, $M = 0.05$, $\alpha = 0.005^\circ$.
The function of interest is the real part of the rightmost eigenvalue $\lambda_r$ of the linearized RANS--SA operator at the steady base flow, which is positive at this Reynolds number and quantifies the instability that drives the K\'arm\'an vortex street.
The angle of attack is offset slightly from $0^\circ$ on purpose: at $\alpha = 0^\circ$ the base flow is exactly symmetric about $y = 0$, so the gradient components for vertically--mirrored design variables are equal in magnitude with opposite sign by symmetry alone, and the verification cannot distinguish a correct adjoint from a one that merely reproduces the symmetry; the small $0.005^\circ$ offset breaks this exact symmetry and produces gradients that differ in magnitude pairwise, providing a stricter test.
The cylinder mesh is parameterized by a free--form deformation (FFD) volume of $2\times 2\times 2 = 8$ control points enclosing the cylinder, shown in~\Cref{fig:ffd_cylinder}; the in--plane $y$--displacement of each of the eight control points is taken as a design variable, giving $n_x = 8$, which is small enough that the central finite--difference reference is tractable.

The CFD adjoint derivative is computed using~\Cref{alg.derivative_mm}: the top--level adjoint admits the analytic solution of~\Cref{eq:analytic_top_sol_1} via the left eigenvector of $\mb{J}_\mathrm{mod}$, the bottom--level right--hand side is assembled from~\Cref{eq:mm_bm,eq:mm_b} using one CDRAD evaluation per eigenvector component, and the bottom--level adjoint is solved with ADflow's parallel KSP at relative tolerance $10^{-11}$.
The finite--difference reference uses central differences with step size $h = 10^{-6}$, requiring $1 + 2 n_x = 17$ primal solves and $2 n_x = 16$ generalized eigenvalue solves; the adjoint requires a single primal solve, a single eigenvalue solve, and a single bottom--level adjoint solve regardless of $n_x$.

The component--wise comparison is reported in~\Cref{tab:cfd_adjoint_verify}.
The two derivatives agree to a relative error in the range $2.9$--$5.0\times 10^{-5}$ on each component, and the norm of the difference is $\lVert\nabla_{\!x}\lambda_r^{\mathrm{adj}} - \nabla_{\!x}\lambda_r^{\mathrm{FD}}\rVert_2 = 1.05\times 10^{-7}$ against a gradient norm of $\lVert\nabla_{\!x}\lambda_r^{\mathrm{FD}}\rVert_2 = 2.84\times 10^{-3}$, giving a relative norm error of $3.7\times 10^{-5}$.
The residual error is consistent with the truncation error of central differences at $h = 10^{-6}$ on a quantity of order $10^{-3}$, indicating that the linear stability coupled adjoint is correct to the limit of what finite differences can verify and that no further sources of error are present in the CDRAD evaluation, the analytic mass--matrix contribution of~\Cref{eq:mm_bm}, or the bottom--level adjoint solve.
The near--mirror pattern visible in~\Cref{tab:cfd_adjoint_verify} between components $(0,2)$, $(1,3)$, $(4,6)$, and $(5,7)$ reflects the geometric layout of the FFD about the cylinder centerline at $y = 0$, where the two control points in each pair are mirror images; the small $\alpha = 0.005^\circ$ offset breaks the exact mirror so the paired magnitudes are not identical, and both the adjoint and the finite--difference gradients pick up the same small asymmetry to within the truncation error of the finite--difference reference.

\begin{table}[h]
  \centering
  \caption{Comparison of the adjoint and finite-difference gradients $\d\lambda_r/\d\mb{x}$.}
  \label{tab:cfd_adjoint_verify}
  \begin{tabular}{cccc}
    \toprule
    \multicolumn{1}{l}{Component $i$}
      & \multicolumn{1}{l}{Adjoint $\partial\lambda_r/\partial x_i$}
      & \multicolumn{1}{l}{FD $\partial\lambda_r/\partial x_i$}
      & \multicolumn{1}{l}{Relative difference} \\
    \midrule
    \multicolumn{1}{r}{$0$} & \multicolumn{1}{r}{$-1.18147\times 10^{-3}$} & \multicolumn{1}{r}{$-1.18143\times 10^{-3}$} & \multicolumn{1}{r}{$3.40\times 10^{-5}$} \\
    \multicolumn{1}{r}{$1$} & \multicolumn{1}{r}{$-8.02607\times 10^{-4}$} & \multicolumn{1}{r}{$-8.02640\times 10^{-4}$} & \multicolumn{1}{r}{$4.19\times 10^{-5}$} \\
    \multicolumn{1}{r}{$2$} & \multicolumn{1}{r}{$+1.18159\times 10^{-3}$} & \multicolumn{1}{r}{$+1.18163\times 10^{-3}$} & \multicolumn{1}{r}{$3.38\times 10^{-5}$} \\
    \multicolumn{1}{r}{$3$} & \multicolumn{1}{r}{$+8.02458\times 10^{-4}$} & \multicolumn{1}{r}{$+8.02424\times 10^{-4}$} & \multicolumn{1}{r}{$4.29\times 10^{-5}$} \\
    \multicolumn{1}{r}{$4$} & \multicolumn{1}{r}{$-1.13997\times 10^{-3}$} & \multicolumn{1}{r}{$-1.13994\times 10^{-3}$} & \multicolumn{1}{r}{$2.88\times 10^{-5}$} \\
    \multicolumn{1}{r}{$5$} & \multicolumn{1}{r}{$-8.39036\times 10^{-4}$} & \multicolumn{1}{r}{$-8.39075\times 10^{-4}$} & \multicolumn{1}{r}{$4.64\times 10^{-5}$} \\
    \multicolumn{1}{r}{$6$} & \multicolumn{1}{r}{$+1.13984\times 10^{-3}$} & \multicolumn{1}{r}{$+1.13987\times 10^{-3}$} & \multicolumn{1}{r}{$2.94\times 10^{-5}$} \\
    \multicolumn{1}{r}{$7$} & \multicolumn{1}{r}{$+8.38973\times 10^{-4}$} & \multicolumn{1}{r}{$+8.38931\times 10^{-4}$} & \multicolumn{1}{r}{$5.00\times 10^{-5}$} \\
    \bottomrule
  \end{tabular}
\end{table}

The cost difference between the two methods is the practical motivation for the adjoint.
At $n_x = 8$ design variables, the central--difference reference requires $16$ additional primal CFD solves and $16$ additional generalized eigenvalue solves on top of the unperturbed pair, so its cost scales linearly with $n_x$ as $1 + 2 n_x$ primals and $2 n_x$ eigensolves.
The linear stability coupled adjoint replaces all of these with one bottom--level adjoint linear solve and a single CDRAD evaluation that contains two ADflow reverse Jacobian--vector products, giving a cost that is independent of $n_x$.
For a typical optimization design space of $n_x \sim 100$ FFD control points the finite--difference reference would require on the order of $200$ extra primal solves and $200$ eigensolves per gradient evaluation, which is unaffordable for the OAT15A buffet--constrained optimization problem of~\Cref{sec:crm_optimization}; the adjoint reduces this to a single additional linear solve.

\subsection{Buffet--constrained drag minimization for the OAT15A airfoil}
\label{sec:crm_optimization}

\begin{figure}[H]
  \centering
  \includegraphics[width=0.85\textwidth]{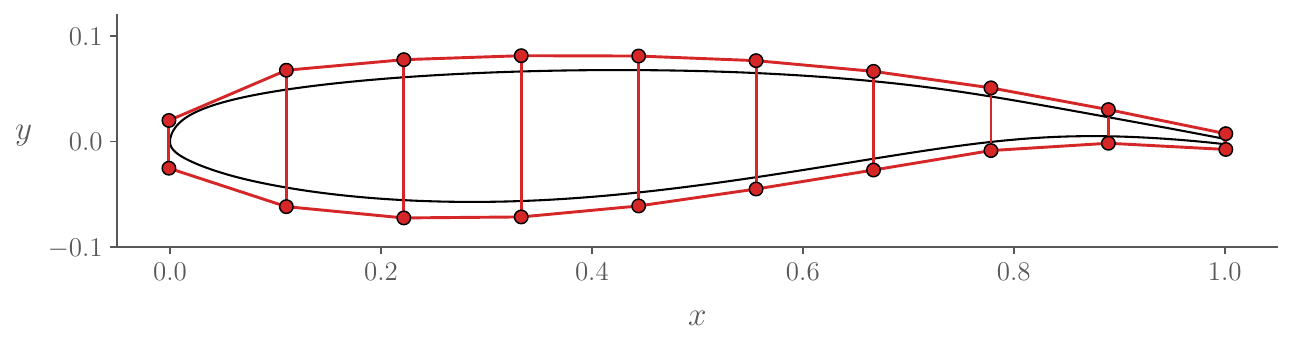}
  \caption{Front view of the $10\times 2\times 2$ FFD lattice that parameterizes the OAT15A airfoil for the single--point buffet--constrained drag minimization.
  The 20 in--plane control points (red filled circles) have $y$--displacement design variables; the $z$--direction is fixed.}
  \label{fig:oat15a_opt_ffd}
\end{figure}

We close the results section with a single--point gradient--based shape optimization of the OAT15A airfoil at the same flight condition used for the LST analysis ($M = 0.73$, $Re = 3.2\times 10^{6}$, $T_\infty = 300$ K), with the buffet--stability constraint $\mathrm{Re}(\lambda) \le 0$ supplied by the linear stability coupled adjoint of~\Cref{sec:ls_adjoint}.
The aerodynamic surface is parameterized by a $10\times 2\times 2$ FFD volume of $40$ control points whose in--plane $y$--displacements are taken as shape design variables; the FFD lattice and the airfoil it encloses are shown in~\Cref{fig:oat15a_opt_ffd}.
The angle of attack $\alpha$ is added as a $41^\mathrm{st}$ design variable so that the lift target can be balanced through both shape and incidence.

The optimization problem is summarized in~\Cref{tab:oat15a_opt_problem}.
The objective is the drag coefficient $C_D$ scaled by $10^{4}$ (drag counts), the lift target is $C_L^* = 0.98$ entered as an equality constraint $C_L - C_L^* = 0$, and the buffet--stability inequality $\mathrm{Re}(\lambda) \le 0$ is enforced through the LST adjoint.
The geometric constraints are the volume bound $V/V_0 \in [1.0, 1.08]$, $200$ thickness--at--station inequalities $t \ge 0.1\,t_\mathrm{baseline}$, and four linear leading--edge / trailing--edge and spanwise symmetry constraints, all evaluated by pyGeo's DVConstraints.
SLSQP from pyOptSparse drives the iteration with default tolerance $\mathrm{acc} = 10^{-6}$; the optimizer terminated at iteration $72$ with SLSQP exit code $\mathrm{Inform} = 8$ (``positive directional derivative for linesearch'').
The total wall time was $17{,}926$ s ($\sim 5$ hr) on $120$ MPI ranks.
The post--optimization Arnoldi verification at three independent Cayley step sizes (presented later in this subsection) independently confirms that $\mathrm{Re}(\lambda) < 0$ at the iter--72 design irrespective of the optimizer's internal Jacobian.

% [x] TODO HS-: usually it is considered as 20 variables due to symmetruy?
\begin{table}[h]
  \centering
  \caption{Single--point buffet--constrained drag minimization formulation for the OAT15A airfoil at $M = 0.73$, $Re = 3.2\times 10^{6}$, $T_\infty = 300$ K.}
  \label{tab:oat15a_opt_problem}
  \begin{tabular}{ll}
    \toprule
    Quantity & Specification \\
    \midrule
    Objective              & minimize $C_D$ \\
    Design variables       & $40$ FFD shape DVs ($20$ effective with $z$--spanwise symmetry, \\
                           & $y$--displacements, bounds $\pm 0.05$) + $\alpha \in [3^\circ, 5^\circ]$ \\
    Lift constraint        & $C_L = C_L^* = 0.98$ (equality) \\
    Buffet constraint      & $\mathrm{Re}(\lambda) \le 0$ (LST adjoint, single $\Delta t = 5$~s in eigsolve) \\
    Volume constraint      & $V/V_0 \in [1.0, 1.08]$ \\
    Thickness constraints  & $200$ stations, each $t \ge 0.1\,t_\mathrm{baseline}$ \\
    Symmetry constraints   & LE/TE pairing and $z$--spanwise symmetry (linear) \\
    Optimizer              & SLSQP via pyOptSparse, $\mathrm{acc} = 10^{-6}$, $\mathrm{maxiter} = 500$ \\
    \bottomrule
  \end{tabular}
\end{table}

\begin{figure}[H]
  \centering
  \includegraphics[width=0.85\textwidth]{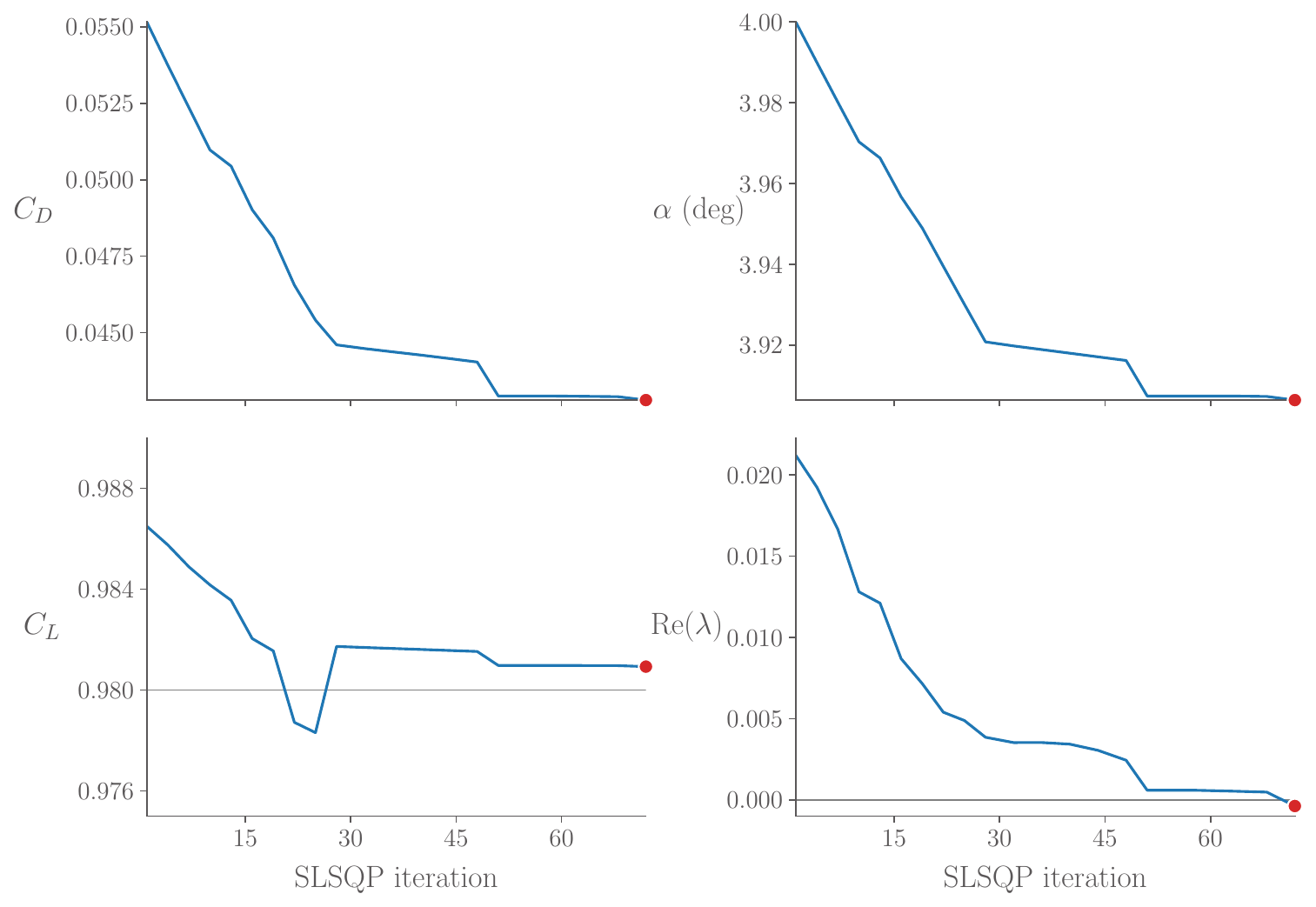}
  \caption{Convergence history of the buffet--constrained drag minimization at the $20$ SLSQP gradient--evaluation iterations.
  Top--left: drag coefficient $C_D$. Top--right: angle of attack $\alpha$. Bottom--left: lift coefficient $C_L$ with the equality target $C_L^* = 0.98$ shown as a horizontal grey line. Bottom--right: real part of the dominant eigenvalue $\mathrm{Re}(\lambda)$ with the buffet--stability threshold $\mathrm{Re}(\lambda) = 0$ shown as a horizontal grey line.
  The red dot marks the converged iterate at iteration $72$.}
  \label{fig:oat15a_opt_iter_history}
\end{figure}

The convergence history of the four objective and constraint quantities at the $20$ SLSQP gradient--evaluation iterations is shown in~\Cref{fig:oat15a_opt_iter_history}.
$C_D$ drops monotonically from $0.05516$ at the baseline to $0.04279$ at iteration $72$, a reduction of $124$ drag counts or $22.4\%$.
The angle of attack $\alpha$ falls from $4.000^\circ$ to $3.906^\circ$ as the optimizer trades incidence for shape lift; $C_L$ stays in a tight band around the equality target $C_L^* = 0.98$ throughout the run, with at most a few thousandths of residual at convergence due to the gradient--Jacobian reasons discussed above.
The growth rate $\mathrm{Re}(\lambda)$ decreases monotonically from $+0.02121$ at the baseline (unstable) through the optimization, becomes active around iteration $50$, and settles at $-3.74\times 10^{-4}$ at iteration $72$ (stable).
The buffet constraint is therefore active at the optimum, and the resulting design lies on the buffet--stability boundary.
% [] TODO HS-: is the 1e-4 within tolerance of 0? if so it is numerically active in general for inequality constraint to be active the equality needs to hold. If it is active, the opt w/o buffet is meaningful and should be shown.

% [x] TODO HS-: CL figure is not autozoned? the range style is diffeerent from the rest figures.

\begin{figure}[H]
  \centering
  \includegraphics[width=0.85\textwidth]{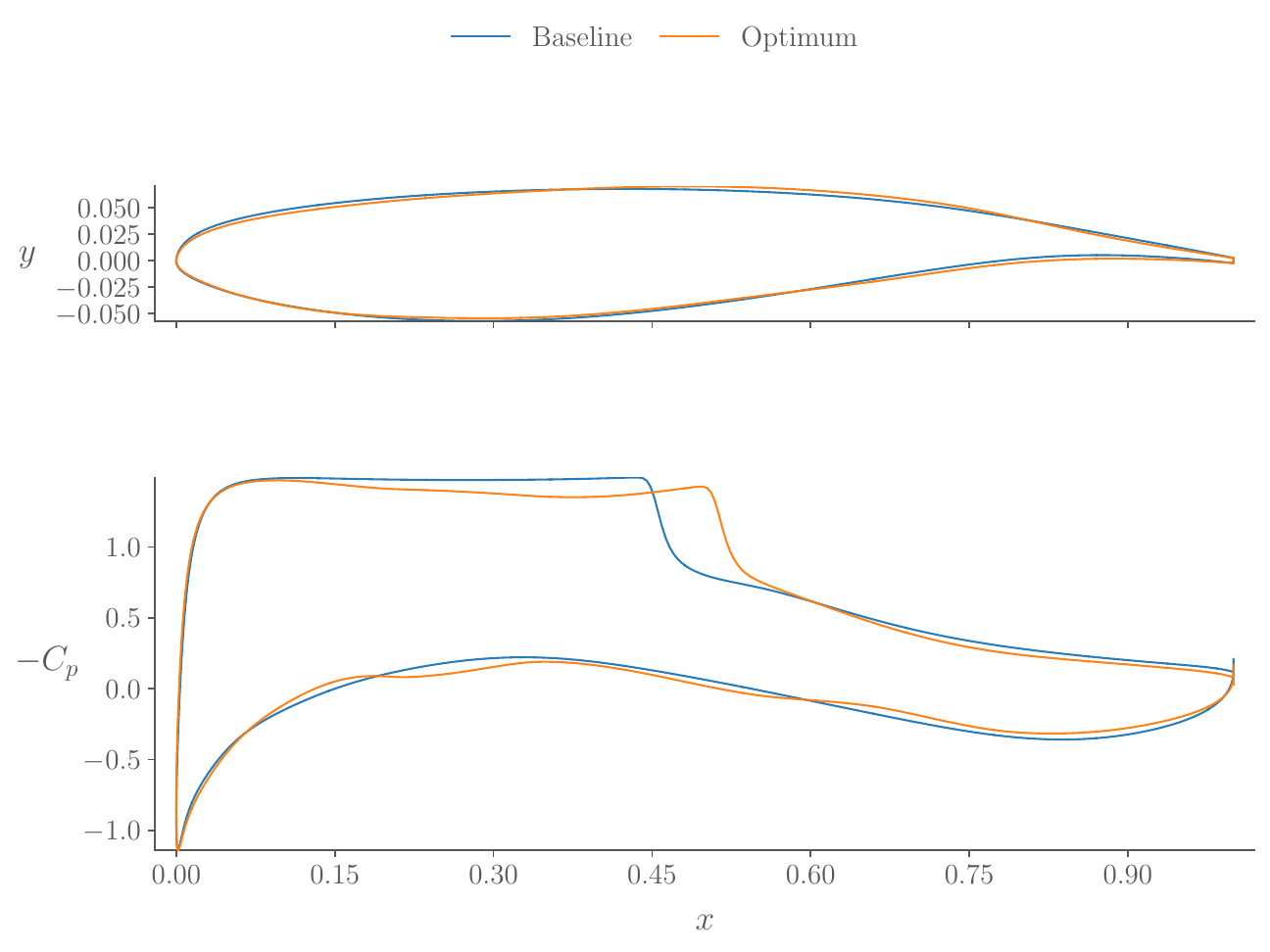}
  \caption{Baseline (blue) versus optimum (orange) at $M = 0.73$, $Re = 3.2\times 10^{6}$.
  Top: airfoil shape. Bottom: surface $-C_p$ at the $z = 0.5$ slice.
  The shock on the suction side moves slightly aft and weakens at the optimum, consistent with the drag reduction.}
  \label{fig:oat15a_opt_shape_cp}
\end{figure}

The geometric and aerodynamic changes from baseline to optimum are shown in~\Cref{fig:oat15a_opt_shape_cp}.
The shape change is small in magnitude ($\lvert \Delta y\rvert$ peaks below $0.005$ chord) but redistributes thickness aft of $x \approx 0.4$ to relax the suction--side curvature where the baseline shock sits.
The corresponding surface pressure in the lower panel shows the shock weakening: the baseline pressure trace has a sharp $-C_p$ drop from about $1.4$ to $0.6$ at $x \approx 0.45$, while the optimum's drop is shallower, smeared across $0.50 \le x \le 0.55$, and starts from a lower upstream plateau.
This shock--weakening pattern is the canonical mechanism behind buffet--onset push--back in transonic supercritical airfoils.

\begin{figure}[H]
  \centering
  \includegraphics[width=0.95\textwidth]{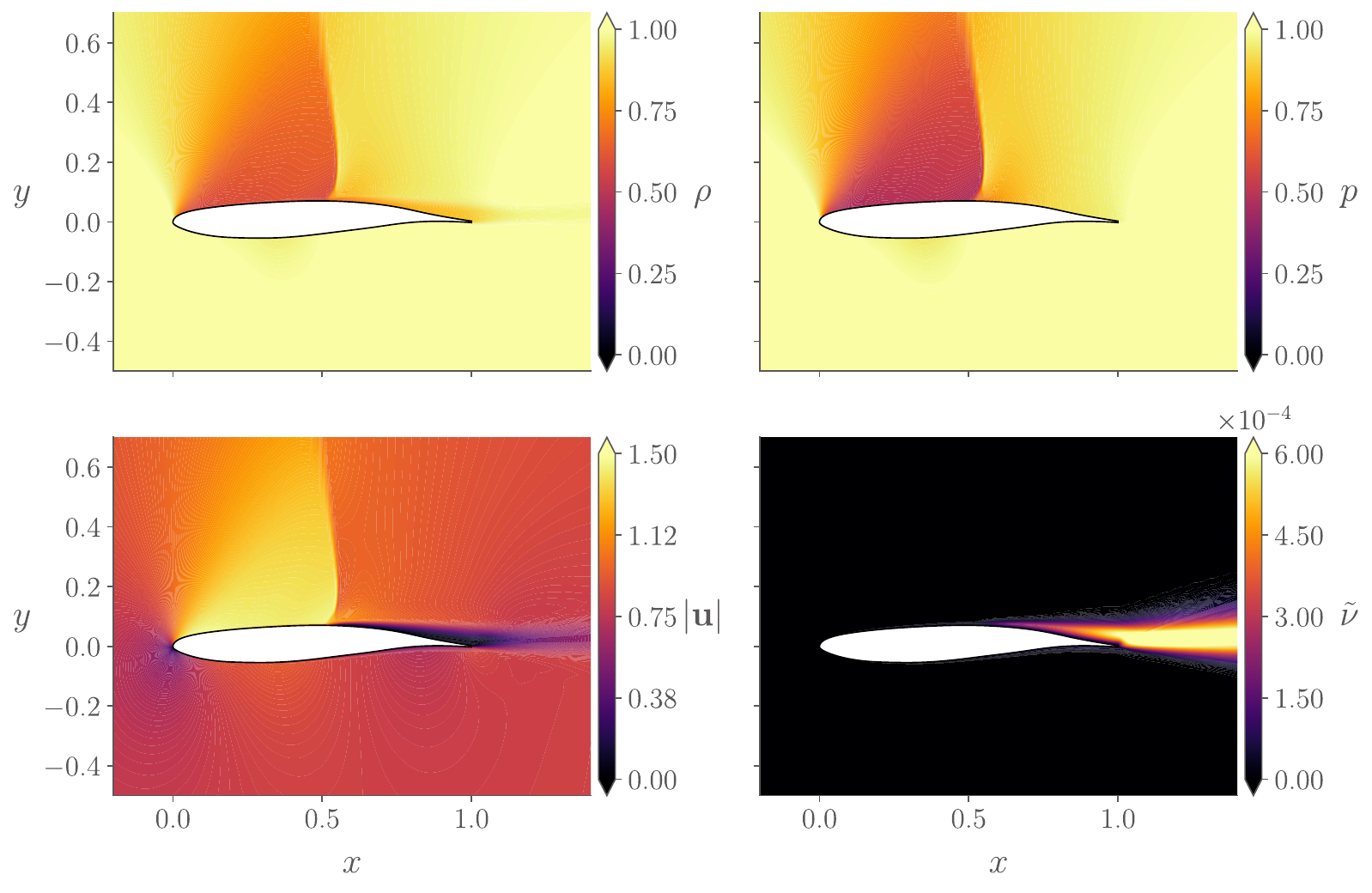}
  \caption{Steady--state contours of density $\rho$, pressure $p$, velocity magnitude $\lvert\mathbf{u}\rvert$, and Spalart--Allmaras working variable $\tilde{\nu}$ at $z = 0.5$ for the optimized OAT15A airfoil (iteration 72) at $M = 0.73$, $Re = 3.2\times 10^{6}$, $\alpha = 3.906^\circ$.
  Colormaps and limits match the baseline tile in~\Cref{fig:oat15a_steady_contour}, so the two figures can be compared at a glance.}
  \label{fig:oat15a_opt_steady_contour}
\end{figure}

\begin{figure}[H]
  \centering
  \includegraphics[width=\textwidth]{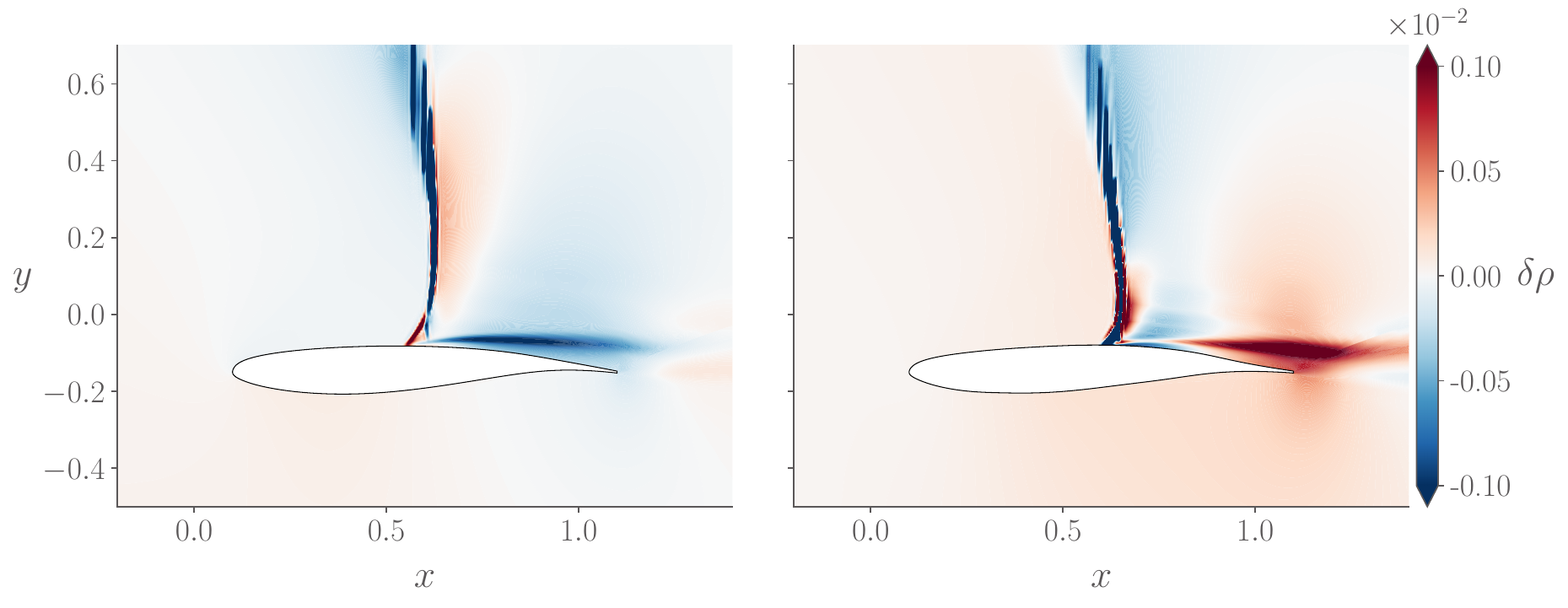}
  \caption{Real part of the density component $\delta\rho$ of the dominant LST eigenmode at $z = 0.5$ for the baseline (left, $\lambda = 0.021206 + 0.391287\,\mathrm{j}$) and the optimum (right, $\lambda = -3.74\times 10^{-4} + 0.4823\,\mathrm{j}$).}
  \label{fig:oat15a_opt_evec_compare}
\end{figure}

The same shock--weakening trend is visible in the steady--state field tile of the optimized airfoil shown in~\Cref{fig:oat15a_opt_steady_contour}, which presents density, pressure, velocity magnitude, and Spalart--Allmaras working variable on the same colormaps and limits as the baseline buffet tile of~\Cref{fig:oat15a_steady_contour}; the reader is referred to that earlier figure for the head--to--head comparison against the unoptimized OAT15A buffet base flow.
Compared to the baseline, the supersonic pocket on the suction side is smaller, the peak velocity magnitude over the optimized airfoil is reduced, the post--shock pressure recovery is gentler, and the turbulent--SA signature in the separated wake is contracted, consistent with the surface $-C_p$ change in~\Cref{fig:oat15a_opt_shape_cp}.
The corresponding LST density eigenmode in~\Cref{fig:oat15a_opt_evec_compare} shows the same physical change in eigenvector space: the strong baseline dipole at the shock foot and the extended signature in the separated boundary layer aft of the shock are both attenuated and shifted aft at the optimum, consistent with the eigenvalue having moved from the right half--plane to the left half--plane.

The optimum design is summarized in~\Cref{tab:oat15a_opt_summary}.
$C_D$ drops by $22.4\%$ ($124$ drag counts) and the dominant eigenvalue's real part flips sign from $+0.02121$ to $-3.74\times 10^{-4}$, with the imaginary part shifting from $0.39129$ to $0.4823$ (the buffet frequency increases by about $23\%$).
The $\alpha$ change is $-0.094^\circ$ and $C_L$ is slightly above target.

\begin{table}[h]
  \centering
  \caption{Baseline (iter 1) versus optimum (iter 72) for the buffet--constrained drag minimization at $M = 0.73$, $Re = 3.2\times 10^{6}$.}
  \label{tab:oat15a_opt_summary}
  \begin{tabular}{lccc}
    \toprule
    Quantity
      & \multicolumn{1}{c}{Baseline}
      & \multicolumn{1}{c}{Optimum}
      & \multicolumn{1}{c}{$\Delta$} \\
    \midrule
    $C_D$                  & \multicolumn{1}{r}{$0.05516$}    & \multicolumn{1}{r}{$0.04279$}            & \multicolumn{1}{r}{$-22.4\%$ ($-124$ counts)} \\
    $C_L$                  & \multicolumn{1}{r}{$0.9865$}     & \multicolumn{1}{r}{$0.9809$}             & \multicolumn{1}{r}{$-0.6\%$} \\
    $\alpha$               & \multicolumn{1}{r}{$4.000^\circ$}& \multicolumn{1}{r}{$3.906^\circ$}        & \multicolumn{1}{r}{$-0.094^\circ$} \\
    $\mathrm{Re}(\lambda)$ & \multicolumn{1}{r}{$+0.02121$}   & \multicolumn{1}{r}{$-3.74\times 10^{-4}$}& \multicolumn{1}{r}{sign flip (stable)} \\
    $\mathrm{Im}(\lambda)$ & \multicolumn{1}{r}{$+0.39129$}   & \multicolumn{1}{r}{$+0.48230$}           & \multicolumn{1}{r}{$+23\%$} \\
    \bottomrule
  \end{tabular}
\end{table}

To independently confirm that the optimum is genuinely buffet--stable, we re--ran the time--stepper Arnoldi eigensolver of~\Cref{sec:cayley} at the iter--72 design with three different Cayley step sizes $\Delta t \in \{3, 5, 10\}$ s, no inverse--iteration polish, and no LR fallback (pure Arnoldi targeting the largest--modulus mode of the Cayley transform).
All three returns $\mathrm{Re}(\lambda) \approx -3.08\times 10^{-4}$ with $\mathrm{Im}(\lambda) \approx 0.4851$, agreeing to five significant digits in the real part and seven in the imaginary part across the three independent solves.
This independent re--evaluation rules out the possibility that the optimizer's reported negative $\mathrm{Re}(\lambda)$ is an artifact of any single eigensolve setting and confirms that the optimum has no unstable eigenmode in the rightmost spectrum.

\begin{figure}[H]
  \centering
  \includegraphics[width=0.85\textwidth]{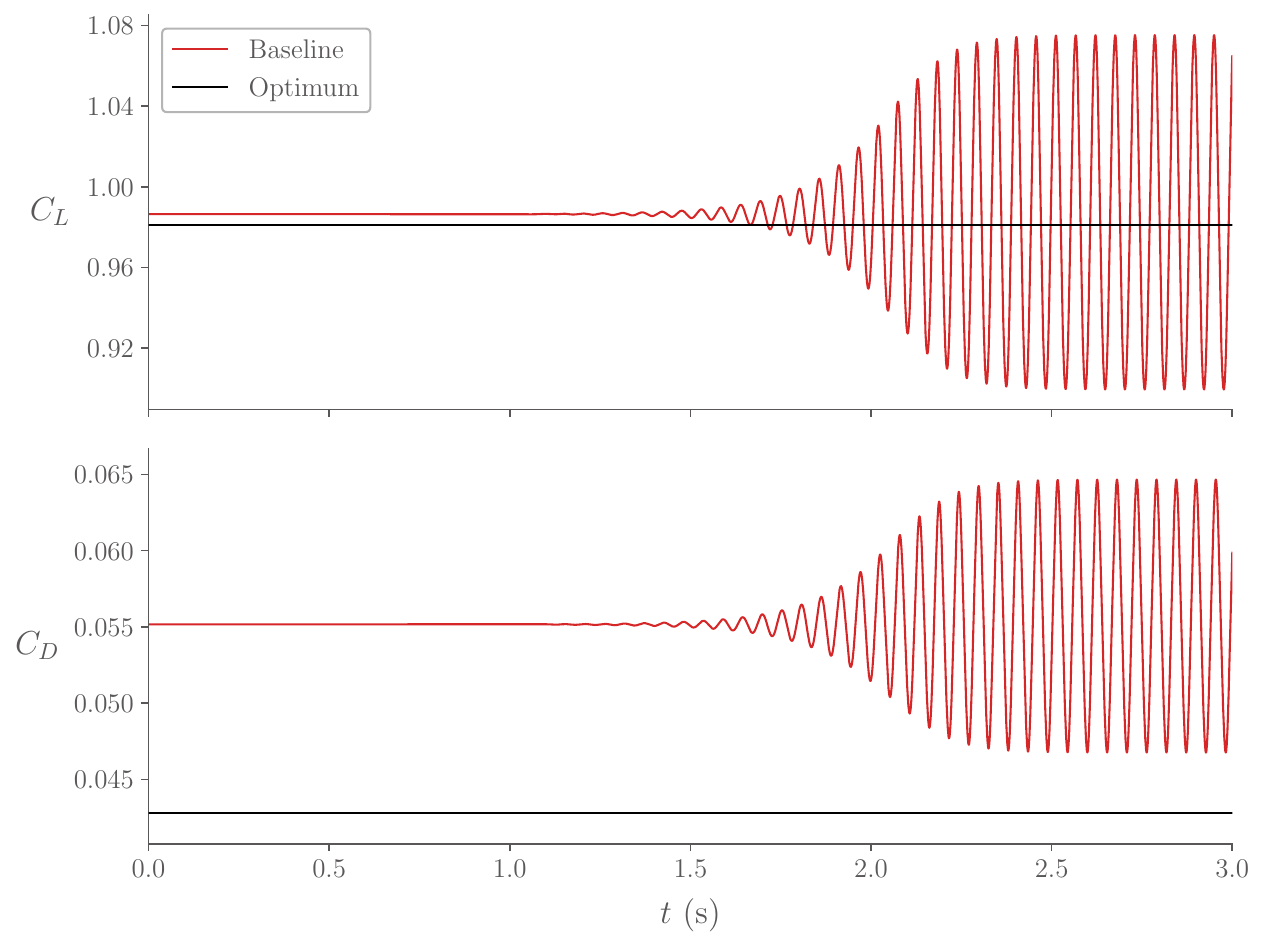}
  \caption{Baseline (red) versus optimum (black) URANS histories at $M = 0.73$, $Re = 3.2\times 10^{6}$, $\Delta t = 10^{-4}$ s, both cold--started from the steady RANS solution.
  Top: $C_L$ vs.\ time. Bottom: $C_D$ vs.\ time.
  The baseline develops the buffet limit cycle, while the optimum holds its initial steady--state values across the plotting window, confirming that the design is a genuinely stable equilibrium.}
  \label{fig:oat15a_opt_unsteady}
\end{figure}

%%% URANS cross-check: baseline LCO vs optimum flatline
As a final cross--check, we ran a cold--started URANS simulation on the optimized OAT15A design at the same flight condition ($M = 0.73$, $Re = 3.2\times 10^{6}$, $\Delta t = 10^{-4}$ s) using ADflow's BDF2 dual--time scheme, with the steady RANS solution as the initial condition.
\Cref{fig:oat15a_opt_unsteady} compares the resulting $C_L$ and $C_D$ histories against the baseline URANS run started from the same kind of initial condition.
The baseline trace develops the buffet limit--cycle oscillation by $t \approx 1$ s and saturates with $C_L$ swinging in $[0.90, 1.08]$ and $C_D$ in $[0.047, 0.065]$, whereas the optimum trace shows no growth and no limit--cycle oscillation across the plotting window: the cluster URANS holds $C_L = 0.9809$ and $C_D = 0.04279$ to within numerical noise across thousands of timesteps.
This is the time--domain manifestation of $\mathrm{Re}(\lambda) < 0$ at the optimum, and confirms that the design is a genuinely stable equilibrium rather than an artifact of the LST eigensolver.
The optimum's flat $C_D$ trace also lies entirely below the baseline LCO drag envelope, consistent with the $22.4\%$ drag reduction reported in~\Cref{tab:oat15a_opt_summary}.

\section{NASA CRM buffet}
\label{sec:crm_buffet}

\begin{figure}[H]
  \centering
  \includegraphics[width=0.85\textwidth]{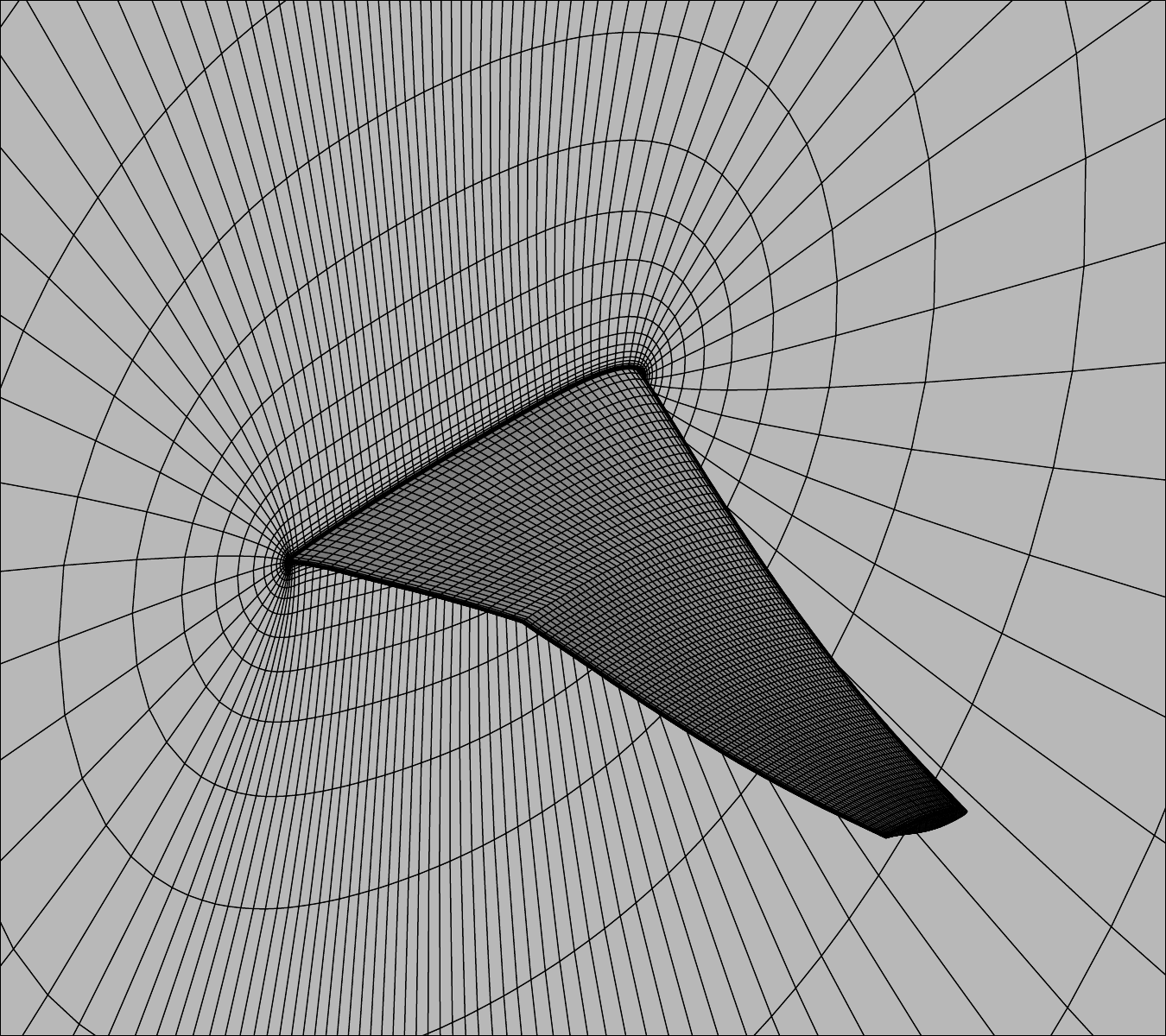}
  \caption{Structured multiblock mesh for the wing--only CRM~\cite{Lacy2016} configuration used in the present study.}
  \label{fig:crm_mesh}
\end{figure}

We next present preliminary CFD results on a three--dimensional configuration: the wing--only variant of the NASA CRM~\cite{Lacy2016}, a representative transonic--transport geometry.
No linear stability analysis is performed for this case; the present results are limited to CFD simulations and serve as a stepping stone toward future three--dimensional buffet--stability studies.
The freestream conditions are $M = 0.85$ and $Re = 5\times 10^{6}$, and we sweep the angle of attack from $\alpha = 3.5^\circ$ to $\alpha = 4.2^\circ$.
\Cref{fig:crm_mesh} shows the structured multiblock mesh used for the CRM--wing simulations, with approximately $450{,}000$ cells.

We first run a converged steady--state RANS solution at each angle of attack in the sweep, and these steady solutions are then used as the initial condition for the unsteady runs.
The unsteady simulations cover the full angle--of--attack sweep $\alpha \in [3.5^\circ, 4.2^\circ]$ in $0.1^\circ$ increments, integrated in URANS mode with ADflow's BDF2 dual--time scheme at a fixed time step of $\Delta t = 10^{-4}$ s.

\begin{figure}[H]
  \centering
  \includegraphics[width=0.8\textwidth]{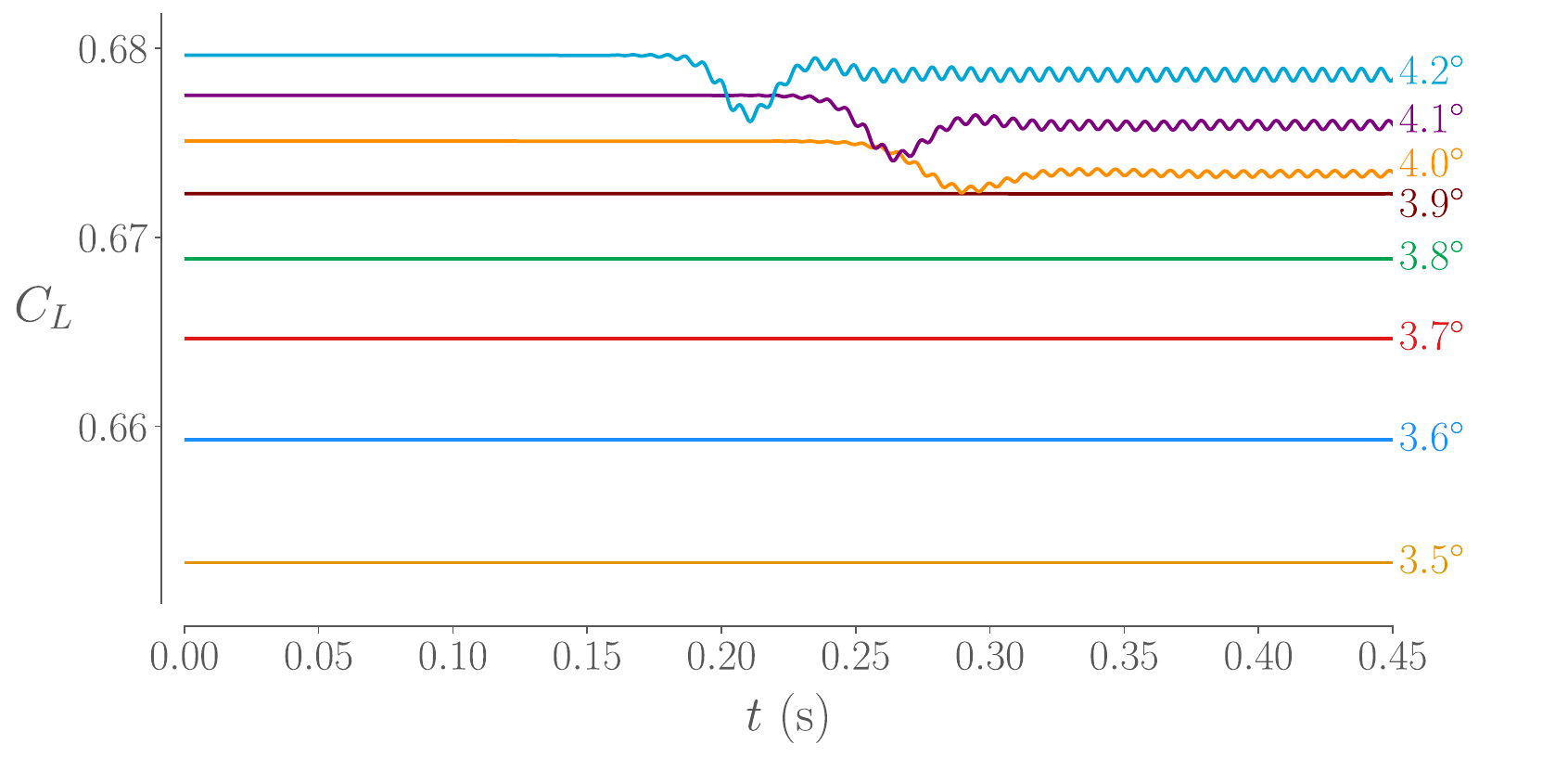}
  \caption{Unsteady lift coefficient $C_L$ vs time for the wing--only CRM at $M = 0.85$, $Re = 5\times 10^{6}$, swept across $\alpha \in [3.5^\circ, 4.2^\circ]$ in $0.1^\circ$ increments. Each run is warm--started from the steady--state RANS solution at its angle of attack and integrated with $\Delta t = 10^{-4}$ s.}
  \label{fig:crm_cl_vs_time}
\end{figure}

The resulting unsteady lift coefficient histories are shown in~\Cref{fig:crm_cl_vs_time}.
The lower--$\alpha$ traces remain nearly constant at their steady--state value across the simulated window, while the higher--$\alpha$ traces develop visible oscillations indicating that the steady solution is no longer a stable fixed point of the URANS dynamics; the oscillation amplitude grows with $\alpha$, consistent with progressive crossing of the buffet--onset boundary on this configuration.
From~\Cref{fig:crm_cl_vs_time} we observe buffet onset at approximately $\alpha \approx 4.0^\circ$ for the specified freestream conditions, with the $\alpha \le 3.9^\circ$ traces remaining quiescent and the $\alpha \ge 4.0^\circ$ traces developing sustained oscillations.
The buffet offset --- the angle of attack at which the limit cycle ceases on the descending branch --- was not computed in the present study and is left as future work.

\begin{figure}[H]
  \centering
  \includegraphics[width=\textwidth]{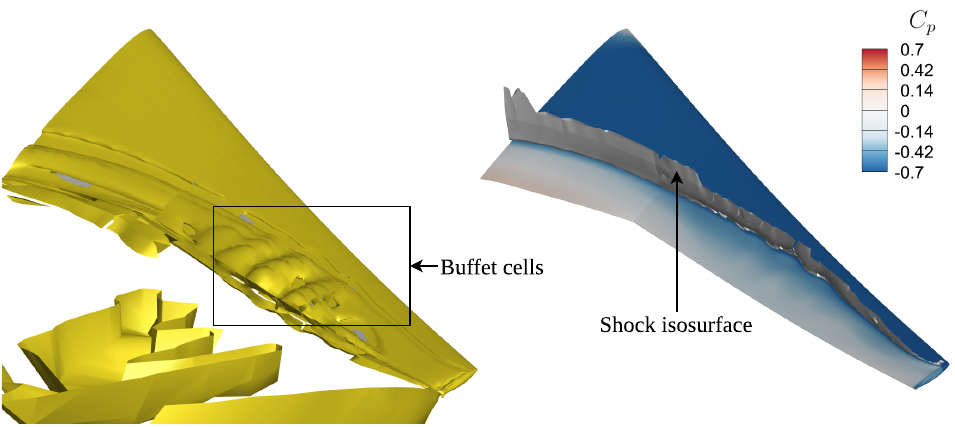}
  \caption{Same instant ($t = 0.45$ s, $\alpha = 4.2^\circ$) of the saturated limit--cycle oscillation viewed through two complementary diagnostics. Left: $Q$--criterion isosurface at $Q = 0.015$. Right: surface pressure coefficient $C_p$ with the shockwave isosurface from ADflow's shock sensor.}
  \label{fig:crm_lco_qcrit_cp}
\end{figure}

\begin{figure}[H]
  \centering
  \includegraphics[width=0.95\textwidth]{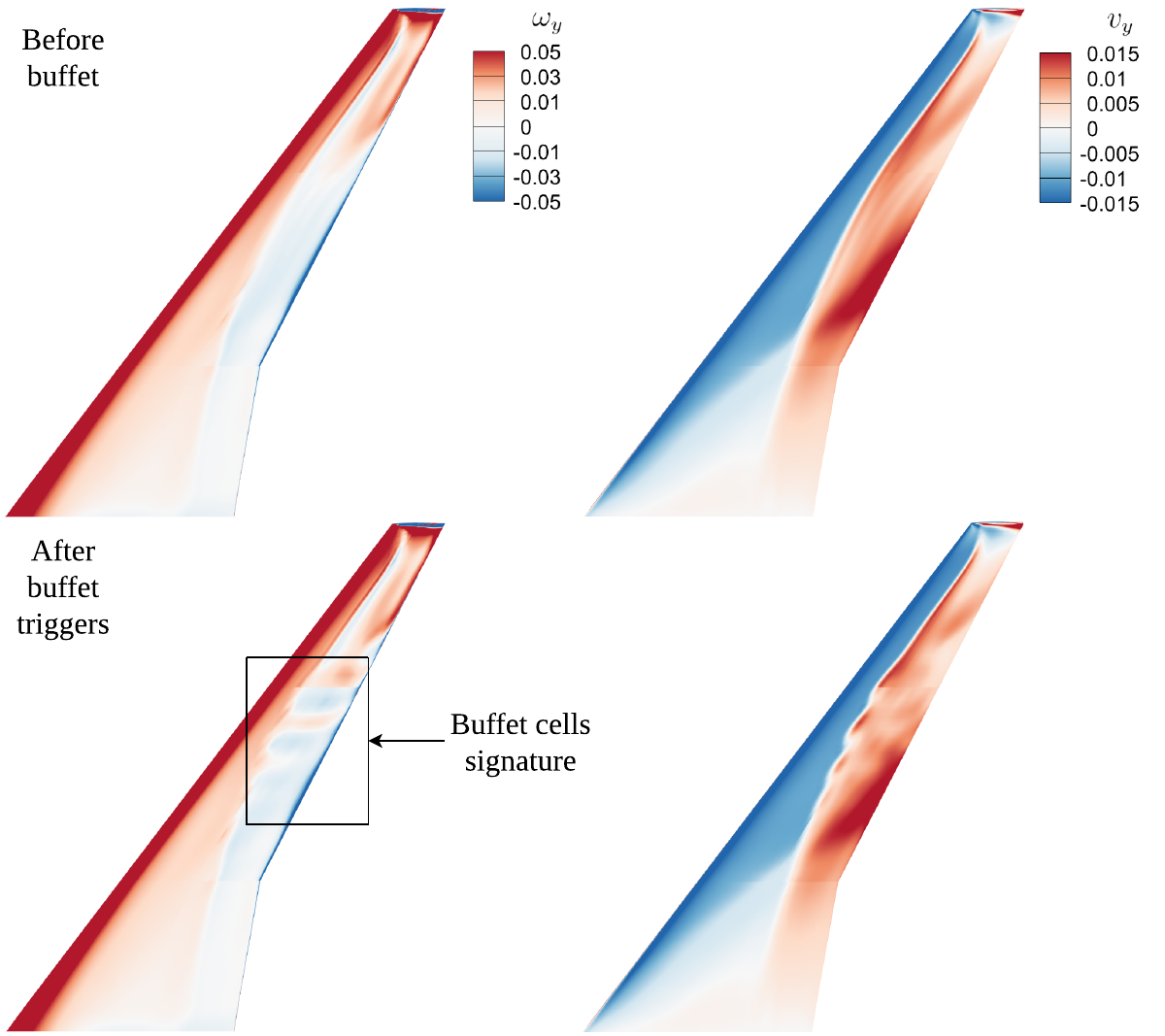}
  \caption{Surface--mounted contours on the wing--only CRM at $M = 0.85$, $Re = 5\times 10^{6}$, $\alpha = 4.2^\circ$, comparing a pre--buffet snapshot (top, before the limit cycle has developed) with a post--buffet snapshot (bottom, $t = 0.45$ s, inside the saturated limit cycle). Left column: spanwise vorticity $\omega_y$. Right column: spanwise velocity $v_y$. The boxed region in the post--buffet row highlights the discrete buffet--cell signature that propagates spanwise toward the wing tip.}
  \label{fig:crm_lco_omegay_vy}
\end{figure}

We zoom in on the highest--angle case, $\alpha = 4.2^\circ$, which is well inside the saturated limit--cycle oscillation regime by the end of the run.
\Cref{fig:crm_lco_omegay_vy} compares surface--mounted contours of the spanwise vorticity component $\omega_y$ and the spanwise velocity component $v_y$ on the wing, with $y$ taken as the spanwise direction, at a pre--buffet snapshot (top row, before the limit cycle has developed) and at a post--buffet snapshot (bottom row, $t = 0.45$ s along the $C_L$ trace of~\Cref{fig:crm_cl_vs_time}, well inside the saturated limit cycle).\footnote{\label{fn:crm_video}\url{https://sichenghe.github.io/assets/videos/transonic-buffet-3d-crm.mp4}}

% [x] TODO HS-: remove the black arrows, show the box for both figures

The post--buffet panels show a pronounced spanwise modulation that is absent from the pre--buffet panels, with discrete buffet cells highlighted by the box overlay; this case showcases buffet cells that propagate spanwise toward the wing tip~\cite{Ohmichi2026}.
The $v_y$ snapshot in addition reveals a streamwise/chordwise oscillation pattern that, locally on the upper surface, looks essentially identical to the periodic shock motion captured by the OAT15A density contours of~\Cref{fig:cl_plot_contour_plot}.
\Cref{fig:crm_lco_qcrit_cp} presents the same instant from two complementary diagnostic angles: a $Q$--criterion isosurface at $Q = 0.015$ that traces the coherent vortical structures shed and reformed during the cycle (left) and the surface $C_p$ field with the shockwave isosurface from ADflow's shock sensor (right), both showing the spanwise variation of the buffet pattern across the wing.
A video that animates the full saturated cycle for this case is available online~\footref{fn:crm_video}.
% [x] TODO HS-: footnote shown in the prev page?

\section{Conclusions}
\label{sec:conclusions}
% [x] TODO HS-: remove refs. in general no refs in the conclusion.
%%% Method and first-principles framing
% [x] TODO HS-: be bold. put the key info to the first sentences of the paras and put key contribution to this paragraph. stress what is first time done. highlight it.
We presented the {\emph{\textbf{first}}} aerodynamic shape optimization in which buffet onset is enforced through a first--principles linear--stability eigenvalue constraint computed by a fully discrete adjoint, replacing the separation metrics, lift--curve--break offsets, and surrogate stability proxies used in all prior buffet--constrained shape optimization work. %[x] TODO HS-: move this up front.
The enabling contribution is a linear stability coupled adjoint that computes, in a single back--substitution sweep, the sensitivity of the dominant buffet eigenvalue to a large number of shape and operating--condition design variables; this extends earlier eigenvalue--sensitivity frameworks, which were restricted to base--flow modifications, to the full aerodynamic shape--optimization setting.
The method exposes both eigenvalue and eigenvector derivatives and computes the Jacobian--vector products by a mixed algorithmic--differentiation and complex--step formulation that returns near--machine--precision derivatives, eliminating the per--design--variable finite--difference cost that has so far limited eigenvalue--constrained shape optimization to a handful of design parameters.

%%% Verification and validation
The implementation was verified end--to--end on three benchmarks of increasing complexity. On the canonical cylinder vortex--shedding benchmark the solver recovered the Hopf bifurcation to within a few hundredths of the published critical Reynolds number and tracked the published Strouhal--number curve.
On the OAT15A transonic airfoil the LST eigenspectrum reproduced the published buffet--bifurcation crossing between three and a half and four degrees of incidence, and the linear--growth phase of an independent unsteady RANS run agreed with the LST prediction to within five percent in both growth rate and frequency.
The CFD adjoint derivatives matched central finite differences across eight design variables on a perturbed cylinder geometry with high accuracy.

%%% Optimization demonstration and three-dimensional preview
Applied end--to--end to the OAT15A at its buffet flight condition, the framework cut drag by twenty--two percent (one hundred and twenty--four counts) while turning an unstable buffet mode into a stable equilibrium, with the eigenvalue constraint, rather than any heuristic, driving the design onto the buffet--stability boundary.
The stabilization is independently confirmed by a post--optimization eigensolve at three different solver settings and by an independent unsteady RANS run from the optimized steady state that exhibits no growth and no limit cycle, providing the first end--to--end demonstration that a first--principles eigenvalue constraint can deliver a verified buffet--stable optimum.
As a stepping stone toward three--dimensional buffet--stability studies, we also exercised the CFD pipeline on the wing--only NASA CRM at the cruise flight condition, recovering buffet onset at approximately four degrees of incidence with a clear spanwise buffet--cell signature inside the saturated limit cycle; extending the proposed adjoint to this configuration is left as future work.

\bibliographystyle{new-aiaa}
\bibliography{bib/mdolab,bib/references,bib/agi}

\appendix

\section{Real Formulation of the Eigenvalue Problem}
\label{app:real_formulation}

The complex eigenvalue problem Eqs.~\eqref{eq:eigen_govern} and \eqref{eq:normalization} can be decomposed into real and imaginary components.
The resulting system of equations is written in terms of real numbers only as
\begin{equation}
\hat{\mb{r}}(\mb{v}; \mb{w}_0, \mb{x})=
\begin{bmatrix}
\hat{\mb{r}}_r\\
\hat{\mb{r}}_i\\
\hat{r}_m \\
\hat{r}_p \\
\end{bmatrix}
=
\begin{bmatrix}
\mb{J} \mb{q}_r - \lambda_r \mb{q}_r + \lambda_i \mb{q}_i\\
\mb{J} \mb{q}_i - \lambda_r \mb{q}_i - \lambda_i \mb{q}_r\\
\mb{q}_r^\intercal \mb{q}_r + \mb{q}_i^\intercal \mb{q}_i - 1\\
\mb{e}_k^\intercal \mb{q}_i
\end{bmatrix}, \quad
\mb{v}
=
\begin{bmatrix}
\mb{q}_r\\
\mb{q}_i\\
\lambda_r \\
\lambda_i \\
\end{bmatrix}.
\end{equation}
The subscripts $r$ and $i$ denote the real and imaginary parts of the eigenvalue equation, respectively; the subscripts $m$ and $p$ denote the magnitude and the phase residual, respectively.
More details on the eigenvalue problem setup are provided by He et al.~\cite{He2022ua}.

\section{Top-Level Adjoint Details}
\label{app:top_adjoint}

The coefficient matrix for the top-level adjoint equation Eq.~\eqref{eq:cadjoint_top} is
\begin{equation}
\f{\p \hat{\mb{r}}}{\p \mb{v}}^\intercal =
\begin{bmatrix}
\mb{J}  - \lambda_r \mb{I} & \lambda_i \mb{I} & - \mb{q}_r & \mb{q}_i \\
- \lambda_i \mb{I} & \mb{J}  - \lambda_r \mb{I} & - \mb{q}_i & - \mb{q}_r \\
2 \mb{q}_r^\intercal & 2 \mb{q}_i^\intercal & 0 & 0 \\
0 & \mb{e}_k^\intercal & 0 & 0
\end{bmatrix}^\intercal.
\end{equation}

When $f = \lambda_r$, we have ${\partial f}/{\partial \mb{v}} = [0, 0, 1, 0]$.
When a function of the eigenvector is used, such as Eq.~\eqref{eq:obj_eigvec}, the partial derivative is ${\partial f}/{\partial \mb{v}} = [\tilde{\mb{q}}_r^\intercal, \tilde{\mb{q}}_i^\intercal, 0, 0]$.

For $f = \lambda_r$, the analytic adjoint solution is given by Eq.~\eqref{eq:analytic_top_sol_1} in the main text.
If $\tilde{\mb{u}}$ is any (unnormalized) left eigenvector, then enforcing $\mb{u}^* \mb{q}=-1$ is achieved by the scaling
\begin{equation}
\mb{u}=\frac{\tilde{\mb{u}}}{-\overline{\tilde{\mb{u}}^* \mb{q}}}.
\end{equation}

Similarly, when $f = \lambda_i$ (relevant when certain frequencies need to be avoided~\cite{Zhu2022}), the adjoint solution is
\begin{equation}
\label{eq:analytic_top_sol_2}
\pmb{\psi}_{\hat{\mb{r}}} =
\begin{bmatrix}
{\mb{u}^{(2)}_r}^\intercal &
{\mb{u}^{(2)}_i}^\intercal &
0 &
0
\end{bmatrix}^\intercal,
\end{equation}
where $\mb{u}^{(2)}$ satisfies $\mb{J}^\intercal \mb{u}^{(2)} = \lambda^* \mb{u}^{(2)}$ with normalization ${\mb{u}^{(2)}}^* \mb{q} = i$.

\section{Bottom-Level Adjoint Details}
\label{app:bot_adjoint}

The key term in the bottom-level adjoint equation Eq.~\eqref{eq:cadjoint_bot} is the partial derivative $\left({\partial \hat{\mb{r}}}/{\partial \mb{w}_0}\right)^\intercal\pmb{\psi}_{\hat{\mb{r}}}$, which can be expanded as
\begin{equation}
\label{eq:RHS_partial}
\frac{\partial \hat{\mb{r}}}{\partial \mb{w}_0}^\intercal\pmb{\psi}_{\hat{\mb{r}}} = \frac{\partial \mb{q}_r^\intercal{\mb{J}}^\intercal\pmb{\psi}_{\hat{\mb{r}},r}}{\partial \mb{w}_0} + \frac{\partial \mb{q}_i^\intercal{\mb{J}}^\intercal\pmb{\psi}_{\hat{\mb{r}},i}}{\partial \mb{w}_0},
\end{equation}
where $\pmb{\psi}_{\hat{\mb{r}},r}, \pmb{\psi}_{\hat{\mb{r}},i} \in \mathbb{R}^n$ are sub-vectors of
\begin{equation}
\pmb{\psi}_{\hat{\mb{r}}} =
\begin{bmatrix}
\pmb{\psi}_{\hat{\mb{r}},r}^\intercal &
\pmb{\psi}_{\hat{\mb{r}},i}^\intercal &
{\psi}_{\hat{\mb{r}},m} &
{\psi}_{\hat{\mb{r}},p}
\end{bmatrix}^\intercal,
\end{equation}
and ${\psi}_{\hat{\mb{r}},m}, {\psi}_{\hat{\mb{r}},p}\in \mathbb{R}$ are scalars related to the magnitude and phase residuals of Eq.~\eqref{eq:normalization}.

Applying the FDRAD formula Eq.~\eqref{eq:jac_der_FD} to Eq.~\eqref{eq:RHS_partial} yields
\begin{equation}
\frac{\partial \hat{\mb{r}}}{\partial \mb{w}_0}^\intercal \pmb{\psi}_{\hat{\mb{r}}}
\approx
\left(\frac{\left(\frac{\partial \mb{r}(\mb{w}_0 + h \mb{q}_r)}{\partial \mb{w}}\right)^\intercal \pmb{\psi}_{\hat{\mb{r}},r}
- \left(\frac{\partial \mb{r}(\mb{w}_0)}{\partial \mb{w}}\right)^\intercal \pmb{\psi}_{\hat{\mb{r}},r}}{h}\right)^\intercal
+
\left(\frac{\left(\frac{\partial \mb{r}(\mb{w}_0 + h \mb{q}_i)}{\partial \mb{w}}\right)^\intercal \pmb{\psi}_{\hat{\mb{r}},i}
- \left(\frac{\partial \mb{r}(\mb{w}_0)}{\partial \mb{w}}\right)^\intercal \pmb{\psi}_{\hat{\mb{r}},i}}{h}\right)^\intercal.
\end{equation}
The terms $\left({\p\mb{r}(\mb{w}_0 + h \mb{r}_2)}/{\p \mb{w}}\right)^\intercal \mb{r}_1$ and $\left({\p\mb{r}(\mb{w}_0)}/{\p \mb{w}}\right)^\intercal \mb{r}_1$ can be computed using RAD.
The CDRAD and CSRAD expansions follow analogously.

\section{Mass--Matrix Form of the Linearized Unsteady Operator}
\label{app:mass-matrix}

ADflow stores the flow state in the primitive--like vector
\begin{equation}
\mb{w} = [\rho,\; u,\; v,\; w,\; \rho E]^\intercal \in \mathbb{R}^{n},
\end{equation}
where $\rho$ is the density, $(u,v,w)$ are the Cartesian velocity scalars, $\rho E$ is the volumetric total energy, and $n$ is the number of degrees of freedom; the corresponding vector of conservative variables is
\begin{equation}
\mb{u}_c(\mb{w}) = [\rho,\; \rho u,\; \rho v,\; \rho w,\; \rho E]^\intercal \in \mathbb{R}^{n}.
\end{equation}
The semi--discrete equations consistent with ADflow's residual scaling read
\begin{equation}
\f{\mathrm{vol}}{\mathrm{volRef}}\,\f{\p \mb{u}_c(\mb{w})}{\p t} + \f{1}{\mathrm{volRef}}\,\mb{r}_s(\mb{w}) = \mb{0},
\end{equation}
where $\mb{r}_s\in\mathbb{R}^{n}$ is the spatial residual and $\mathrm{vol}/\mathrm{volRef}$ is the cell--volume scaling.
Linearizing about a steady base state $\mb{w}_0\in\mathbb{R}^{n}$ with perturbation $\mb{w}'\in\mathbb{R}^{n}$ and assuming $\mb{w}'(t)=\mb{q}\,\mathrm{e}^{\lambda t}$ with eigenvector $\mb{q}\in\mathbb{C}^{n}$ and eigenvalue $\lambda\in\mathbb{C}$ yields the generalized eigenproblem
\begin{equation}
\mb{J}\,\mb{q} = -\lambda\,\mb{M}\,\mb{q},
\qquad
\mb{J} \coloneqq \f{1}{\mathrm{volRef}}\left.\f{\p \mb{r}_s}{\p \mb{w}}\right|_{\mb{w}_0},
\qquad
\mb{M} \coloneqq \f{\mathrm{vol}}{\mathrm{volRef}}\left.\f{\p \mb{u}_c}{\p \mb{w}}\right|_{\mb{w}_0},
\end{equation}
with $\mb{J},\mb{M}\in\mathbb{R}^{n\times n}$.
Upon expanding $\p\mb{u}_c/\p\mb{w}$ for the five mean--flow variables one obtains
\begin{equation}
\f{\p \mb{u}_c}{\p \mb{w}}
=
\begin{bmatrix}
1 & 0 & 0 & 0 & 0\\
u & \rho & 0 & 0 & 0\\
v & 0 & \rho & 0 & 0\\
w & 0 & 0 & \rho & 0\\
0 & 0 & 0 & 0 & 1
\end{bmatrix},
\end{equation}
so $\mb{M}\neq\mb{I}$ in compressible flow with primitive variables; the Spalart--Allmaras working variable enters $\mb{w}$ and $\mb{u}_c$ identically and contributes an identity sub--block under the same $\mathrm{vol}/\mathrm{volRef}$ scaling.
The off--diagonal entries between $\rho$ and the momenta encode the conservative time derivatives $\p(\rho u)/\p t = \rho\,\p u/\p t + u\,\p\rho/\p t$ (and analogously for the other two components), and these couplings are negligible only when the density perturbation $\rho'$ is small (e.g., low--Mach cylinder flow); they are significant in transonic buffet, where density fluctuations are nontrivial.
In the incompressible limit, $\rho$ is constant and not a dynamic variable, so $\rho'=0$ and the continuity equation enforces $\nabla\cdot\mb{u}'=\mb{0}$ on the velocity perturbation $\mb{u}'$; the conservative time derivative collapses to a constant diagonal mass operator that scales out of the eigenproblem, which is why a standard EVP with $\mb{M}=\mb{I}$ recovers low--Mach cylinder shedding eigenvalues to within plotting accuracy of the generalized form.

The same dynamics can also be written directly in conservative variables.
Introducing the conservative perturbation $\mb{u}_c' = \mb{M}\,\mb{w}' \in \mathbb{R}^{n}$ and viewing $\mb{r}_s$ locally as $\mb{r}_s(\mb{w}(\mb{u}_c))$, the linearized dynamics in $\mb{u}_c'$ become
\begin{equation}
\f{\p \mb{u}_c'}{\p t} = \mb{J}_2\,\mb{u}_c',
\qquad
\mb{J}_2 \coloneqq -\left.\f{\p \mb{r}_s}{\p \mb{u}_c}\right|_{\mb{u}_{c,0}},
\end{equation}
where $\mb{u}_{c,0}=\mb{u}_c(\mb{w}_0)$.
The corresponding primitive--variable system, obtained by left--multiplying $\mb{M}\p\mb{w}'/\p t + \mb{J}\mb{w}'=\mb{0}$ by $\mb{M}^{-1}$, is
\begin{equation}
\f{\p \mb{w}'}{\p t} = \mb{J}_1\,\mb{w}',
\qquad
\mb{J}_1 \coloneqq -\mb{M}^{-1}\mb{J}.
\end{equation}
Upon differentiating $\mb{u}_c'=\mb{M}\mb{w}'$ in time and substituting the primitive evolution, one finds $\mb{J}_2 = \mb{M}\mb{J}_1\mb{M}^{-1}$, so $\mb{J}_1$ and $\mb{J}_2$ are related by a similarity transformation: they share the same eigenvalues and their eigenvectors are connected by $\mb{u}_c' = \mb{M}\mb{w}'$.
The two formulations are therefore equivalent representations of the same linearized physics, valid wherever the primitive--to--conservative mapping is locally invertible.
The primitive-variable form introduces the matrix \(\mathbf{M} = \partial \mathbf{U}/\partial \mathbf{w}\), while the conservative-variable form absorbs this transformation into the system operator.
As a result, the two linearizations are mathematically equivalent and differ only in the chosen state representation.

\section{Linear Stability Analysis with a Mass Matrix}
\label{sec:lst_mass_matrix}

This appendix specializes the coupled adjoint of~\Cref{sec:ls_adjoint} to the LST problem produced by ADflow, where the semi--discrete system carries a non--trivial mass matrix $\mb{M}=\p\mb{u}_c/\p\mb{w}$ relating the conservative state $\mb{u}_c$ to the primitive state $\mb{w}$ (both in $\mathbb{R}^{n}$), as derived in~\Cref{app:mass-matrix}.
Including the Spalart--Allmaras working variable, the cell--local block of $\mb{M}$ takes the lower--triangular form
\begin{equation}
\label{eq:mm_block}
\mb{M}_\mathrm{cell}
=
\begin{bmatrix}
1 & 0 & 0 & 0 & 0 & 0 \\
u & \rho & 0 & 0 & 0 & 0 \\
v & 0 & \rho & 0 & 0 & 0 \\
w & 0 & 0 & \rho & 0 & 0 \\
0 & 0 & 0 & 0 & 1 & 0 \\
0 & 0 & 0 & 0 & 0 & 1
\end{bmatrix},
\end{equation}
where $u,v,w$ are the Cartesian velocity scalars and $\rho$ is the density.
This block has determinant $\rho^3 \neq 0$ for physically meaningful flows, so $\mb{M}^{-1}$ is available cell--locally in closed form at $\mathcal{O}(n)$ cost.
Upon absorbing $\mb{M}^{-1}$ into the Jacobian via the modified operator
\begin{equation}
\label{eq:mm_jmod}
\mb{J}_\mathrm{mod} \coloneqq -\mb{M}^{-1}\mb{J} \in \mathbb{R}^{n\times n},
\end{equation}
the generalized eigenvalue problem reduces to the standard form $\mb{J}_\mathrm{mod}\,\mb{q}=\lambda\,\mb{q}$ with the same eigenvalues $\lambda\in\mathbb{C}$ and right eigenvectors $\mb{q}\in\mathbb{C}^{n}$, and the coupled adjoint of~\Cref{sec:ls_adjoint} can be reused verbatim with $\mb{J}_\mathrm{mod}$ in place of $\mb{J}$ and the standard normalization $\mb{q}^*\mb{q}=1$ in place of the $\mb{M}$--weighted one.

For the function of interest $f=\lambda_r$ (the real part of the eigenvalue), the top--level adjoint admits the analytic solution of~\Cref{eq:analytic_top_sol_1} once the left eigenvector $\mb{u}\in\mathbb{C}^{n}$ of $\mb{J}_\mathrm{mod}$ is recovered from the standard left eigenproblem
\begin{equation}
\label{eq:mm_left_eig}
\mb{J}_\mathrm{mod}^\intercal \mb{u} = \lambda^* \mb{u}, \quad \mb{u}^* \mb{q} = -1;
\end{equation}
neither a generalized left--eigenvector solve nor an $\mb{M}$--weighted normalization is needed.

The bottom--level adjoint right--hand side requires more care, because the CDRAD identity~\Cref{eq:jac_der_CD} applies only to true Jacobians and $\mb{J}_\mathrm{mod}=-\mb{M}^{-1}(\mb{w})\,\mb{J}(\mb{w})$ is not one.
Let $\mb{r}_1,\mb{r}_2\in\mathbb{R}^{n}$ denote two generic real vectors that arise as the seed and weight of a typical Jacobian--vector form $\mb{r}_2^\intercal\mb{J}_\mathrm{mod}^\intercal\mb{r}_1$.
Defining the auxiliary base--state vector $\pmb{\eta}_0\coloneqq\mb{M}^{-\intercal}(\mb{w}_0)\,\mb{r}_1$ and applying the product rule, the term $\p(\mb{r}_2^\intercal\mb{J}_\mathrm{mod}^\intercal\mb{r}_1)/\p\mb{w}_0$ splits into a CDRAD--friendly piece $\mb{b}_\mathrm{J}$ and an analytic mass--matrix piece $\mb{b}_\mathrm{M}$.
The first reuses ADflow's reverse--mode Jacobian--vector product with $\pmb{\eta}_0$ held fixed,
\begin{equation}
\label{eq:mm_bj}
\mb{b}_\mathrm{J}
\approx
-\left(\f{\mb{J}(\mb{w}_0+h\mb{r}_2)^\intercal \pmb{\eta}_0 - \mb{J}(\mb{w}_0-h\mb{r}_2)^\intercal \pmb{\eta}_0}{2h}\right)^\intercal,
\end{equation}
where $h>0$ is the user--chosen complex--step / finite--difference step size.
The second piece, after using the eigenvalue relation $\mb{J}(\mb{w}_0)(\mb{q}_r+\mathrm{j}\mb{q}_i)=-\lambda\mb{M}(\mb{w}_0)(\mb{q}_r+\mathrm{j}\mb{q}_i)$ (with $\mb{q}_r,\mb{q}_i\in\mathbb{R}^{n}$ the real and imaginary parts of the eigenvector and $\lambda=\lambda_r+\mathrm{j}\lambda_i$) to eliminate $\mb{M}^{-1}\mb{J}\mb{q}_{r,i}$, closes into
\begin{equation}
\label{eq:mm_bm}
\begin{aligned}
\mb{b}_\mathrm{M}[k]
=
&-\lambda_r \bigl[\pmb{\eta}_{0,r}^\intercal (\p \mb{M}/\p w_k)\,\mb{q}_r + \pmb{\eta}_{0,i}^\intercal (\p \mb{M}/\p w_k)\,\mb{q}_i\bigr] \\
&+\lambda_i \bigl[\pmb{\eta}_{0,r}^\intercal (\p \mb{M}/\p w_k)\,\mb{q}_i - \pmb{\eta}_{0,i}^\intercal (\p \mb{M}/\p w_k)\,\mb{q}_r\bigr],
\end{aligned}
\end{equation}
where $\pmb{\eta}_{0,r}\coloneqq\mb{M}^{-\intercal}(\mb{w}_0)\pmb{\psi}_{\hat{\mb{r}},r}$, $\pmb{\eta}_{0,i}\coloneqq\mb{M}^{-\intercal}(\mb{w}_0)\pmb{\psi}_{\hat{\mb{r}},i}$ are the real and imaginary parts of the top--level adjoint vector $\pmb{\psi}_{\hat{\mb{r}}}\in\mathbb{C}^{n}$ pulled back through $\mb{M}^{-\intercal}$, the index $k=1,\ldots,n$ labels the entries of $\mb{w}_0$, and each bilinear form $\mb{a}^\intercal(\p\mb{M}/\p w_k)\mb{b}$ for $\mb{a},\mb{b}\in\mathbb{R}^{n}$ is evaluated cell--locally in closed form from~\Cref{eq:mm_block}.
The full bottom--level right--hand side then assembles as
\begin{equation}
\label{eq:mm_b}
\mb{b}
=
\mb{b}_\mathrm{M}(\mb{q}_r,\mb{q}_i;\,\pmb{\eta}_{0,r},\pmb{\eta}_{0,i})
+ \mb{b}_\mathrm{J}(\mb{q}_r;\,\pmb{\eta}_{0,r})
+ \mb{b}_\mathrm{J}(\mb{q}_i;\,\pmb{\eta}_{0,i}).
\end{equation}
Throughout, transposed products with the modified operator must be evaluated as $\mb{J}_\mathrm{mod}^\intercal\mb{v}=-\mb{J}^\intercal(\mb{M}^{-\intercal}\mb{v})$, applying $\mb{M}^{-\intercal}$ first cell--locally and then feeding the result as the adjoint seed to ADflow's reverse Jacobian--vector product; the reverse order is not equivalent because $\mb{M}^{-1}$ and $\mb{J}$ do not commute.

The bottom--level adjoint equation $(\p\mb{r}/\p\mb{w}_0)^\intercal\pmb{\psi}_{\mb{r}}=\mb{b}$, with bottom--level adjoint $\pmb{\psi}_{\mb{r}}\in\mathbb{R}^{n}$, uses the same operator as the mass--free case and reuses ADflow's existing solver; the mass--matrix dependence is confined to the assembly of $\mb{b}$.
The total derivative with respect to the design vector $\mb{x}\in\mathbb{R}^{n_x}$ is
\begin{equation}
\label{eq:mm_total}
\f{\d \lambda_r}{\d \mb{x}}
=
-\pmb{\psi}_{\mb{r}}^\intercal \f{\p \mb{r}}{\p \mb{x}}
-\pmb{\psi}_{\hat{\mb{r}}}^\intercal \f{\p \hat{\mb{r}}}{\p \mb{x}},
\end{equation}
with the second term evaluated by central finite differences of the scalar $s(\mb{x})=-\mb{q}_r^\intercal\mb{J}_\mathrm{mod}^\intercal\pmb{\psi}_{\hat{\mb{r}},r}-\mb{q}_i^\intercal\mb{J}_\mathrm{mod}^\intercal\pmb{\psi}_{\hat{\mb{r}},i}$ at $(\mb{w}_0,\mb{x}\pm h\mb{e}_i)$, where $\mb{e}_i\in\mathbb{R}^{n_x}$ is the $i$--th canonical basis vector of design space and $\mb{M}^{-\intercal}(\mb{w}_0)$ is held fixed; no explicit $\p\mb{M}/\p\mb{x}$ term arises because $\mb{M}$ depends on $\mb{w}_0$ alone.
The complete procedure is given in~\Cref{alg.derivative_mm}; it is structurally identical to~\Cref{alg.derivative} except that every transposed Jacobian--vector product is evaluated with $\mb{M}^{-\intercal}$ applied first to the adjoint seed, and the bottom--level right--hand side adds the analytic $\p\mb{M}/\p\mb{w}_0$ contribution $\mb{b}_\mathrm{M}$ on top of the pure--$\mb{J}$ CDRAD contribution $\mb{b}_\mathrm{J}$.

\begin{algorithm}[H]
\begin{spacing}{1.5}
\caption{Local stability derivative using $\mb{J}_\mathrm{mod}=-\mb{M}^{-1}\mb{J}$.}
\label{alg.derivative_mm}
\begin{algorithmic}[1]
\Function{$\mb{g}_\mathrm{stab}^{\,\mathrm{mod}}$}{$\mb{x}$}
\State $\f{\p \lambda_r}{\p \mb{v}} = \begin{bmatrix} 0 & 0 & 1 & 0 \end{bmatrix}^\intercal$ \Comment{Set the RHS for the top-level adjoint equation.}
\State $\pmb{\psi}_{\hat{\mb{r}}}\leftarrow \f{\p \hat{\mb{r}}}{\p \mb{v}}^\intercal \pmb{\psi}_{\hat{\mb{r}}}=\f{\p \lambda_r}{\p \mb{v}}$ \Comment{Solve via the analytic left eigenvector of $\mb{J}_\mathrm{mod}$, \cref{eq:mm_left_eig}.}
\State $\mb{b} \leftarrow \f{\p \hat{\mb{r}}}{\p \mb{w}_0}^\intercal \pmb{\psi}_{\hat{\mb{r}}}$ \Comment{Assemble $\mb{b}_\mathrm{M}+\mb{b}_\mathrm{J}$ using \cref{eq:mm_bm,eq:mm_b}.}
\State $\pmb{\psi}_{\mb{r}} \leftarrow \f{\p \mb{r}}{\p \mb{w}_0}^\intercal \pmb{\psi}_{\mb{r}}=\mb{b}$ \Comment{Solve the bottom-level adjoint, \cref{eq:cadjoint_bot}.}
\State $\f{\d \lambda_r}{\d \mb{x}} \leftarrow -\pmb{\psi}_{\mb{r}}^\intercal \f{\p \mb{r}}{\p \mb{x}} - \pmb{\psi}_{\hat{\mb{r}}}^\intercal \f{\p \hat{\mb{r}}}{\p \mb{x}}$ \Comment{Total derivative via~\cref{eq:mm_total}.}\\
\Return $\f{\d \lambda_r}{\d \mb{x}}$
\EndFunction
\end{algorithmic}
\end{spacing}
\end{algorithm}

\section{Eigenvalue Units and Non--dimensionalization}
\label{app:eig_conversion}

The eigenvalues returned by ADflow are nondimensional under the convention $t \to t\,\sqrt{R T_\infty}/L$, where $R = 287\ \mathrm{J/(kg\,K)}$ is the specific gas constant, $T_\infty$ is the freestream static temperature, and $L = 1\ \mathrm{m}$ is the reference length used in the simulation.
The dimensional growth rate $\sigma$ in $1/\mathrm{s}$ and the dimensional angular frequency $\omega$ in $\mathrm{rad/s}$ map to our nondimensional eigenvalue components as
\begin{equation}
\label{eq:our_nondim}
\sigma_a = \sigma\,\f{L}{\sqrt{R T_\infty}},
\qquad
\omega_a = \omega\,\f{L}{\sqrt{R T_\infty}}.
\end{equation}
\citet{Sartor2015} report the growth rate $\sigma_\mathrm{paper}$ in $1/\mathrm{s}$ and the frequency as a Strouhal number $S_L = f\,c/U_\infty$ based on chord $c$ and freestream velocity $U_\infty$.

The clean quantity to compare between the two conventions is the convective nondimensional growth rate $\sigma_c \coloneqq \sigma\,L/U_\infty$.
Substituting the first part of~\Cref{eq:our_nondim} into this definition and using $U_\infty = M\sqrt{\gamma R T_\infty}$ gives
\begin{equation}
\label{eq:sigma_c_us}
\sigma_c = \f{\sigma_a}{M\sqrt{\gamma}},
\end{equation}
while the same convective scaling applied to the paper's growth rate gives $\sigma_{c,\mathrm{paper}} = \sigma_\mathrm{paper}\,c/U_\infty$.
Equating $\sigma_c = \sigma_{c,\mathrm{paper}}$ yields the conversion
\begin{equation}
\label{eq:sigma_conversion}
\sigma_a = \sigma_\mathrm{paper}\,\f{c}{U_\infty}\,M\sqrt{\gamma}.
\end{equation}
For Sartor's flight conditions ($c = 0.23\ \mathrm{m}$, $U_\infty = 240.93\ \mathrm{m/s}$, $M = 0.73$, $\gamma = 1.4$), the conversion factor evaluates to $\sigma_a / \sigma_\mathrm{paper} = 8.246\times 10^{-4}$.

For the Strouhal number, equating the dimensional angular frequency from~\Cref{eq:our_nondim} with $\omega = 2\pi f$ and Sartor's $S_L = f\,c/U_\infty$ under a consistent unit--chord normalization gives
\begin{equation}
\label{eq:omega_conversion}
\omega_a = 2\pi\,M\sqrt{\gamma}\,S_L,
\end{equation}
so $\omega_a / S_L = 5.428$ for the same flight conditions.
The two relations~\Cref{eq:sigma_conversion,eq:omega_conversion} are applied to every computed eigenvalue before plotting in~\Cref{fig:oat15a_lst_spectrum}.

\section{Eigenvalue Verification from the Unsteady Linear--Growth Phase}
\label{app:lst_unsteady_verification}

The dominant LST eigenvalue can be recovered from a URANS time history of the same base state during its initial linear--growth phase, before nonlinear saturation sets in.
This appendix details the peak--fit procedure used to produce~\Cref{fig:oat15a_lst_validation} and to obtain the unsteady $\sigma_\mathrm{unsteady}$ and $f_\mathrm{unsteady}$ values reported in~\Cref{sec:results}.

Let $\mb{w}_0$ denote a marginally unstable equilibrium and $\lambda = \sigma + i\omega$ the eigenvalue with the largest real part of the linearized operator at $\mb{w}_0$, with corresponding eigenvector $\mb{q}$.
For a small perturbation $\mb{w}'(0)$ that has a non--vanishing projection onto $\mb{q}$, the linear evolution is
\begin{equation}
\label{eq:lin_growth_state}
\mb{w}'(t) = \mathrm{Re}\!\left[c\,\mb{q}\,\mathrm{e}^{\lambda t}\right] + \text{(decaying components)},
\end{equation}
where $c\in\mathbb{C}$ is set by the initial projection.
Any real--valued linear functional of the perturbation, in particular the lift fluctuation $C_L^{\prime}(t) \coloneqq C_L(t) - \overline{C_L}$, inherits the same time dependence,
\begin{equation}
\label{eq:cl_growth}
C_L^{\prime}(t) = A_0\,\mathrm{e}^{\sigma t}\cos(\omega t + \phi),
\end{equation}
where $A_0$ and $\phi$ depend on the initial condition.

The successive local maxima $t_1 < t_2 < \cdots$ of $C_L^{\prime}$ satisfy $\cos(\omega t_n + \phi) = 1$, so the upper envelope at the peaks reads
\begin{equation}
\label{eq:envelope}
C_L^{\prime}(t_n) = A_0\,\mathrm{e}^{\sigma t_n}.
\end{equation}
Taking the natural logarithm of~\Cref{eq:envelope} gives a linear relation
\begin{equation}
\label{eq:log_envelope}
\ln \lvert C_L^{\prime}(t_n)\rvert = \ln A_0 + \sigma\,t_n,
\end{equation}
so a least--squares fit of $\ln\lvert C_L^{\prime}(t_n)\rvert$ against $t_n$ recovers the unsteady growth rate $\sigma_\mathrm{unsteady}$ as the slope.
The successive peak intervals $\Delta t_n = t_{n+1}-t_n$ all equal $T = 2\pi/\omega$ in the linear regime, so the angular frequency is recovered as $\omega_\mathrm{unsteady} = 2\pi/\overline{\Delta t}$ and the physical frequency as $f_\mathrm{unsteady} = 1/\overline{\Delta t}$.

The fit is performed only on the linear--growth window where $\sigma$ is approximately constant; outside this window the dynamics are either dominated by initial--transient decay (early times) or by nonlinear saturation as the limit cycle is approached (late times).
The window is selected automatically by sweeping the candidate start time and length, fitting~\Cref{eq:log_envelope} on each candidate, and choosing the window with the highest coefficient of determination $R^2$ subject to a minimum peak count and a slope--stability check across $\pm$ small shifts in the start time.
For the OAT15A run of~\Cref{fig:oat15a_cl_vs_time} with $\Delta t = 5\times 10^{-5}$ s, this procedure selects $t \in [1.4001,\,1.8001]$ s, identifies seven peaks at $R^2 = 1.0000$, and returns $\sigma_\mathrm{unsteady} = 6.16\ \mathrm{s}^{-1}$ and $f_\mathrm{unsteady} = 18.37$ Hz; the demeaned signal with the detected peaks and the log--linear fit are shown in~\Cref{fig:oat15a_lst_validation}.
Converting the L0 LST eigenvalue $\lambda = 0.021206 + 0.391287\,\mathrm{j}$ to physical units via~\Cref{eq:our_nondim} of~\Cref{app:eig_conversion} gives $\sigma_\mathrm{LST} = 6.23\ \mathrm{s}^{-1}$ and $f_\mathrm{LST} = 18.27$ Hz, matching the unsteady values to within $1.1\%$ in growth rate and $0.5\%$ in frequency.

\clearpage
\end{document}